\newcommand{\chunk}[2]{%
\fcolorbox{black}{yellow}{\bfseries\sffamily\scriptsize#1}%
{$\blacktriangleright$#2$\blacktriangleleft$}%
}

\newcommand{\aitor}[1]{\chunk{Aitor}{\textbf{\textcolor{red}{\textsl{#1}}}}}
\newcommand{\pablo}[1]{\chunk{Pablo}{\textbf{\textcolor{blue}{\textsl{#1}}}}}

\documentclass[acmsmall,screen]{acmart}
\usepackage{graphicx} % Required for inserting images
\usepackage{amsmath,amsfonts}
\usepackage{multirow}
\usepackage{algorithmic} 

\usepackage{graphicx}
\usepackage{textcomp}
\usepackage{xcolor}
\usepackage{booktabs}
\usepackage{adjustbox}
\usepackage{subcaption}
\usepackage{float}
\usepackage{graphicx}
\usepackage{caption}
\usepackage{subcaption}
\usepackage{tcolorbox}
\usepackage{makecell}

\usepackage{hyperref}

\usepackage[linesnumbered,algoruled,boxed,lined]{algorithm2e}
\usepackage{xcolor}
\usepackage{subcaption}
\usepackage[utf8]{inputenc}
\usepackage{flushend}
\SetKwRepeat{Do}{do}{while}
\let\oldnl\nl% Store \nl in \oldnl
\newcommand{\nonl}{\renewcommand{\nl}{\let\nl\oldnl}}% Remove line number for one line
\usepackage{float}
\newtcolorbox{custombox}[1]{
	colback=gray!10,
	colframe=gray!20,
	left=1mm,
	right=1mm,
	top=1mm,
	bottom=1mm,
	fonttitle=\bfseries,
	arc=2mm,
	leftrule=0mm,
	rightrule=.5mm,
	toprule=0mm,
	bottomrule=.5mm,
	notitle,
	before=\par\smallskip\noindent,
	before upper={\textbf{#1: } },
}

\newsavebox\CBox

\newcommand{\revision}[1]{{\color{black}#1}}
\title{Delta Debugging for Cyber-Physical Systems with Flaky Test Executions}
\author{Pablo Valle}
\email{pvalle@mondragon.edu}
\orcid{0000-0002-0588-316X}
\affiliation{%
  \institution{Mondragon University}
  \streetaddress{Goiru 2}
  \city{Mondragon}
  \state{Gipuzkoa}
  \country{Spain}
  \postcode{20500}
}

\author{Shaukat Ali}
\email{shaukat@simula.no}
\orcid{0000-0002-9979-3519}
\affiliation{%
  \institution{Simula Research Laboratory}
  \city{Oslo}
  \country{Norway}
}

\author{Aitor Arrieta}
\email{aarrieta@mondragon.edu}
\orcid{0000-0001-7507-5080}
\affiliation{%
  \institution{Mondragon University}
  \streetaddress{Goiru 2}
  \city{Mondragon}
  \state{Gipuzkoa}
  \country{Spain}
  \postcode{20500}
}

\date{January 2024}

\begin{document}
\begin{abstract}
Simulation-based testing is widely used to validate Cyber-Physical Systems (CPSs), yet modern CPS simulators frequently exhibit non-deterministic (``flaky'') behavior, making failures difficult to reproduce and debug. Although delta debugging has proven effective for deterministic systems, its underlying assumptions do not hold in stochastic environments. This paper presents three delta debugging algorithms that combine statistical failure analysis, repeated executions, and environment-aware reduction \revision{to produce reduced test inputs that preserve the target failure in stochastic CPSs}. We evaluate the proposed techniques on \revision{three} complementary case study systems: an industrial elevator dispatching system employing stochastic optimization, an autonomous mobile robot exhibiting simulator-induced non-determinism, \revision{and a VLA-controlled robotic manipulation system with stochastic action generation during policy inference}. The results show that the proposed approaches substantially reduce debugging time while preserving the original failure behavior. More importantly, we observe that minimizing failure-inducing test inputs frequently increases failure reproducibility compared with the original executions. By eliminating execution segments that introduce incidental stochastic effects, the reduced test inputs isolate the causal conditions of the failure and consistently reproduce it with higher probability. These findings suggest that delta debugging not only simplifies failure analysis but also mitigates execution flakiness, providing a practical foundation for debugging CPSs.

\end{abstract}




\maketitle

\section{Introduction}

Cyber-Physical Systems (CPSs) integrate computational components with physical processes through sensing, communication, and control~\cite{derler2011modeling,baheti2011cyber,alur2015principles}. Accordingly, considerable research effort has been devoted to the verification and testing of CPSs, including simulation-based test generation~\cite{arrieta2017employing,arrieta2017search,matinnejad2016automated,matinnejad2018test}, regression test optimization~\cite{arrieta2016test,arrieta2019pareto} mutation testing~\cite{bartocci2023property,lee2026fuzzing,cornejo2021mutation}, and runtime monitoring~\cite{stocco2020misbehaviour,ayerdi2024marmot}. In comparison, considerably less attention has been paid to the automated debugging of failures once they have been detected. This imbalance is problematic because CPS failures commonly emerge from long interactions between software components, controllers, physical processes, and environmental conditions, leaving engineers with extensive execution traces and complex test inputs that are difficult to inspect manually.

Simulation-based testing is central to the development of CPSs because it enables engineers to exercise operational conditions that would be expensive, time-consuming, or unsafe to reproduce using physical prototypes. However, repeated executions of the same test input do not necessarily produce identical outcomes. Sources of stochasticity may originate from the System Under Test (SUT), the simulation infrastructure, or their interaction. For example, CPS controllers may employ randomized optimization algorithms, such as genetic algorithms, while simulators may introduce variability through physics computations, floating-point operations, concurrency, and event scheduling. Additional non-determinism may arise from perception pipelines, variable inference times, and communication middleware such as the Robot Operating System (ROS). Consequently, an identical test input may produce different trajectories, quantitative outcomes, or even different pass/fail verdicts across repeated executions.

Recent empirical evidence indicates that such flakiness is not exceptional. Amini et al.~\cite{amini2024evaluating}, for example, observed substantial variation across several autonomous-driving test configurations, with soft flakiness affecting between 4\% and 68\% of the generated tests and hard flakiness reaching up to 74\% in some configurations. They further showed that rerunning test inputs can materially change both the fitness values obtained by randomized testing algorithms and the number of failures detected. These findings demonstrate that simulator flakiness can undermine not only testing outcomes but also any subsequent debugging activity that assumes deterministic failure reproduction. A reduced input may appear to preserve a failure in one execution and fail to reproduce it in the next, making it difficult to determine whether the reduction removed a necessary condition or whether the observed difference is merely caused by stochastic execution variability. 

One established strategy for facilitating debugging is to reduce a failure-inducing test input to a smaller input that still triggers the original failure. Delta Debugging~\cite{zeller2002simplifying} systematically removes parts of an input and repeatedly executes the resulting candidates to isolate a minimal failure-inducing subset. In CPSs, this can substantially reduce the amount of information that engineers must inspect. For example, a full-day elevator traffic scenario containing thousands of passenger events may be reduced to the comparatively small set of events immediately preceding an abnormal dispatching decision~\cite{valle2023applying}. Similarly, a long robotic trajectory may be reduced to the segment required to reproduce a lane-departure failure. However, conventional Delta Debugging assumes that the outcome of each candidate input can be determined from a single execution. This assumption does not hold for stochastic CPSs.

A direct adaptation would be to execute every candidate input multiple times and statistically determine whether it preserves the original failure. Although this strategy improves confidence in the reduction decision, it can make Delta Debugging prohibitively expensive. CPS test executions often involve computationally demanding simulators and long-running scenarios~\cite{arrieta2016test,arrieta2019pareto}, and Delta Debugging already requires the evaluation of multiple candidate inputs. Repeating every candidate many times therefore multiplies an already substantial computational cost. The central challenge is thus to preserve the statistical reliability required by stochastic executions while avoiding unnecessary test reruns.

To address this challenge, we first establish a stochastic Delta Debugging baseline that replaces the deterministic pass/fail decision with repeated executions, failure clustering, and statistical comparison. While this baseline enables reliable reduction under execution variability, its computational cost is prohibitive for long-running CPS simulations. We therefore propose two techniques that improve its practicality. Optimized Stochastic Delta Debugging ($DD_{OS}$) reduces execution time by speculatively evaluating candidate reductions, storing intermediate solutions, and performing repeated statistical validation only when necessary. Environment-Wise Optimized Stochastic Delta Debugging ($EWDD_{OS}$) further exploits stable states of the CPS environment to identify promising starting points for reduction, further accelerating the debugging process. Instead of requiring identical execution outcomes, both techniques determine whether a candidate reproduces the same class and statistical characteristics of the original failure.
We evaluate the approaches using \revision{three} complementary CPSs (an industrial CPS case study system \revision{and two open-soruce case studies}). 

Our results show that the proposed techniques can substantially reduce failure-inducing test inputs and that the optimized variants can decrease debugging cost compared with conservatively validating every reduction. More importantly, the results reveal that minimization increases the failure reproduction ratio of the resulting test inputs, thereby reducing flakiness. Thus, the proposed techniques produce test inputs that are not only smaller, but often more reliable debugging artifacts.

The main contributions of this paper are:

\begin{itemize}
    \item We formulate Delta Debugging for CPSs tested under test flakiness \revision{within the existing reduction framework}, replacing deterministic failure preservation with a statistical assessment based on repeated executions, failure clustering, and distributional comparison.

    \item We formulate a stochastic baseline \revision{($DD_S$)} for Delta Debugging based on repeated executions and statistical failure comparison, and propose two practical extensions \revision{($DD_{OS}$ and $EWDD_{OS}$)} that reduce its computational cost through \revision{speculative reduction, deferred validation with rollback,} and environment-aware reduction.
    
    \item We conduct an empirical evaluation on an industrial elevator-dispatching system, on a DNN-controlled autonomous mobile robot \revision{, and on a VLA-controlled robotic manipulation system,} covering multiple scenarios and distinct sources of stochasticity.

    \item We show that the proposed techniques effectively reduce failure-inducing inputs while decreasing the execution cost of debugging. Most notably, we find that minimized inputs are able to reproduce the target failure more reliably than the original inputs, suggesting that test-input reduction can mitigate the effects of execution flakiness.

    \item We provide a replication package including both a virtual machine with the ROS environment for reproducing the experiments for the Leo Rover case study~\cite{valle2026Replication} and the scripts fro analyzing the results from both case studies~\cite{valle2026DD4CPSs}.
\end{itemize}

The rest of the paper is structured as follows: We give a general background of testing and failure isolation in the context of CPSs in Section~\ref{sec:background}. Section~\ref{sec:caseStudy} introduces our industrial and open-source case studies and which are the sources of randomness on each of them. In Section~\ref{sec:Approach} we present our approach. In Sections~\ref{sec:Evaluation} and~\ref{sec:analysisi of the results} we present our empirical evaluation and we analyze the results. We draw conclusions from the results and highlight key lessons learned in Section~\ref{sec:Discussion}. Finally, in Section~\ref{sec:RelatedWork} we position our work with existing works and we conclude the paper presenting our future work in Section~\ref{sec:Conclusions}.

\section{Background}
\label{sec:background}
This section introduces key aspects related to testing and failure isolation in the context of CPSs.

\subsection{Simulation-based testing}
Simulation-based testing provides a versatile environment to test CPSs across different development stages. In this context, testing is structured into different levels that progressively increase the fidelity of the simulation environment. At the highest level of abstraction, model-in-the-loop (MIL) testing~\cite{arrieta2017automatic} emphasizes on verifying the correctness of mathematical models and control algorithms. This is succeed by software-in-the-loop (SIL) testing~\cite{matinnejad2016automated}, in which the real control software operates within a simulated setting to assess its integration with dynamic models. This is the level in which this study is developed. Finally, at the lowest level of abstraction, hardware-in-the-loop (HIL) testing~\cite{eidson2011distributed, kane2014monitor} combines physical hardware elements with simulation models to accurately replicate real-world interactions. This tiered approach helps identify and address possible problems early in the development process, balancing development costs, risks, and accuracy.

As simulation-based testing allows engineers to represent both the physical dynamics and the control software of the CPS~\cite{stocco2020misbehaviour, khatiri2024simulation,arrieta2020seeding, hou2015simulation}, it facilitates the exploration of a wide range of operational scenarios without the need for expensive physical prototypes. Tools like Simulink, Modelica, and Gazebo are commonly employed to model the dynamic behavior of physical processes and their interactions with CPS controllers. This method not only reduces the development costs and risks but also allows for the simultaneous execution of large-scale experiments, which is crucial due to the wide input space of CPSs~\cite{nejati2019evaluating,menghi2020approximation,matinnejad2016automated, birchler2023teaser}.

However, as simulators become increasingly realistic, ensuring deterministic executions becomes progressively more challenging~\cite{amini2024evaluating}. Small differences in floating-point computations, numerical integration, event scheduling, thread interleavings, and timing accumulate throughout the simulation, causing identical test inputs to produce different execution traces and, in some cases, different failure outcomes. This execution variability makes failures difficult to reproduce and considerably complicates debugging~\cite{amini2024evaluating}. Consequently, debugging techniques for modern CPSs should explicitly account for stochastic execution behavior rather than assuming deterministic test outcomes.

%However, as the accuracy of simulations improves, the difficulty of ensuring determinism also rises~\cite{amini2024evaluating}. Numerical uncertainties\aitor{what are these?}, timing variations, and model simplifications can lead the simulator to generate stochastic (i.e., "flaky") results. These uncertainties make it more difficult to replicate and identify failures, a problem acknowledged by~\citet{amini2024evaluating}. As a result, although simulation is an essential tool, its limitations require robust and automated debugging techniques that can manage variations in simulation results. 

\subsection{Flaky Simulators}
Deterministic simulation has historically served as the foundation for testing CPSs, providing a framework for reproducible evaluation under controlled conditions~\cite{khatiri2024simulation, arrieta2019search, briand2016testing}. Simulators like Carla (developed using Unreal Engine), MetaDrive and Gazebo are specifically designed to provide virtual settings where environmental factors can be accurately controlled. This control is essential for replicating scenarios that would be dangerous or impractical to evaluate in the physical world. In theory, these simulators facilitate reliable testing by guaranteeing that multiple executions of a test input produce the same behavior of the SUT by seed controlled randomization and the deterministic nature of the physics engines. Such determinism is crucial not only for simple testing techniques but also for sophisticated testing methods such as regression testing~\cite{leung1989insights}, mutation testing~\cite{cornejo2021mutation}, safety validation~\cite{corso2021survey} and test case generation~\cite{arrieta2017search}. 

However, as pointed out by~\citet{osikowicz2024empirically} and~\citet{amini2024evaluating} various commonly used CPS simulators face significant non-determinism. This is due to several factors such as artifacts of the physics engine, discrete-time integration techniques and floating-point estimations in numerical solvers. In addition, the physics computation and enhancements to the rendering of the images in such simulators also could lead to non-deterministic behavior. Concurrency also contributes to the occurrence of flaky simulation results. 

%Modern simulators utilize multi-core processing for real-time capabilities, but this parallelism leads to race conditions in managing events and thread scheduling. This race conditions, affected by variations in hardware setups and thread scheduling can result in big differences in simulation results \aitor{this sentence requires citations}. Moreover, sensor modeling imperfections exacerbate these problems. Simulated sensors like lidar, radar or a camera may introduce subtle variations in the sampling time that carry through the perception-processing-actuation cycle, particularly when interacting with machine learning elements, creating another layer of non-determinism. For instance, in our second case study system, involving a mobile robot, %the case of the Leo Rover\aitor{at this stage, LeoRover has not been introduced so far}, a camera interacts with a neural network to guide the rover.

In this work, we consider \revision{three} complementary case study systems that exhibit stochastic behavior for different reasons. The first is an industrial elevator-dispatching system developed by Orona. Although the simulation environment itself is deterministic, the System Under Test (SUT) employs a Genetic Algorithm to optimize elevator assignments in large search spaces. Consequently, repeated executions of the same test input may produce different dispatching decisions and, therefore, different system behaviors. The second case study system is an autonomous mobile robot based on the Leo Rover platform\footnote{\url{https://www.leorover.tech/}}. In this case, stochasticity primarily originates from the execution platform rather than the controller itself. The robot is evaluated in the Gazebo simulator and relies on the Robot Operating System (ROS) for communication between its software components. Variability introduced by the physics engine, floating-point computations, event scheduling, image acquisition and processing, execution-time fluctuations of the DNN-based controller, and communication delays in ROS may cause identical test inputs to produce different execution traces and failure outcomes. \revision{The third case study system is a robotic manipulation system controlled by the GR00T N1.7 Vision-Language-Action (VLA) model and evaluated in the LIBERO and SimplerEnv simulation environments. In this case, stochasticity originates primarily from the policy's generative action head, which transforms sampled noise into robot actions conditioned on visual observations, task instructions, and robot-state information. Consequently, repeated executions with the same task instruction and initial environmental configuration may produce different action sequences. These differences propagate through the closed perception--action loop, potentially leading to different trajectories and task outcomes. Together, the three case study systems cover variability arising from randomized optimization, simulation and execution infrastructure, and generative policy inference, allowing us to examine the proposed techniques across different task and control settings.}

%\aitor{I would add another paragraph saying this:} In the context of this study, two systems are employed, an industrial and an open-source case study system. In the case of the industrial system, encompassing the dispatching algorithm of elevators, while the simulator does not involve any source of flakiness, Orona, our industrial partner, has stochastic algorithms for optimally routing elevators when their search space is huge. Specifically, in the context of this paper, a Genetic Algorithm is used. On the other hand, in the case of the open-source case study system, a mobile robot from LeoRover\footnote{https://www.leorover.tech/}, Gazebo is used as simulator. The system also employs ROS nodes. Therefore, this system has several sources for producing flaky simulations \aitor{mention some/all}.

\subsection{Debugging of CPSs}

Debugging Cyber-Physical Systems (CPSs) is inherently more challenging than debugging conventional software because faults often emerge from the interaction between software controllers, communication middleware, sensors, actuators, and the physical environment~\cite{zampetti2022empirical,timperley2024robust}. Consequently, the root cause of a failure may not correspond to a single software defect, but instead to a particular sequence of events or environmental conditions that collectively drive the system into an unsafe state. This challenge is further exacerbated in stochastic CPSs, where identical test inputs may lead to different execution traces due to randomized algorithms, simulator artifacts, timing variations, or communication delays. As a result, failures can be difficult to reproduce consistently, substantially increasing the cost of debugging.

A major obstacle during debugging is the size of the execution that precedes a failure. CPS test inputs frequently consist of long event sequences or trajectories~\cite{valle2023applying}. For example, in our industrial case study, a single simulation may represent an entire day of passenger traffic~\cite{valle2023applying}, comprising thousands of passenger arrivals and elevator movements before a failure is observed (Section~\ref{sec:ElevatorCaseStudy}). Identifying which subset of these events is actually responsible for the failure is therefore a difficult and time-consuming task.

Delta Debugging~\cite{zeller2002simplifying} addresses this problem by systematically reducing a failure-inducing test input while preserving the original failure. By removing execution segments that are not required to trigger the failure, the resulting minimized input substantially reduces the amount of information that engineers must inspect during fault localization. Our previous work~\cite{valle2023applying} demonstrated the effectiveness of Delta Debugging for deterministic elevator dispatching systems. However, its underlying assumption that the same test input always reproduces the same failure no longer holds for stochastic CPSs, motivating the techniques proposed in this paper.

\subsection{Delta Debugging}

Delta Debugging, introduced by Zeller and Hildebrandt~\cite{zeller2002simplifying}, is an automated input-reduction technique for isolating the elements of a test input that are necessary to reproduce a failure. Let \(t\) denote an input that triggers a failure. Delta Debugging repeatedly partitions \(t\) into smaller subsets and evaluates both these subsets and their complements. Whenever a smaller candidate still reproduces the failure, that candidate becomes the new input to be minimized. If neither the subsets nor their complements preserve the failure, the algorithm increases the partition granularity and continues the search. This process terminates when no individual element can be removed without losing the failure, yielding a \emph{1-minimal} failure-inducing input. A 1-minimal input is not necessarily the globally smallest possible input, but none of its individual elements can be removed while preserving the observed failure.

By eliminating input elements that are not required to trigger the failure, Delta Debugging reduces the amount of information that developers must inspect during fault localization. It has consequently been applied to a broad range of failure-inducing artifacts, including tree-structured inputs such as XML documents~\cite{misherghi2006hdd,artho2011iterative,wang2021probabilistic}, requests and interactions in microservice systems~\cite{zhou2018delta}, formulas processed by Satisfiability Modulo Theories solvers~\cite{brummayer2009fuzzing}, and test inputs for Cyber-Physical Systems~\cite{valletowards,valle2023applying}. Although the structure of these inputs differs, the underlying principle remains the same: progressively remove parts of the original input while retaining the failure of interest.

In CPSs, the inputs subjected to reduction frequently consist of long sequences of operational events, commands, trajectories, or environmental conditions rather than conventional software artifacts. For example, a test of an elevator-dispatching system may encode an entire day of operation, including thousands of passenger arrivals, destination requests, and elevator movements~\cite{valle2023applying,valle2023automated,ayerdi2020qos,ayerdi2021generating}. When such a test reveals a failure, only a small subset of those events may be necessary to trigger it. Applying Delta Debugging can therefore produce a substantially shorter scenario that is easier to replay, inspect, and analyze.

Our previous work~\cite{valle2023applying} adapted Delta Debugging to the CPS domain through an environment-aware reduction strategy for elevator-dispatching systems. The approach observes the state of the simulated environment and identifies stable operational states from which the reduction process can be started. By avoiding the repeated simulation of prefixes that do not influence the failure, the method can reduce both the size of the resulting test input and the computational cost of the reduction process.

However, classical Delta Debugging and our previous environment-aware adaptation assume deterministic test executions. Under this assumption, a candidate input either consistently reproduces the target failure or consistently does not. This binary decision is fundamental to the reduction process, as each observation determines which parts of the input are retained or discarded. In stochastic CPSs, however, repeated executions of the same candidate may produce different traces or failure outcomes. A candidate may reproduce the target failure in some executions but not in others, making a single pass/fail observation insufficient and potentially causing the algorithm to discard relevant input elements or retain irrelevant ones. Addressing this limitation requires replacing deterministic failure preservation with repeated execution and statistical comparison, while controlling the substantial computational cost introduced by such validation. This challenge motivates the stochastic Delta Debugging techniques presented in this paper.

\section{Case Study Systems} \label{sec:caseStudy}
\revision{In this paper, we consider three case study systems: an industrial elevator dispatching algorithm and two open-source robotic systems, namely an autonomous robot based on the Leo Rover platform and a VLA-enabled autonomous robot controlled by GR00T N1.7.}

\subsection{Industrial Case Study System -- Elevator Dispatching Algorithm}\label{sec:ElevatorCaseStudy}

\subsubsection{Overview of the CPS}
\begin{figure}[ht]
\centering
\includegraphics[width=0.85\textwidth]{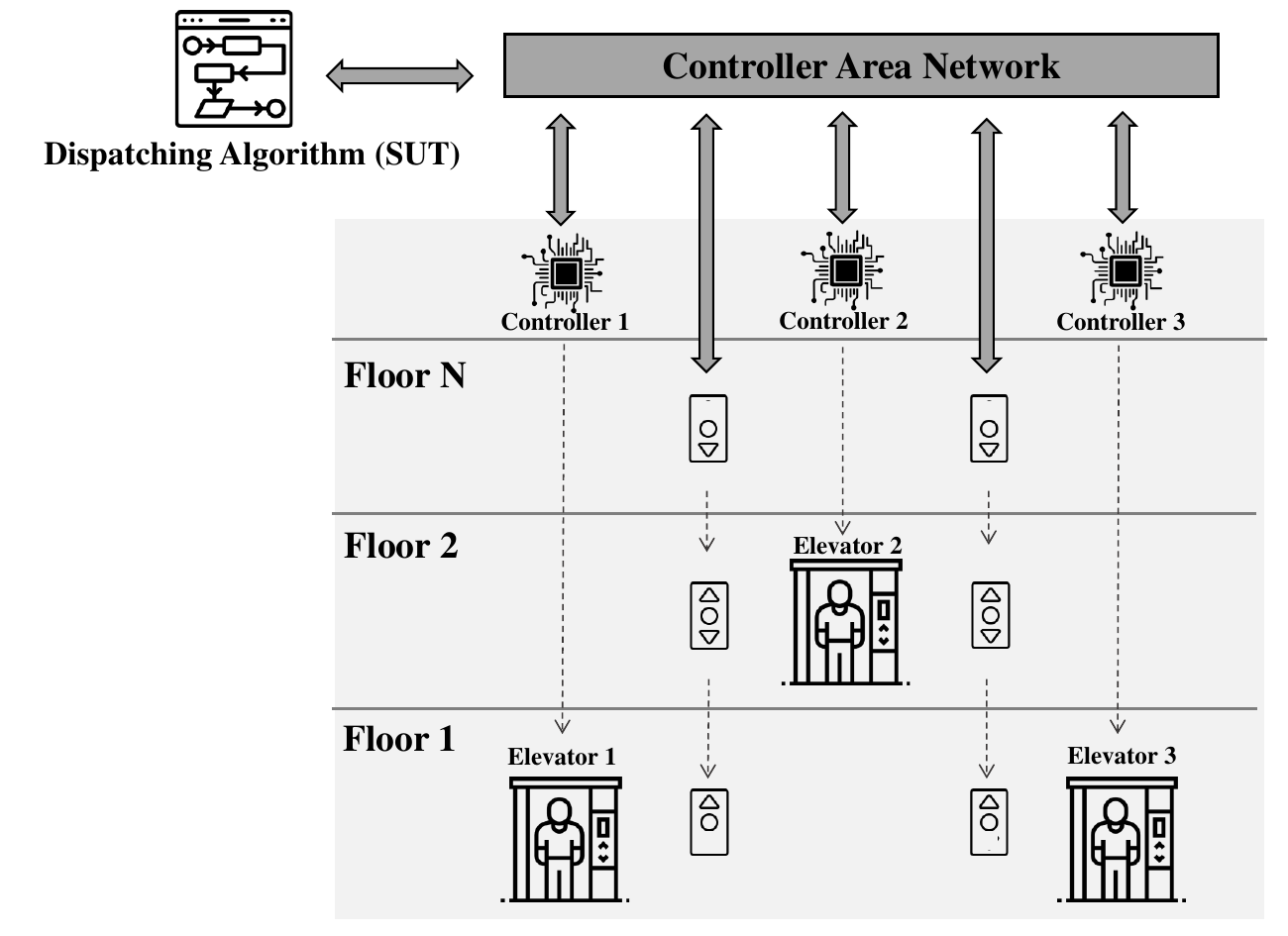}
\caption{Overview of our industrial case study}
\label{fig:Elevator}
\end{figure}

Figure \ref{fig:Elevator} shows an overview of our industrial case study system provided by Orona~\footnote{\url{https://www.orona-group.com/es-es/}}, one of the worldwide leading companies in the elevation domain. This system involves a complex CPS where different computational units, communication protocols and mechanical and electrical components interact among them to transport passengers from one floor to another. Every time a passenger enters the system, the passenger makes a call through a button. This call is communicated to the dispatching algorithm through a Controller Area Network (CAN) bus. The dispatching algorithm determines which elevator is attending each call. This selection is based on different criteria, such as reducing passengers' waiting times or reducing energy consumption. When the dispatching algorithm selects which elevator should be assigned to a call, this is communicated through the CAN bus to the elevator controller. This controller makes the necessary moves to safely attend to the queued passengers.

\subsubsection{The System Under Test} In our industrial case study, the System Under Test (SUT) is the traffic dispatching algorithm, which is the one that selects the elevator to attend each call. While Orona has a large suite of dispatching algorithms, we used one whose core algorithm is a Genetic Algorithm~\citet{beamurgia2016modified} (as it is a stochastic algorithm, and therefore a good subject for our study). The dispatching algorithm takes information about the environment, such as the number of passengers each elevator has or the position of each elevator. Based on it, the algorithm returns a solution to assign an elevator to each call. However, the algorithm may operate with incomplete information, %\aitor{to me it is unclear why we want to motivate the case study with uncertainty}, 
for example, the destination of the passenger, the weight of the passengers, or the number of passengers behind each call. Additionally, there may be cases where the system faces unforeseen situations for which the algorithm is not prepared or configured for~\cite{han2022elevator,han2022uncertainty,valle2023automated}. When that occurs, the situation should be isolated as much as possible to propose a patch or adjustment.

\subsubsection{Test Execution Platform} To test the traffic dispatching algorithms, different simulation test levels are conducted in Orona~\cite{ayerdi2020towards}. The first level refers to the Software-in-the-Loop (SiL) test level, where we integrate our technique. In such level, the traffic dispatching algorithm is integrated with Elevate\footnote{\url{https://elevate.helpdocsonline.com/home}\label{elevate}}, a commercial simulation tool. Elevate takes as input two files: (1) the installation of the elevators (including data like the number of elevators, speed of each of them, number of floors, etc.) and (2) the test input, which includes a set of passengers traveling through the installation. With these two input files, Elevate simulates the physical components of the elevator system (e.g., elevator speed, engines) and provides a file with several pieces of information (e.g., the time each passenger had to wait, which elevator attended each passenger, energy consumption). Then, an oracle parses this file to raise a test verdict (i.e., pass or fail). Orona also conducts tests in later test levels, including at the Hardware-in-the-Loop (HiL) test level as well as in operation. However, since our approach does not include this level, we do not explain it in detail. The reader can refer to a previous study to better understand the entire testing process conducted by Orona to test their dispatching algorithms~\cite{ayerdi2020towards}.

%\aitor{I would add another subsection named "sources of randomness" or similar. Explain why the tests are stochastic. In this case, it is because we use a Genetic Algorithm.}

\subsubsection{Sources of randomness}
The key source of randomness in this industrial case study system is its algorithm. The selected traffic dispatching algorithm is a genetic algorithm, which is employed in buildings with a large number of floors and where a large number of passengers is expected.  Genetic algorithms are stochastic in nature, from its initial population generator, to later genetic operators (e.g., mutation, crossover, selection). In this case, the simulator (i.e., Elevate), is deterministic, as all the communication between the dispatching algorithm and the rest of controllers (e.g., individual elevator controllers) are simulated. % Moreover, the roulette wheel selection method, based on fitness scores, includes additional randomness when deciding which individuals are selected for reproduction. Finally, both the crossover and mutation processes apply random changes to the genetic structure of individuals, thereby adding additional non-determinism to the execution of the test case. 

\subsection{Open-source Case Study System -- Autonomous robot based on the Leo Rover platform}

\subsubsection{Overview of the CPS} 
\begin{figure}[h]
    \centering
    \includegraphics[width=0.85\linewidth]{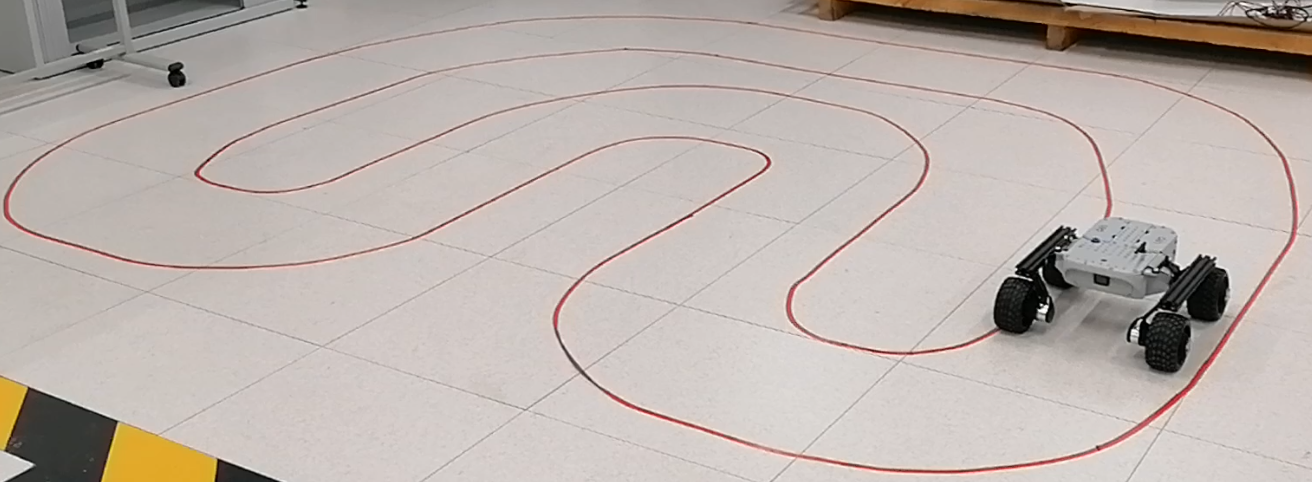}
    %\vspace{-5pt}
    \caption{LeoRover in our laboratory}
    \label{fig:Leo Rover}
\end{figure}

Figure \ref{fig:Leo Rover} provides an overview of our \textcolor{black}{first} open-source case study system, the LeoRover~\footnote{\url{https://github.com/LeoRover}} mobile robot. The Leo Rover is a mobile autonomous robot where different computational units, communication protocols, and mechanical and electrical components interact to enable effective navigation and task execution. The rover captures images that are processed by a Deep Neural Network (DNN) model. This model provides two outputs: (1) the linear velocity reference and (2) the angular velocity reference. These two outputs are communicated to a lower-level controller for translating them into engine speeds through a Robot Operating System (ROS)-based communication system. ROS is responsible for managing real-time interactions between the components and ensures that the commands are executed effectively and timely, allowing the rover to navigate its environment based on specific criteria (e.g., obstacle avoidance, path optimization, and task completion).

\subsubsection{The System Under Test} For our \textcolor{black}{first} open-source case study system, the SUT is the entire Leo Rover itself, an autonomous robotic platform composed of multiple interconnected subsystems, including a Deep Neural Network (DNN)–based navigation controller. Although the Leo Rover is designed to operate reliably under normal conditions and can adapt to previously unseen scenarios through its DNN-based controller, it remains susceptible to failures. These failures can stem from poorly trained DNN models, incorrect sensor calibration, or unexpected environmental variations. In such cases, isolating the failure is crucial to determine which component of the Leo Rover is responsible for the observed error and effectively identify the root cause.

\subsubsection{Test Execution Platform} Leo Rover’s control algorithm is connected to a simulation environment, specifically the Gazebo simulator~\cite{koenig2004design}. Gazebo simulates the physical aspects of the rover’s operation, including dynamics, sensor data, and interaction with the environment. Gazebo takes the environmental setup as the primary input, which includes information about the environment surroundings of the rover (e.g., terrain type, obstacles, lines to be followed by the rover, and the starting position and orientation of the rover). These inputs allow the simulator to replicate real-world conditions. The output from Gazebo includes data such as the rover’s position, the path taken, and any collisions or errors found during navigation. This output is then analyzed using an oracle that verifies whether the rover is within the lines of the circuit. Based on this data, the algorithms proposed in this work are able to identify a failure and minimize the test inputs needed to reproduce that failure reliably, thus helping localize the source of the observed failure and identify the components responsible for these failures.

\subsubsection{Sources of randomness}
The behavior of the Leo Rover across simulators can vary due to three different sources of randomness: (1) the simulator, (2) the processing of images, and (3) the communication framework. Firstly, the Gazebo Simulator, which is crucial for our Software-in-the-Loop (SiL) testing, exhibits inherent non-determinism. Differences in its physics engine, event timing, and occasional unreliable performance in mimicking real-world dynamics create minor inconsistencies in sensor readings, collision detections, and general environmental interactions.

In addition, the image-based navigation system adds randomness. The time to capture and process an image may vary across different runs because of different computational demands and delays in the navigation process. For instance, the image processing time of the DNN can vary due to differences in the system load and resource distribution. These timing variations can result in taking different images across runs, resulting in different outputs of the navigation system and different trajectories across runs. Furthermore, the ROS-based communication employed to manage data transfer between the Leo Rover's subsystems (i.e., sensors and controllers) may experience random delays. These delays, combined with the aforementioned randomness sources, make the Leo Rover behave stochastically.
%\subsection{Importance of isolating failure-inducing test inputs}

\revision{

\subsection{Open-source Case Study System -- VLA-enabled Autonomous Robot}

\subsubsection{Overview of the CPS}
Figure~\ref{fig:Libero} shows an overview of our second open-source case study, which considers an autonomous robotic manipulation system controlled by NVIDIA Isaac GR00T N1.7~\cite{nvidia2025gr00tn1openfoundation}, a Vision-Language-Action (VLA) model. The system integrates visual observations, natural-language task instructions, and robot-state information to perform manipulation tasks in its environment. GR00T N1.7 combines a vision-language backbone with a generative action head to predict continuous robot actions. These predictions are processed by an embodiment-specific interface and passed to a lower-level controller, which translates them into commands for the robot's actuators. Depending on the selected robot configuration, these actions describe changes in the end-effector pose and gripper commands. After executing a sequence of actions, the system collects updated observations and generates subsequent actions, establishing a closed perception--action loop. This interaction enables the robot to perform language-conditioned tasks, such as grasping an object and placing it in a specified location.

\subsubsection{The System Under Test}
The SUT comprises the robotic control stack, including observation preprocessing, the GR00T N1.7 policy, action postprocessing, and the interfaces connecting the policy to the robot controller. The test execution platform provides the simulated robot and its physical environment. Although GR00T N1.7 supports language-conditioned manipulation, its successful execution of a task depends on the observations, instruction, environmental configuration, and compatibility between the policy and the robot embodiment. Failures may arise from incorrect visual interpretation, inadequate grounding of the instruction in the scene, unsuitable action predictions, or errors in observation and action transformations. For instance, the robot may select the wrong object, fail to establish a stable grasp, or release an object outside the intended target region. Reducing the inputs that trigger these failures helps identify relevant failure-inducing conditions and supports subsequent investigation of the responsible components.

\subsubsection{Test Execution Platform}
The SUT is evaluated through Software-in-the-Loop (SiL) testing using two open-source simulation environments: LIBERO~\cite{liu2023libero} and SimplerEnv~\cite{li24simpler}. LIBERO, provides language-conditioned robotic manipulation tasks through environments built on RoboSuite and MuJoCo. SimplerEnv provides simulated manipulation environments corresponding to real robotic setups, using SAPIEN-based environments. These platforms supply the physical dynamics, rendered observations, robot-state information, and task-evaluation mechanisms required to execute the GR00T N1.7 policy. Each test input specifies a task instruction and an environmental configuration, including the initial robot state, object positions and orientations, and task-relevant scene properties. During execution, the selected environment supplies observations to the policy and advances the simulation according to the resulting actions. The policy inputs and outputs are adapted to the observation format, action conventions, and robot embodiment of each environment.

\begin{figure}[h]
    \centering
    \includegraphics[width=0.85\linewidth]{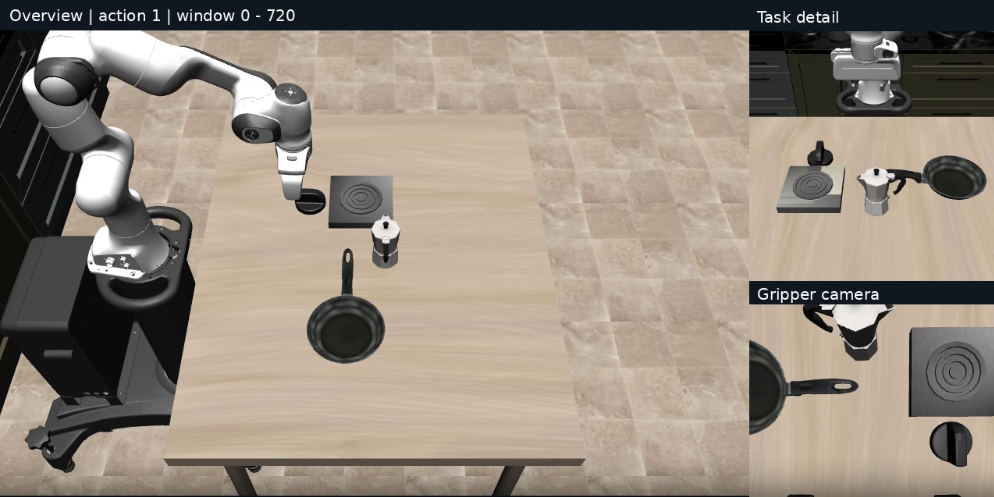}
    %\vspace{-5pt}
    \caption{LIBERO simulation environment}
    \label{fig:Libero}
\end{figure}

Execution produces a trajectory containing robot and object states, executed actions, and task-completion information. A task-specific oracle uses the environment's success conditions to determine whether the instructed goal is achieved within a predefined execution horizon. For example, a placement task requires the designated object to reach the specified target region. The algorithms proposed in this work use these outcomes to identify failing tests and minimize the configurable inputs needed to reproduce them. Candidate reductions must preserve task validity and the relevant failure condition. Repeated executions can then assess whether the reduced input reproduces the failure consistently.

\subsubsection{Sources of Randomness}
The main source of randomness in this case study is the GR00T N1.7 policy itself. Its generative action head uses a flow-matching process that transforms sampled Gaussian noise into robot actions, conditioned on visual observations, the task instruction, and the robot state. Different noise samples can therefore lead to different action sequences even when the policy receives identical inputs. This introduces randomness during inference, without requiring changes to the model parameters or the initial environmental configuration. In our experiments, the initial environmental state and task instruction remain unchanged across repeated executions. Nevertheless, variability in the generated actions can produce different robot trajectories. These differences propagate through the closed control loop since executing different actions changes the robot's interaction with the environment and, consequently, the observations used to predict subsequent actions. During manipulation, differences in gripper alignment or object contact can lead to different outcomes, such as successfully grasping an object in one execution and failing in another. This stochastic behavior makes the VLA-enabled robot a suitable subject for our study, since the same test input may produce both passing and failing executions. Consequently, a single execution is insufficient to establish whether a candidate input consistently reproduces a failure. This case study system allows us to evaluate whether the proposed algorithms can isolate and minimize failure-inducing inputs while accounting for variability in the SUT's behavior.

}

\section{Approach}\label{sec:Approach}

In this section, we present our approach and its technical contributions.

\subsection{Formalization}\label{sec:Formalization}
\revision{In the context of debugging CPSs, let ($\mathcal{T}$) represent the space of possible test inputs, where each test input ($TI\in\mathcal{T}$) consists of a collection or sequence of elements, such as passengers entering a building or guidance points. Let ($f(TI)\in{\mathrm{pass},\mathrm{fail}}$) denote the oracle verdict for an individual execution. Under stochastic executions, repeated evaluations of the same input may produce different verdicts or failures with different characteristics. Therefore, observing a failure in a single execution is insufficient to determine whether a reduced input preserves the original failure. Given an original failure-inducing input $TI_0$, we identify a target failure class $c$ through clustering of the observed failures across repeated executions. In this work, failure preservation means that a reduced input reproduces this target class and satisfies the acceptance criteria for its reproduction frequency and measured failure characteristics, as described in Section 4.3.1. This operational definition allows the target failure to become more reproducible after reduction; it does not require identical execution traces or an unchanged reproduction probability. The objective of delta debugging is to reduce $TI_0$ within the admissible input space while maintaining failure preservation under these criteria. The reduction procedure returns a reduced test input that satisfies the configured failure-preservation criteria based on repeated executions. The result depends on the explored candidates, the execution budget, and the acceptance thresholds. Because the observations can vary across runs, the procedure may accept different reductions for the same original input. Therefore, this definition makes no claim of global minimality or 1-minimality. Acceptance provides empirical evidence that the reduced input reproduces the selected failure class and meet the configured criteria for reproduction frequency and measured failure characteristics.}

\subsection{Running example}\label{sec:runningExample}
As a running example for Delta Debugging (Table \ref{tab:waypoints}), let us consider a Test input ($TI$). For the Leo Rover case study, the test input is composed of the starting point of the simulation and the circuit, which generates an output of 20 waypoints. Each waypoint (e.g., $w_1$, $w_2$, ..., $w_{20}$) includes key attributes like (1) timestamp, (2) coordinates (x,y) and (3) heading and (4) failure, which indicates whether the Leo Rover is outside the circuit ($failure \to 1$) or it is inside ($failure \to 0$). When the execution finishes, the oracle takes the output and evaluates whether a failure occurred, i.e., $f(TI)$ as \textit{fail}. The objective of the Delta Debugging algorithm is to reduce this input (i.e., reduce the distance between the starting point of the simulation and the failure point) to a \revision{reduced} subsequence of waypoints $TI'$ so that the same failure still occurs, i.e., $f(TI') = \mathsf{\textit{fail}}$. 

To do so, the algorithm starts by removing all the waypoints after the failure occurs since these waypoints don't affect the behavior of the SUT. Let's imagine that the Leo Rover went off the circuit at 8:10; consequently, $w_{18}$ did not affect the failure, since the Leo Rover went out at $w_{17}$, when the failure was first detected. Therefore, the algorithm discards waypoints $w_{18}$, and $w_{20}$ from the possible set of starting points. Then, the algorithm splits by half the sequence of remaining waypoints and executes the SUT and selects the last half of them as possible Starting points, i.e., $TI' = \{w_9, w_{10}, ..., w_{17}\}$, so that $w_9$ is the new starting point. That is, in the simulation, the rover starts at the position (X and Y) and heading (roll, pitch, yaw) of $w_9$. Suppose that the failure still persists with this $TI'$; this subset is then further divided by two, so that $TI' = \{w_{13}, w_{14}, w_{15}, w_{16}, w_{17}\}$ and the starting point is $w_{13}$. This process continues iteratively until the reduced test input is not able to reproduce the original failure, i.e., $f(TI') = \mathsf{\textit{Pass}}$. For instance, if $TI' = \{w_{15}, w_{16}, w_{17}\}$ is not able to reproduce the failure, a part of the removed waypoints (i.e., $\{w_{13}, w_{14}\}$) must be added. \revision{The search continues until its stopping criterion is met. In this example, the returned input is the contiguous range $\{w_{14}, w_{15}, w_{16}, w_{17}\}$, with $w_{14}$ as the starting point. Several contiguous waypoint ranges may preserve the target failure; this example illustrates one possible outcome. Repeated runs on the same original input may follow different reduction paths because execution observations and validation outcomes can vary. The algorithms therefore do not guarantee a unique or globally smallest reduced input.} This sequence could represent a sharp turn, a sudden change in speed, or conflicting heading directions that the CPS cannot handle correctly.

\begin{table}[h!]
\centering
\caption{Output waypoints for a Leo Rover navigating a circuit.}
\label{tab:waypoints}
\begin{tabular}{lcrrrrrr}
\toprule
ID  & Time  & X (m)   & Y (m)   & Roll (rad) & Pitch (rad) & Yaw (rad) & Failure\\ \cmidrule{1-8}
$w_1$  & 08:00:00   & 0   & 0    & 0.00          & 0.00           & 1.57      & 0\\ 
$w_2$  & 08:01:00   & 5   & 0    & 0.00          & 0.00           & 1.57      & 0\\ 
$w_3$  & 08:01:30   & 10  & 0    & 0.00          & 0.00           & 1.57      & 0\\ 
$w_4$  & 08:01:45   & 15  & 0    & 0.00          & 0.00           & 1.57      & 0\\ 
$w_5$  & 08:02:20   & 20  & 0    & 0.00          & 0.00           & 1.57      & 0\\ 
$w_6$  & 08:03:25   & 20  & 5    & 0.00          & 0.00           & 0.00       & 0  \\ 
$w_7$  & 08:03:50   & 20  & 10   & 0.00          & 0.00           & 0.00       & 0  \\ 
$w_8$  & 08:04:35   & 15  & 10   & 0.00          & 0.00           & 4.71      & 0\\ 
$w_9$  & 08:05:40   & 10  & 10   & 0.00          & 0.00           & 4.71      & 0\\ 
$w_{10}$ & 08:06:15   & 5   & 10   & 0.00          & 0.00           & 4.71     & 0 \\ 
$w_{11}$ & 08:06:40   & 0   & 10   & 0.00          & 0.00           & 4.71     & 0 \\ 
$w_{12}$ & 08:07:55   & 0   & 5    & 0.00          & 0.00          & 3.14      & 0\\ 
$w_{13}$ & 08:08:30   & 0   & 0    & 0.00          & 0.00           & 3.14     & 0 \\ 
$w_{14}$ & 08:08:45   & 5   & 0    & 0.00          & 0.00           & 1.57      & 0\\ 
$w_{15}$ & 08:09:10   & 10  & 0    & 0.00          & 0.00           & 1.57     & 0 \\ 
$w_{16}$ & 08:09:25   & 15  & 0    & 0.00          & 0.00           & 1.57     & 0 \\ 
$w_{17}$ & 08:10:20   & 20  & 0    & 0.00          & 0.00           & 1.57     & 1 \\ 
$w_{18}$ & 08:11:00   & 20  & -5   & 0.00          & 0.00           & 0.00      & 1   \\ 
$w_{19}$ & 08:11:30   & 20  & -10  & 0.00          & 0.00           & 0.00      & 1   \\ 
$w_{20}$ & 08:12:00   & 15  & -10  & 0.00          & 0.00           & 4.71     & 1 \\ \bottomrule
\end{tabular}

\end{table}

\subsection{Delta Debugging for Stochastic Processes in CPSs}% \aitor{maybe it is better not to put ``adaption'', as we target journal first. We can put  Delta Debugging for Stochastic Processes in CPSs}} 

This section introduces the proposed delta debugging algorithm to deal with the stochastic nature of CPSs tested on flaky simulators. 
To this end, we implemented three versions of the algorithm: (1) Stochastic Delta Debugging, (2) Optimized Stochastic Delta Debugging, and (3) Environment-Wise Optimized Stochastic Delta Debugging. The former refers to an adaptation of the delta debugging algorithm for stochastic CPSs. In the second, we propose an enhanced version that addresses the challenge that CPS testing needs to deal with, i.e., long test execution time when needing to minimize the failure inducing test input. In the latter, we add awareness of the environment to the delta debugging algorithm. 

\subsubsection{Stochastic Delta Debugging (DD\textsubscript{S})}\label{sec:SDD}
%\cite{amini2024evaluating} \aitor{shall we call "stochastic CPSs"? or the process (i.e., simulation) itself is stochastic? i.e., the simulation is stochastic and this can be because the CPS internals are stochastic or because other issues}\pablo{I guess we can say that the process is stochastic due to the CPS internals, therefore we can name it as "stochastic processes in CPSs, as it is named the title of Sec 4.3}\aitor{I think it is not only due to the CPS being stochastic, simulators can also be and our approach is generic to that. Check this paper and we should refer to it, it discusses these things: Amini, M. H., Naseri, S., \& Nejati, S. (2024). Evaluating the impact of flaky simulators on testing autonomous driving systems. Empirical Software Engineering, 29(2), 47.}.
Algorithm \ref{alg:BaseStochasticDeltaDebugging} describes the Delta Debugging algorithm for CPSs tested on flaky simulators. As inputs, it receives the System Under Test (SUT), the initial failure inducing Test Input ($TI$) and its Failing Time ($F_t$), which indicates the simulation time at which the oracle detected the failure. As an output, it provides $TI'$, which corresponds to the \revision{reduced} failure-inducing Test Input for $TI$. This is obtained by following the next procedures.

\begin{figure}[ht]
\centering
\begin{subfigure}[!hhtp]{0.49\textwidth}{\label{fig:Scenario1}\includegraphics[width=\textwidth, trim= 40 0 140 0]{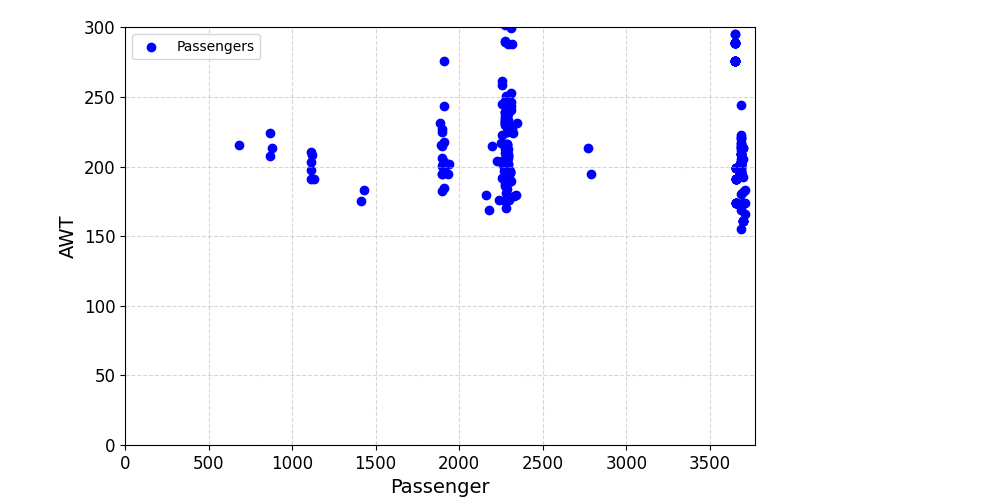}}
\end{subfigure}
\
\begin{subfigure}[!hhtp]{0.49\textwidth}{\label{fig:Gaussian Mixture}\includegraphics[width=\textwidth, trim= 40 0 140 0]{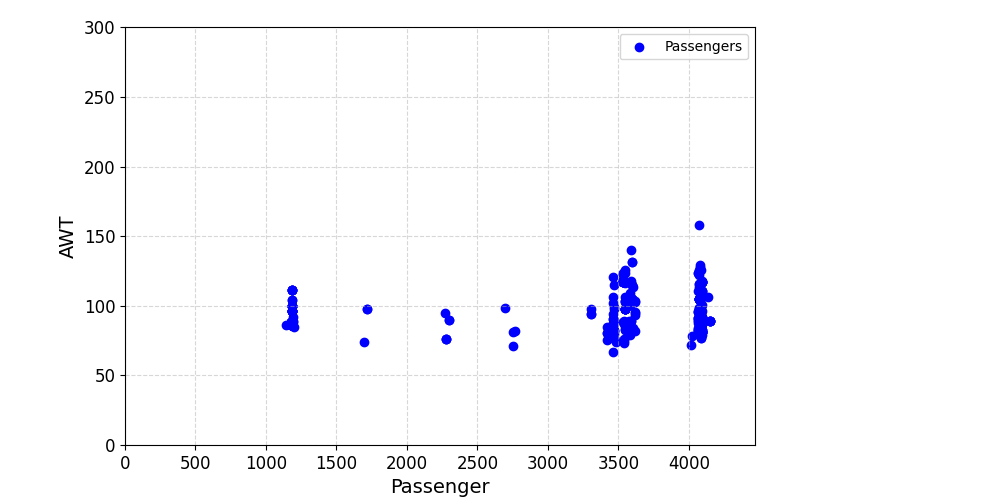}}
\end{subfigure}

\caption{Failure distribution across multiple runs in Scenario 1 and Scenario 6 for Orona's case study.}
\label{fig:failuresOrona}

\end{figure}

First, the algorithm categorizes the failures into clusters (Line 1, Algorithm \ref{alg:BaseStochasticDeltaDebugging}) to consider the chance of triggering different failures across the executions of the original test input. As Figure~\ref{fig:failuresOrona} depicts, the same test input may produce different behavior of the SUT and also trigger different failures. Similar to other research studies~\cite{dickinson2001finding, arunajadai2004failure, digiuseppe2012concept}, we used clusters to identify different instances of failures. By means of \textsc{ClusterFailures} function, \revision{the algorithm selects the cluster containing the largest number of observed failures as the default target for reduction. This frequency-based policy provides an automatic selection criterion; however, it can be adapted by the user to select a desired group of failures, e.g., a group of specific failures}. Then, the algorithm splits the test input to stop once the selected failure is triggered (Line 2, Algorithm \ref{alg:BaseStochasticDeltaDebugging}). This is performed because if the SUT fails at certain point, there is no need to carry out executing the remaining test. %\aitor{I think that we need a section of preliminaries where we explain these kind of concept. We can call it $t_f$ or something like that}\pablo{That could be Section 4.1 and Section 4.2, where we explain how is delta debugging computed and explain the basis of the delta debugging algorithm}
For instance, if the rover goes off the road after 50 seconds, the algorithm does not need to continue the simulation once this event happens. Afterward, the algorithm calculates the reduction range to be applied for minimization and maximization purposes (Line 3), followed by the first reduction process (Line 4). This process is carried out by invoking Algorithm \ref{alg:Passenger_Split_Minimizing}, which takes as input (i) $TI'$ and (ii) $reduction$. %\aitor{probably all variables should be among \$-s} 
Therefore, the function splits the test input $TI'$ by half, returning the exact position of the rover (i.e., point $TI'_{n/2}$).

\begin{algorithm}[ht]
\caption{Stochastic Delta Debugging Algorithm}\label{alg:BaseStochasticDeltaDebugging}
   % \begin{algorithmic}
    \KwIn{$SUT$ \tcp*{\small System Under Test} \\
    \nonl $F_t$ \tcp*{\small Failing time}\\
    \nonl $TI$ = $\{p_1, p_2,..., p_{n}\}$   \tcp*{\small Initial failure inducing test input} \\}
    \KwOut{TI'= $\{p'_1, p'_2,..., p'_{n}\}$ \tcp*{\small Minimized failure inducing test input} \\}
    $cluster$ $\gets$ \textsc{ClusterFailures}($TI,F_t$) ; \\
    $TI'$ $\gets$ \textsc{Split}($TI,F_t$); \\ 
    $reduction$ $\gets$ $TI'.np / 2 $ ;\\
    %it $\gets$ 1;\\
    $TI\textsubscript{NEW}$ $\gets$ \textsc{SplitMin}($TI', reduction$);\\
    \While {reduction $\geq$ 1}{
       
        $Verdict$ $\gets$ \textsc{ExecuteMultTimes}($TI\textsubscript{NEW}, SUT$, $cluster$);\\
        $reduction$ $\gets$ $reduction / 2$;\\
        \eIf{Verdict == Failure}{
            $TI'$ $\gets$ $TI\textsubscript{NEW}$;\\
            $TI\textsubscript{NEW}$  $\gets$ \textsc{SplitMin}($TI\textsubscript{NEW}, reduction$);\\
        }{
            $TI\textsubscript{NEW}$ $\gets$ \textsc{SplitMax}($TI\textsubscript{NEW}$,$TI', reduction$);\\
        }
        
    }
    
   % \end{algorithmic}
  % \vspace{-0.3cm}
\end{algorithm}

After the initial minimization, the algorithm enters a while loop (Lines 5-14) that aims to further minimize the failing test input. \revision{Within this loop, the candidate test input ($TI_{\mathrm{NEW}}$) is evaluated using the \emph{ExecuteMultipleTimes} function. This function performs ($n_{\mathrm{rep}}$) executions of the candidate and applies the statistical failure-comparison procedure described below. The parameter ($n_{\mathrm{rep}}$) therefore controls the execution budget of each repeated-execution validation. We use $n_{\mathrm{rep}}=30$ as the reference configuration in RQ1 and RQ2, and investigate its sensitivity in RQ3.}

\revision{The repeated-execution procedure assesses failure preservation through two complementary comparisons: 1) how frequently $TI\textsubscript{NEW}$ reproduces the target failure class and 2) whether the measured characteristics of the reproduced failures differ from those of $TI$. The procedure first determines how many observations belong to the selected failure cluster. Cluster membership identifies the target class according to the selected failure representation, while the subsequent comparisons assess reproduction frequency and failure characteristics within that class. These comparisons use the odds ratio returned by the Fisher's exact test routine~\cite{upton1992fisher} and the Mann-Whitney U test~\cite{mann1947test}. For reproduction frequency, the algorithm constructs a contingency matrix containing the numbers of observations assigned and not assigned to the selected failure cluster for $TI$ and $TI\textsubscript{NEW}$. The Fisher's exact test returns an odds ratio and a p-value. \revision{The reproduction-frequency criterion uses the estimated odds ratio. The p-value returned by Fisher's exact test does not enter the acceptance decision.} With the odds ratio defined as the odds of target-cluster membership for $TI$ divided by those for $TI\textsubscript{NEW}$, a value at or below 1.68 satisfies the frequency-comparison criterion. This threshold is motivated by the effect-size interpretation in~\cite{chen2010big} and serves as a practical acceptance heuristic. 

The algorithm then applies the Mann-Whitney U test to the failure characteristics of the observations assigned to the selected cluster for $TI$ and $TI\textsubscript{NEW}$. This rank-based comparison avoids assuming normally distributed observations. We use 0.05 as the nominal significance level for each Mann--Whitney U comparison. The candidate satisfies the characteristic-comparison criterion when all configured comparisons yield p-values greater than 0.05. This means that the tests did not detect a difference in the measured characteristics with the available observations. However, the comparisons may have limited power when few executions reproduce the target failure. The algorithm accepts $TI\textsubscript{NEW}$ as preserving the target failure when it satisfies both the frequency and characteristic criteria and the required observations are available.}

\begin{algorithm}[ht]
\caption{\textsc{SplitMinEvent}: Split Minimizing}\label{alg:Passenger_Split_Minimizing}
   % \begin{algorithmic}
    \KwIn{ $TI\textsubscript{NEW}$=$\{p_1, p_2,..., p_{n}\}$ \tcp*{\small Minimized test input previously selected}\\
    \nonl $splitSize$ \tcp*{\small \# of points to remove}    \\
    }
    \KwOut{$TI\textsubscript{MIN}$ \tcp*{\small Minimized test input} \\}
    \For{$i\gets splitSize$ \KwTo $TI\textsubscript{NEW}.np$}{
        
        $TI\textsubscript{MIN}$ $\leftarrow$ $TI\textsubscript{MIN} \cup p_i$;\\
        
    }
   % \end{algorithmic}
  % \vspace{-0.3cm}
\end{algorithm}

\begin{algorithm}[ht]
\caption{\textsc{SplitMax}: Split Maximizing}\label{alg:Passenger_Split_Maximizing}
   % \begin{algorithmic}
    \KwIn{ \nonl $TI\textsubscript{NEW}$ \tcp*{\small Current test input} \\
    \nonl $TI'$ = $\{p_1, p_2,..., p_{n}\}$ \tcp*{\small Minimized test input} \\
     \nonl $splitSize$ \tcp*{\small \# of points to add} \\ }
    \KwOut{$TI\textsubscript{MAX}$ \tcp*{\small Maximized test input} \\}
    $toSplit$  $\gets$  $TI.np$-($TI\textsubscript{NEW}.np$ + $splitSize$); \\
    \For{$i\gets toSplit$ \KwTo $TI.np$}{
        \
        $TI\textsubscript{MAX}$ $\leftarrow$ $TI\textsubscript{MAX} \cup p_i$;\\

    }
   % \end{algorithmic}
  % \vspace{-0.3cm}
\end{algorithm}

If the returned verdict is \textit{``failure''}, the minimization procedure can continue (Lines 8-10), as it means that $TI\textsubscript{NEW}$ reproduces the original failure consistently. The test input in $TI\textsubscript{NEW}$ is assigned to $TI'$ (Line 9), and the minimization routine is invoked by means of the \textit{\textsc{SplitMin}} function (line 10). Conversely, if the returned verdict is \textit{``pass''}, it means that $TI\textsubscript{NEW}$ does not reproduce the original failure, therefore, the test input requires to be enlarged (Lines 11-13), by invoking Algorithm \ref{alg:Passenger_Split_Maximizing} (Line 12), which takes as input (i) $TI\textsubscript{NEW}$, (ii) $TI'$ and (iii) $redeuction$. This Algorithm adds the reduction number of points to $TI\textsubscript{NEW}$ from $TI'$, thus enlarging current $TI\textsubscript{NEW}$. %which selects the new starting point in a point between $TI_{NEW}$ and $TI'$\aitor{Hard to understand, because $TI_{NEW}$ and $TI'$ are not points, but test inputs... we need to better explain, or at least formalize a bit}.
This procedure returns the maximized test input in $TI_{NEW}$, which is tested in Line 5, Algorithm \ref{alg:BaseStochasticDeltaDebugging}. This process is repeated until $TI'$ and $TI_{NEW}$ are the same, i.e., the reduction is $\leq$ 1. When this condition is met, the algorithm returns the \revision{reduced} failure-inducing test input.

\subsubsection{Optimized Stochastic Delta Debugging (DD\textsubscript{OS})}\label{sec:ODD}
The optimized version of the Delta Debugging algorithm for CPSs tested on flaky simulators (i.e., Algorithm~\ref{alg:OptimizedStochasticDeltaDebugging}) \revision{reduces cost by deferring repeated validation while exploring promising reductions. Whereas $DD_S$ validates every candidate through repeated executions before continuing, $DD_{OS}$ can proceed provisionally after a single-execution screen. The archive stores candidates and their validation status, supporting rollback and avoiding revalidation of already verified candidates. Recursion organizes the search and validation phases, while the main optimization comes from deferring repeated validation rather than applying it to every intermediate candidate.} The algorithm can be divided into two components: (1) recursion handling (Algorithm~\ref{alg:OptimizedStochasticDeltaDebugging}, Lines 2–12) and (2) the minimization procedure (Algorithm~\ref{alg:OptimizedStochasticDeltaDebugging}, Lines 13-25). 

\begin{algorithm}[ht]

\caption{Optimized Delta Debugging Algorithm}\label{alg:OptimizedStochasticDeltaDebugging}

   % \begin{algorithmic}
    \KwIn{$SUT$ \tcp*{\small System Under Test}
    \nonl $F_t$ \tcp*{\small Failing time (Optional)}
    \nonl $TI$ = $\{p_1, p_2,..., p_{n}\}$   \tcp*{\small Initial failure inducing test input} 
    \nonl $archive$ \tcp*{\small Archive of solutions (Optional)}
    \nonl $cluster$ \tcp*{\small Cluster of failures}}
    \KwOut{$archive$= $\{TI_1, TI_2,..., TI_n\}$ \tcp*{\small Minimized failure inducing test inputs} }

    $TI\textsubscript{NEW}$ $\gets$ $TI$;\\
    \eIf{$archive.n > 0$}{
        %$reduction$ $\gets$ getReductions($TI\textsubscript{NEW}, archive$); 
        $TI\textsubscript{NEW}$, $archive$, $reduction$ $\gets$ \textsc{CheckSolutions}($TI\textsubscript{NEW}$, $archive$, $SUT$, $cluster$);\\
        \eIf{$reduction$ == 0}{
            $archive$ $\gets$ \textsc{saveToArchive}($TI\textsubscript{NEW}, 1$);\\
            \textbf{return} $archive$;\\
        }{
             $archive$ $\gets$ \textsc{saveToArchive}($TI\textsubscript{NEW}, 1$);  
        }        
    
    }{
        
        $TI'$, $archive$, $reduction$ $\gets$ \textsc{InitArchive}($TI,F_t$); 
    }
    
    $TI\textsubscript{NEW}$ $\gets$ \textsc{SplitMin}($TI', reduction$);\\
    \While {reduction $\geq$ 1}{
        $Verdict$ $\gets$ \textsc{Execute}($TI\textsubscript{NEW}, SUT$, $cluster$);\\
        \eIf{Verdict $\neq$ Failure}{
            $TI\textsubscript{NEW}$, $archive$, $reduction$ $\gets$ \textsc{CheckMultTimes}($TI\textsubscript{NEW}$, $archive$, $SUT$, $cluster$);\\
            $archive$ $\gets$ \textsc{SaveToArchive}($TI\textsubscript{NEW}, 1$); \\     
        }{
            $archive$ $\gets$ \textsc{SaveToArchive}($TI\textsubscript{NEW}, 0$);\\
        }
        $reduction$ $\gets$ $reduction / 2$;\\
        $TI\textsubscript{NEW}$ $\gets$ \textsc{SplitMin}($TI\textsubscript{NEW}, reduction$);\\
    }
    $archive$ $\gets$ \textsc{OptimizedDeltaDebugging}($TI\textsubscript{NEW}$,$archive$, $SUT$, $cluster$) ;
   % \end{algorithmic}
  % \vspace{-0.3cm}
\end{algorithm}

First, the algorithm starts by determining whether it is running for the first time (i.e., the number of solutions in the archive is 0) or it is operating recursively on an archive that already contains multiple test inputs. In the initial run, the archive is empty, therefore it initializes the archive by adding the original test input to it as well as it splits the test input by the failing time (Line 11). This split is critical because it isolates the test input that is responsible for the failure, thereby setting the first stage for further reduction. Once the archive is initialized, the algorithm splits the test input by half (Line 13) and enters in the reduction procedure (Lines 14-24).

Within the reduction procedure, the algorithm first executes once the current test input by invoking the function $\textsc{ExecuteTest}$ (Line 15). This function executes once the test input and assesses whether the failure is statistically similar to the ones obtained from $TI'$. It does this by first checking if the observed failure lies within the selected cluster of failures. Then, it computes the Z-score~\cite{wackerly2008mathematical}, which measures how many standard deviations the failure raised by $TI\textsubscript{NEW}$ is from the mean of the set of failures raised by $TI'$. This test determines how common the result is within the distribution. By using a Z-score range of -1.96 to 1.96, we established a confidence interval of 95\%, which is a common statistical threshold, the same as the one used in the analysis of Fisher's exact test and Mann-Whitney U test. Therefore, if the Z-score falls within the range, it indicates that the failure raised by $TI\textsubscript{NEW}$ is similar to the one raised by $TI'$ (i.e., the verdict is set to \textit{''Failure''}). 

On the contrary, if the verdict is \textit{''Pass''}, the algorithm initiates a more rigorous verification process via the \textsc{CheckMultTimes} function (Line 17). This function first executes multiple times (i.e., by default 30 times) $TI\textsubscript{NEW}$ and assesses statistically whether the failure reproduced is similar to the original one by means of clustering, Fisher's exact test and Mann-Whitney U test as in Algorithm~\ref{alg:BaseStochasticDeltaDebugging}. When these additional tests continue to produce a \textit{''Pass''} verdict, (i.e., $TI\textsubscript{NEW}$ is not able to reproduce the original failure consistently), the algorithm enters in a rollback phase (Lines 2-8, Algorithm~\ref{alg:CheckMultTimes}). Inside this loop, the minimization is undone, specifically, the algorithm reverts the test input to a previous test input stored in the archive (Line 3, Algorithm~\ref{alg:CheckMultTimes}). In addition, the algorithm updates the reduction, since $TI\textsubscript{NEW}$ is not able to reproduce the failure, it does not make sense to continue reducing, therefore, the failure-inducing \revision{reduced} test input could be between the current $TI\textsubscript{NEW}$ and the discarded test input. So, in Line 3, the algorithm gets the difference (i.e., $reduction$) between the current $TI\textsubscript{NEW}$ and the discarded test input. If $TI\textsubscript{NEW}$ has already been statistically assessed, (i.e., $verified$ is\textit{1}) the algorithm exits this loop (Lines 2-8). If not, $TI\textsubscript{NEW}$ is executed multiple times and statistically assessed (Line 7). This process is repeated until one of the test input in the archive statistically reproduces the failure raised by the original test input. 

\begin{algorithm}[ht]
\caption{\textsc{CheckMultTimes}: Evaluate a test input multiple times}\label{alg:CheckMultTimes}
    \KwIn{$TI\textsubscript{NEW}$ \tcp*{\small Test input Under Evaluation}
    \nonl $archive$ \tcp*{\small Archive of solutions}
    \nonl $SUT$ \tcp*{\small System Under Test}
    \nonl $cluster$ \tcp*{\small Cluster of failures}}
    
    \KwOut{$TI\textsubscript{NEW}$ \tcp*{\small Minimized failure inducing test inputs} 
    \nonl $archive$ \tcp*{\small Archive of solutions}
    \nonl $reduction$ \tcp*{\small Test input reduction range}}

    $statisticalVerdict$ $\gets$ \textsc{ExecuteMultTimes}($TI\textsubscript{NEW}, SUT$);
    
    \While{$statisticalVerdict$ == 0}{
    
        $TI\textsubscript{NEW}, verified , archive$, $reduction$ $\gets$ \textsc{UndoMinimization}($TI\textsubscript{NEW}, archive$); \\
        \eIf{verified==1}{
            \textbf{break;}
        }{
            $statisticalVerdict$ $\gets$ \textsc{ExecuteMultTimes}($TI\textsubscript{NEW}, SUT$);
        }
    }
\end{algorithm}

\revision{Once a candidate that reproduces the target failure is identified, the archive is updated accordingly (Lines 18–21). The algorithm saves the candidate test input with a verification flag: a 1 indicates that it has been accepted through repeated-execution validation, whereas a 0 indicates that it has been provisionally retained after a single-execution screen. This candidate is further reduced (Lines 22–23), and the process of execution, evaluation, and potential rollback is repeated until the reduction value becomes less than or equal to one. At this point, the current reduction phase is complete. The subsequent recursive invocation checks whether the selected candidate has already been validated and, if necessary, performs repeated-execution validation or rolls back to a previously accepted candidate before returning the minimized failure-inducing input.}

After each successful reduction (i.e., the algorithm reaches line 25), the algorithm calls itself, using the current archive as input. This recursive invocation is essential to ensure that the latest candidate is re-evaluated in case it was only executed once. At this point, the algorithm instead of going directly to Line 11, it enters in the Lines 2-10 procedures. In this procedure, the algorithm proceeds by invoking the \textsc{CheckSolutions} function. The objective of this function is to verify that the candidate test input in the archive consistently reproduces the failure. To achieve this, the function first checks whether the current solution has already been assessed (Line 1, Algorithm~\ref{alg:CheckSolutions}). If so, the function returns the current solution as the minimized test input with a reduction value of zero. Conversely, the algorithm enters in a while loop (Lines 2-8). It can happen that the solution in the archive has not already been statistically assessed (i.e., executed several times), therefore it is executed multiple times (i.e., 30 times default) and compared to the original results (Line 7). This process continues until finding a solution in the archive that statistically reproduces the original failure (i.e, $verified$ is 1 or $statisticalVerdict$ is one). When this procedure finishes it can happen that between current solution and the previous one there is still a range of minimization (i.e.,$ reduction >0$) or there is no range of minimization (i..e, $reduction == 0$). In this case, the solution is saved to the archive (Line 5, Algorithm~\ref{alg:OptimizedStochasticDeltaDebugging}) and the algorithm returns the current archive. In the former case, the solution is saved to the archive and the minimization procedure starts again.  

\begin{algorithm}[ht]
\caption{\textsc{CheckSolutions}: Check the solutions in the archive} \label{alg:CheckSolutions}
    \KwIn{$TI\textsubscript{NEW}$ \tcp*{\small Test input Under Evaluation}
    \nonl $archive$ \tcp*{\small Archive of solutions}
    \nonl $SUT$ \tcp*{\small System Under Test}
    \nonl $cluster$ \tcp*{\small Cluster of failures}}
    
    \KwOut{$TI\textsubscript{NEW}$ \tcp*{\small Minimized failure inducing test inputs} 
    \nonl $archive$ \tcp*{\small Archive of solutions}
    \nonl $reduction$ \tcp*{\small Test input reduction range}}
    
    $TI\textsubscript{NEW}, verified , archive$, $reduction$ $\gets$ \textsc{UndoMinimization}($TI\textsubscript{NEW}, archive$);
            
    \While{$verified$ == 0}{
        $statisticalVerdict$ $\gets$ \textsc{ExecuteMultTimes}($TI\textsubscript{NEW}, SUT$);\\
        \eIf{$statisticalVerdict$ == 1}{
            \textbf{break;}
        }{
            %$reduction$ $\gets$ getReductions($TI\textsubscript{NEW}, archive$); 
            $TI\textsubscript{NEW}, verified , archive$, $reduction$ $\gets$ \textsc{UndoMinimization}($TI\textsubscript{NEW}, archive$); 
        }
    }
\end{algorithm}

\subsubsection{Environment-Wise Optimized Stochastic Delta Debugging (EWDD\textsubscript{OS})}\label{sec:ODD}

\begin{figure}[h]
    \centering
    \includegraphics[width=\linewidth]{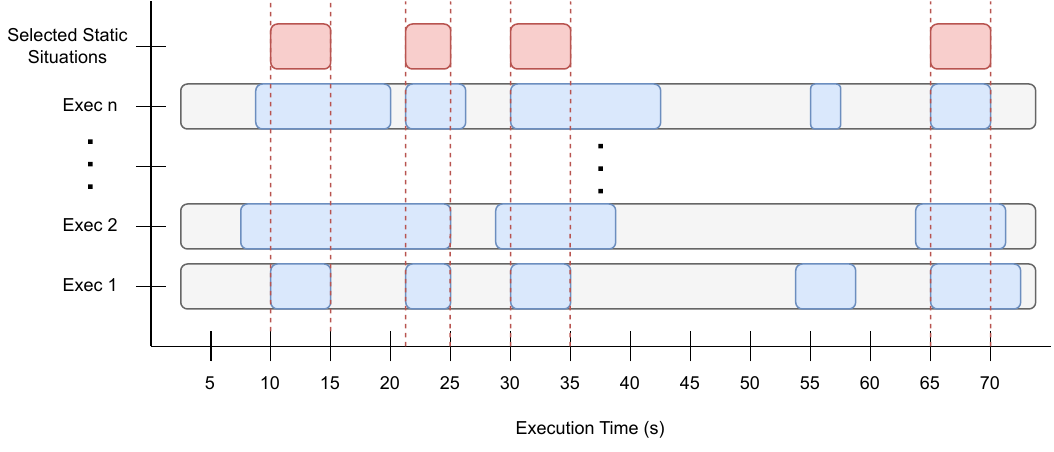}
    %\vspace{-5pt}
    \caption{Static States Selection example: In blue, static states for each execution of the original test input; in red dotted lines, convergence of static states across all executions; and in red, static states in common between all executions.}
    \label{fig:StaticStates}
\end{figure}

We now introduce the proposed Environment-Wise Optimized Delta Debugging algorithm (EWDD\textsubscript{OS})), presented in Algorithm~\ref{alg:Environment-awareDD}. This algorithm extends Algorithm~\ref{alg:OptimizedStochasticDeltaDebugging} by incorporating a preprocessing phase (Algorithm ~\ref{alg:Environment-awareDD} Lines 6-15) that handles static situations and generates an initial approximation of the minimized test input. Before the execution of Algorithm~\ref{alg:Environment-awareDD}, the results from the original test executions are analyzed to identify clusters of failures and the corresponding static situations for each execution of the original test input. Then, all the static situations from all the executions inside the failing cluster (i.e., the cluster of the selected failure) are taken. Due to the inherent non-determinism of the CPS and Simulator, these static situations can vary across different runs, particularly in timing. To address this variability, the algorithm selects a common time slot that is shared among all identified static situations at a given points, as Figure~\ref{fig:StaticStates} depicts.

\begin{algorithm}[ht]

\caption{Environment-wise Optimized Delta Debugging Algorithm}\label{alg:Environment-awareDD}

   % \begin{algorithmic}
    \KwIn{$SUT$ \tcp*{\small System Under Test}
    \nonl $F_t$ \tcp*{\small Failing time (Optional)}
    \nonl $TI$ = $\{p_1, p_2,..., p_{n}\}$   \tcp*{\small Initial failure inducing test input} 
    \nonl $archive$ \tcp*{\small Archive of solutions (Optional)}
    \nonl $cluster$ \tcp*{\small Cluster of failures}
    \nonl $staticSituations$ \tcp*{\small Array of Static Situations}}
    \KwOut{$archive$= $\{TI_1, TI_2,..., TI_n\}$ \tcp*{\small Minimized failure inducing test inputs} }

    $TI\textsubscript{NEW}$ $\gets$ $TI$;\\
    \eIf{$archive.n > 0$}{
        Lines 3-9 Algorithm~\ref{alg:OptimizedStochasticDeltaDebugging}      
    }{
        
        $TI'$, $archive$, $reduction$ $\gets$ \textsc{InitArchive}($TI,F_t$); \\
        \If{$staticSituations.n$ > 0}{
            \While{$verdict$ == 0}{
                $currentStatic$ $\gets$ \textsc{GetStaticSituation}($staticSituations$, $F_t$);\\
                $TI'$ $\gets$ \textsc{ReduceTestInput}($currentStatic$);\\
                \textsc{UpdateEnv}($currentStatic$);\\
                $Verdict$ $\gets$ \textsc{Execute}($TI\textsubscript{NEW}, SUT$, $cluster$);\\
                \If{$verdict$ == 1}{
                    $verdict$ $\gets$ \textsc{ExecuteMultTimes}($TI', SUT, cluster$);\\
                }
                $F_t$ $\gets$ \textsc{GetStaticTime}($currentStatic$);
            }
            $archive$ $\gets$ \textsc{SaveToArchive}($TI', 1$);
        }
        
    }
    
    Lines 13-25 Algorithm~\ref{alg:OptimizedStochasticDeltaDebugging}
   % \end{algorithmic}
  % \vspace{-0.3cm}
\end{algorithm}

Once this result analysis phase is completed, the execution of Algorithm~\ref{alg:Environment-awareDD} begins. If it is the first time that this algorithm is executed the archive is then initialized with the original test input (Line 5). If static situations are present the algorithm enters in a loop (Lines 7-16) that iteratively refines the test input to ensure that it still reproduces the original failure. In each iteration, the algorithm selects the last static situation before the failure (Line 8). Then in Lines 9 and 10, the algorithm reduces the test input and updates the environment according to the static situation (i.e., the elevator positions is adjusted). The reduced test input is then executed once. If that reduced test input is able to reproduce the original failure, then it is assessed multiple times (i.e., 30 times by default) so the results are statistically compared to the original results (Line 13). If the reduced test input consistently reproduces the original failure, the algorithm exits the loop and updates the archive with this reduced test input (Line 17). Otherwise, the algorithm iterates with the preceding static situation. After this minimization phase finishes, the algorithm proceeds with the next steps outlined in Algorithm~\ref{alg:OptimizedStochasticDeltaDebugging} (Lines 13-25).

\section{Empirical Evaluation}\label{sec:Evaluation}

This section empirically evaluates the proposed approach using \revision{one industrial and two open-source case study systems}. We aimed at answering the following research questions (RQs):

\begin{itemize}    
    \item \revision{\textit{\textbf{RQ1 -- How effective and efficient is optimized stochastic Delta Debugging compared with the stochastic baseline?}} We compare $DD_{OS}$ with $DD_S$ across the three case studies to assess speculative reduction, deferred validation, and rollback as a combined strategy. $DD_S$ adapts traditional $DD$ to these systems by validating every candidate through repeated executions, whereas $DD_{OS}$ uses single-execution screening and defers repeated validation. Both share the same failure-preservation criterion and $n_{\mathrm{rep}}=30$ per full validation, while total execution counts may differ. We measure effectiveness using TIRR and TIFR, and efficiency using debugging time.}

    \item \textit{\textbf{RQ2 -- How does the environment affect the performance of the delta debugging algorithm?}} As shown in previous work \cite{valle2023applying}, being aware of the environment improved significantly the efficiency of the delta debugging algorithm. However, the stochastic nature of the SUTs could affect the environment. For instance, for the Orona's case study system we take into account the static situations (i.e., when all the elevators are stopped with the doors opened), which could vary among the simulations of the same test case depending on the decisions of the dispatching algorithm. Therefore, with this RQ we aim at assessing whether for stochastic CPSs being aware of the environment also improves the performance of the delta debugging algorithm.

    \item \revision{\textit{\textbf{RQ3 -- How does the number of repeated executions used to assess failure similarity affect the effectiveness and efficiency of stochastic delta debugging?}} This RQ investigates the sensitivity of the approach to ($n_{\mathrm{rep}}$), the number of executions used in each repeated-execution validation of a candidate test input. We examine how varying this parameter affects debugging time, test-input reduction, and target-failure reproducibility, using ($n_{\mathrm{rep}}=30$) as the reference configuration.}

     %\item \textit{RQ4 -- Among the test input reduction techniques, which one performs better?} This RQ aims at assessing whether the test input reduction technique affects the performance of our approach compared to the traditional one. To this end, we compared the performance of our approach against the traditional delta debugging when reducing the input by events. To do so, we executed both algorithms in the industrial case study system, since for the Leo Rover case study system, there are no events.  \pablo{ This could be interesting to analyze, since we are using genetic algorithms, when removing one passenger it could alter the performance of the system.}

     %\item \textit{RQ5 -- Qualitative analysis of our approach.} \pablo{I think this research question can be removed and added as plain discussion to the evaluation part.}
\end{itemize}
\revision{The comparisons assess the proposed components incrementally. RQ1 compares $DD_S$ with $DD_{OS}$ to evaluate speculative reduction, deferred repeated-execution validation, and rollback as a combined strategy. RQ2 compares $DD_{OS}$ with $EWDD_{OS}$ to assess the additional contribution of environment-aware reduction. RQ3 varies the repeated-execution budget of $EWDD_{OS}$ to assess configuration sensitivity, rather than the effect of removing stochastic validation. Notice that, failure clustering and the repeated-execution acceptance criteria are retained across the algorithm variants for a fair comparison across the different proposed algorithms.}
%\subsection{Experimental Set-up}

\subsection{Test Input Characteristics}
For the Orona's case study system, we used their Genetic Dispatching Algorithm and full-day real traffic data from 2 buildings. As test inputs, we considered 3 passenger profiles for each building, which are detailed in Table \ref{table:characteristicsOrona}.

\begin{table}[ht]
\centering
\caption{Characteristics of the considered installations and test inputs for Orona's case study}
\label{table:characteristicsOrona}
\begin{tabular}{lrrrrr}
\toprule
Building & Test Input & \multicolumn{1}{c}{\# of elevators} & \multicolumn{1}{l}{\# of floors} & \multicolumn{1}{l}{\# of passengers} & \multicolumn{1}{c}{\begin{tabular}[c]{@{}c@{}}Execution Time in \\ speed-up simulation (s)\end{tabular}} \\ \cmidrule{1-6}
\multirow{3}{*}{Building 1} & Test Input 1 & 3 & 12 & 3,769 & 368.88 \\
 & Test Input 2 & 3 & 12 & 3,105 & 352.10 \\
 & Test Input 3 & 3 & 12 & 3,294 & 361.22 \\ \cmidrule{1-6}
\multirow{3}{*}{Building 2} & Test Input 4 & 6 & 10 & 6,558 & 351.87  \\
 & Test Input 5 & 6 & 10 & 5,452 & 250.73 \\
 & Test Input 6 & 6 & 10 & 4,467 & 223.82\\ \bottomrule
\end{tabular}

\end{table}

\begin{figure}[ht]
\centering
\begin{subfigure}[!hhtp]{0.32\textwidth}{\label{fig:BarceloaCircuit}\includegraphics[width=\textwidth, trim= 0 0 0 0]{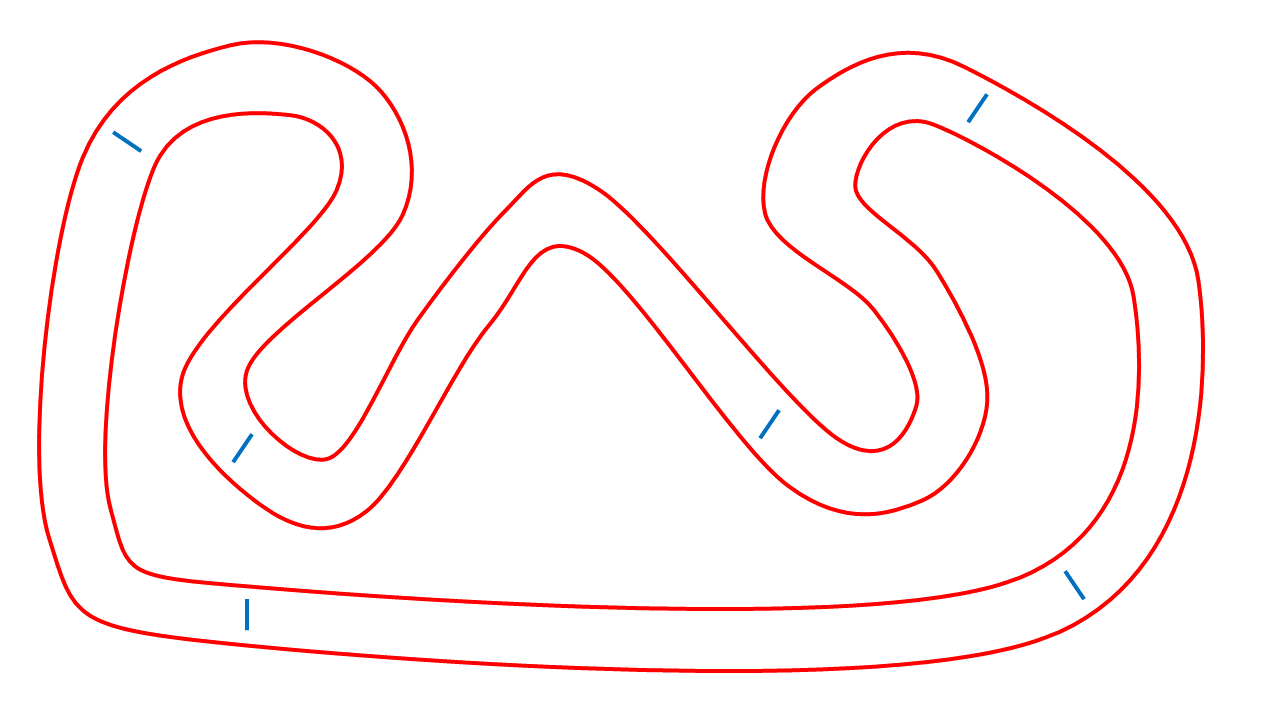}}
\caption{Barcelona Circuit}
\end{subfigure}
\
\begin{subfigure}[!hhtp]{0.32\textwidth}{\label{fig:ImolaCirucit}\includegraphics[width=\textwidth, trim= 0 0 0 0]{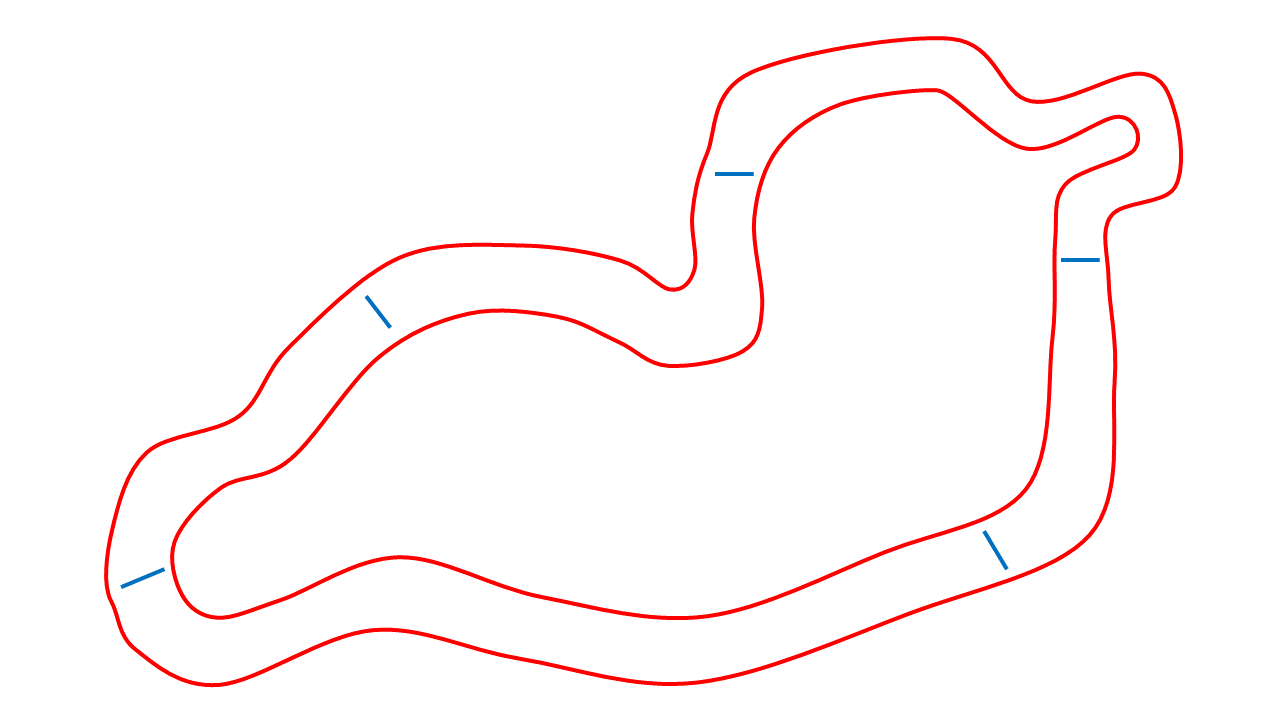}}
\caption{Imola Circuit}
\end{subfigure}
\
\begin{subfigure}[!hhtp]{0.32\textwidth}{\label{fig:ImolaCirucit}\includegraphics[width=\textwidth, trim= 0 0 0 0]{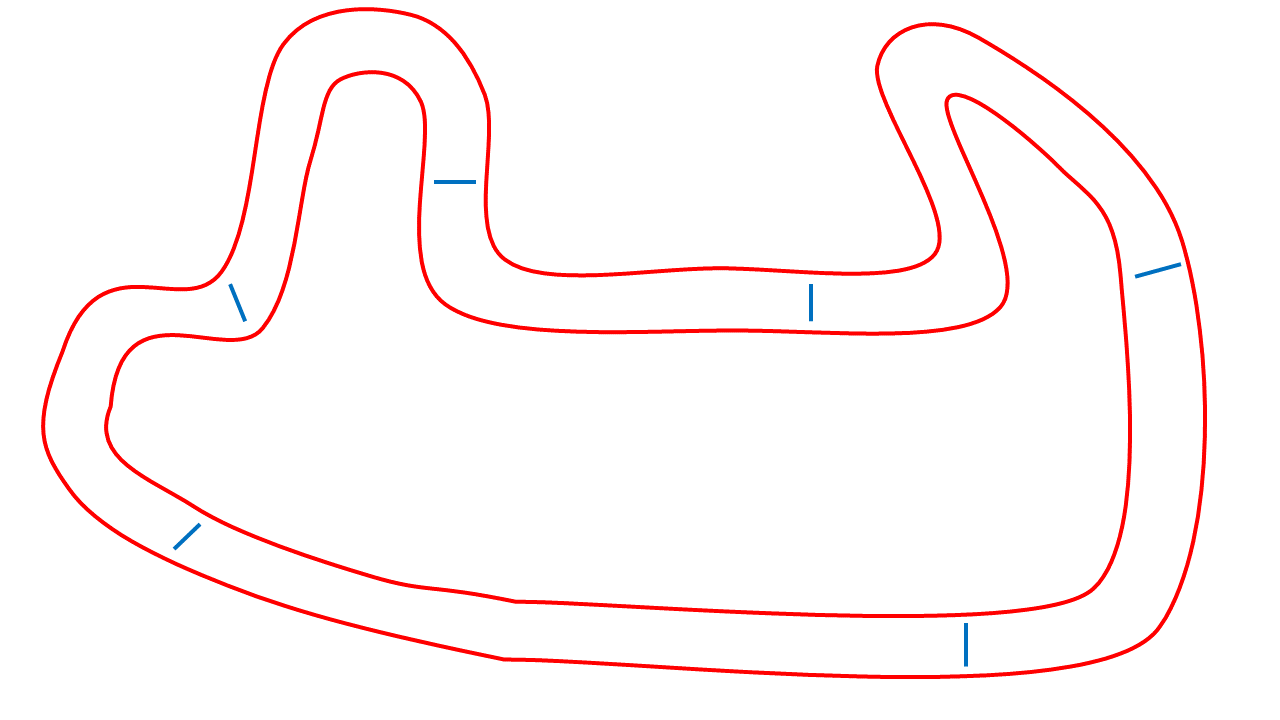}}
\caption{Las Vegas Circuit}
\end{subfigure}

\caption{Realistic F1 circuit scenarios replicated for Leo Rover }
\label{fig:circuits}

\end{figure}

For the Leo Rover case study system, as Figure~\ref{fig:circuits} depicts, we designed 3 scenarios. These scenarios were manually built to replicate real-world F1 circuits, specifically Barcelona, Imola, and Las Vegas circuits. In each circuit, we added small blue lines to simulate events that caused the Leo Rover to stop for a certain period of time (e.g., a pedestrian or another vehicle crossing the road).

\revision{For the VLA case study, we evaluated GR00T N1.7 on five manipulation tasks across two robotic embodiments, as summarized in Table~\ref{table:characteristicsVLA}. The first two tasks, \textit{Pick Up} and \textit{Move Near}, had maximum episode lengths of 80 and 120 action steps, respectively. Both tasks were evaluated in SimplerEnv~\cite{li24simpler} using the simulated Google robot~\footnote{\url{https://everydayrobots.ai/}}, a mobile manipulator. Tasks 3--5 were selected from the LIBERO benchmark~\cite{liu2023libero}. Task 3 required placing two objects into a basket, Task 4 required placing a mug inside a microwave and closing its door, and Task 5 required turning on the stove and then placing a moka pot on it. These three long-horizon tasks were evaluated using a simulated Franka Panda robotic arm, with a maximum episode length of 720 action steps per task.}

\begin{table}[ht]
\color{black}
\centering
\caption{\revision{Characteristics of the considered manipulation tasks for VLA case study}}
\label{table:characteristicsVLA}
\begin{tabular}{lllrrr}
\toprule
Scenario & Environment & Embodiment & \multicolumn{1}{c}{\# of actions} & \multicolumn{1}{c}{\# of atomic tasks} & \multicolumn{1}{l}{\# Task classification}  \\ \cmidrule{1-6}
Task 1 & SimplerEnv & Google Robot & 80 & 1 & short-horizon \\
Task 2 & SimplerEnv & Google Robot & 120 & 2 & medium-horizon \\
Task 3 & Libero & Franka Panda & 720 & 4 & long-horizon \\
Task 4 & Libero & Franka Panda & 720 & 4 & long-horizon \\
Task 5 & Libero & Franka Panda & 720 & 4 & long-horizon \\\bottomrule
\end{tabular}

\end{table}

\subsection{Execution Platform}
On the one hand, we conducted the experiments for Orona's case study on a Windows 11 PC with a dual-core CPU Intel Core i5 $7^{th}$ generation, and 16 GB RAM. As a simulator for executing the tests, we used Elevate 8.19. On the other hand, for the Leo Rover case study, we used an Ubuntu virtual machine running on a Windows 11 PC. The virtual machine comprised 16GB RAM and an 12-core CPU AMD Ryzen 9 7900X with a 16GB NVIDIA GeForce RTX 4080 GPU.

\subsection{Evaluation Metrics}\label{sec:EvalMetrics}
We evaluated our approach from two perspectives: (1) efficiency and (2) effectiveness. The former relates to how fast the algorithm provides the minimized failure-inducing test input, whereas the latter refers to the quality of the provided minimized failure-inducing test inputs. 

\textbf{Efficiency:} To assess the efficiency when comparing both approaches, we measured the execution time required by the algorithms to return the failure-inducing test input. %In addition, we provide the number of times each algorithm executed the test inputs, which provides a more intuitive measure of how efficient one algorithm is compared to the other. 

\textbf{Effectiveness:} Effectiveness relates to the quality of the results obtained by the different algorithms. We assessed the effectiveness of our approaches from two perspectives: (1) Test Input Reduction Ratio and (2) Failure Reproduction Rate. For the first one, we employed the Test Input Reduction Ratio with respect to the failing time ($TIRR_{ft}$), proposed in our previous work \cite{valletowards} which can be calculated as follows:

\begin{equation}
    TIRR_{ft}=1-\dfrac{tet_{fail}(TI')}{tet_{fail}(TI)}
\end{equation}

\noindent where, $tet_{fail}(TI')$ represents the test execution time required to trigger a failure using the minimized test input and $tet_{fail}(TI)$ represents the test execution time required to trigger a failure using the original test input. In addition, to complement the measure of the test input reduction ratio, we provide the simulation time needed to execute the failure for both case studies and the number of passengers of the minimized test input for Orona's case study. This way, we provide an intuitive measure of the effectiveness of our approach.

On the other hand, since the case studies are stochastic, it could be that sometimes the failure is not reproducible with the generated test case, so we measured the failure ratio ($TIFR$). This metric aims at giving confidence about the minimized test inputs to the user by assessing how reproducible the original failure is with the provided minimized test inputs. So, the higher this metric, the more probable it is to reproduce the failure with the minimized test inputs.

\subsection{Experimental runs and statistical tests.} Due to the stochastic nature of the considered algorithms, we executed them multiple times. To answer the RQs, we executed each algorithm 10 times on each scenario. Therefore, in total for the Orona's case study system, we executed 2 (buildings) $\times$ 3 (test inputs) $\times$ 3 (traditional, improved DD and environment-aware improved DD) $\times$ 10 (runs)  = 180 runs . On the other hand, for the Leo Rover case study system, we executed 3 (scenarios) $\times$ 3 (traditional, improved DD and environment-aware improved DD) $\times$ 10 (runs) = 90 runs. In summary, in our evaluation, we made a total of 270 runs, which were parallelized by case study, each one on one computer. Notice that the multiple execution of the Leo Rover case study was not parallelized, since it required real-time simulation and the computational resources required to execute the experiments were high, so we could execute only one instance of it in the computer. For the Orona's case study, we were able to parallelize the executions by the algorithm and building, therefore, the cumulative number of executions for this case study was equivalent to 30 runs, since we could execute 6 instances of Elevate at the same time. 

For all RQs, we assessed the statistical significance of the difference between the results by the different algorithms. We first analyzed how the data was distributed by employing the Shapiro-Wilk test. Since the data in some cases was normally distributed and in other cases it was not normally distributed, we used the ANOVA test and the Wilcoxon rank sum test according to the distribution of the data. We considered that there was statistical significance between the compared techniques when the p-value was below 0.05. In addition, we evaluated the effect sizes through the Vargha and Denaley's \^{A}$_{12}$ value, which according to Romano et al. \cite{romano2006exploring} the effect size of the \^{A}$_{12}$ value can be categorized as \textit{negligible} if $d  < 0.147$, \textit{small} if $d <$ 0.33 , \textit{medium} if $d < 0.474$ and \textit{large} if $d \geq 0.474$ , where $d = 2|$\^{A}$_{12}$$ - 0.5|$.

\subsection{Configuration of the algorithms}
The configurability of our approach is essential, as it enables the adaptation from one case study to another. In addition to the typical modifications required to handle differences in the test inputs and simulation outputs, we had to adapt our approach in two perspectives: (1) the clustering process and (2) the comparison of a single run against the original test inputs. \revision{In addition to the case-specific failure representations and clustering methods, the algorithms are configured through the repeated-execution budget ($n_{\mathrm{rep}}$) and the acceptance criteria described in Section 4.3. For RQ1 and RQ2, we set ($n_{\mathrm{rep}}=30$). RQ3 varies this parameter while retaining the same case-specific failure representation, clustering method, acceptance thresholds, and stopping criteria.}

As mentioned in Section \ref{sec:SDD}, the first step of our approach is to categorize the failures \revision{into different clusters. Incorrect grouping may affect candidate acceptance since merging distinct failures can lead to accepting a different failure, whereas separating manifestations of the same failure can cause valid reductions to be rejected.} Since clustering is an unsupervised method, selecting \revision{an appropriate} number of clusters based on the executions of the initial test input was essential. To select the number of clusters, we used the Bayesian Information Criterion (BIC) \cite{schwarz1978estimating}, which we found more suitable than Silhouette Score \cite{rousseeuw1987silhouettes} and the Akaike Information Criterion (AIC) \cite{akaike1974new}. Although the Silhouette Score is widely used since it measures the cluster cohesion and separation, it does not penalize model complexity, making it less reliable for high-dimensional or noisy data, such as our data. Similarly, AIC evaluates how well a model fits but tends to favor more complex models, increasing the chances of overfitting. In contrast, BIC applies a stronger penalty for model complexity, preventing overfitting and providing a more robust and generalizable clustering solution.

\begin{figure}[ht]
\centering
\begin{subfigure}[!hhtp]{0.49\textwidth}{\label{fig:K-Means}\includegraphics[width=\textwidth, trim= 40 0 40 0]{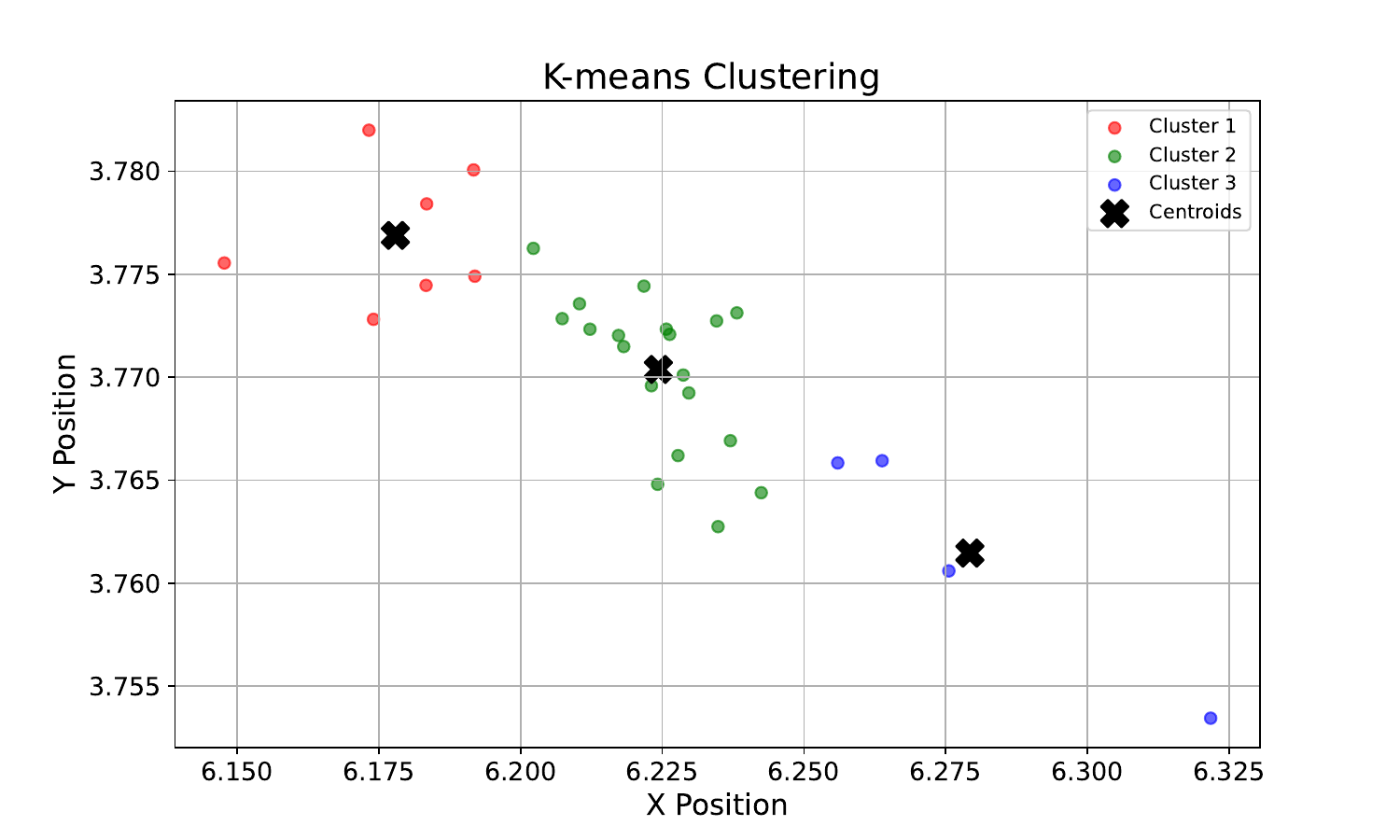}}
\end{subfigure}
\
\begin{subfigure}[!hhtp]{0.49\textwidth}{\label{fig:Gaussian Mixture}\includegraphics[width=\textwidth, trim= 40 0 40 0]{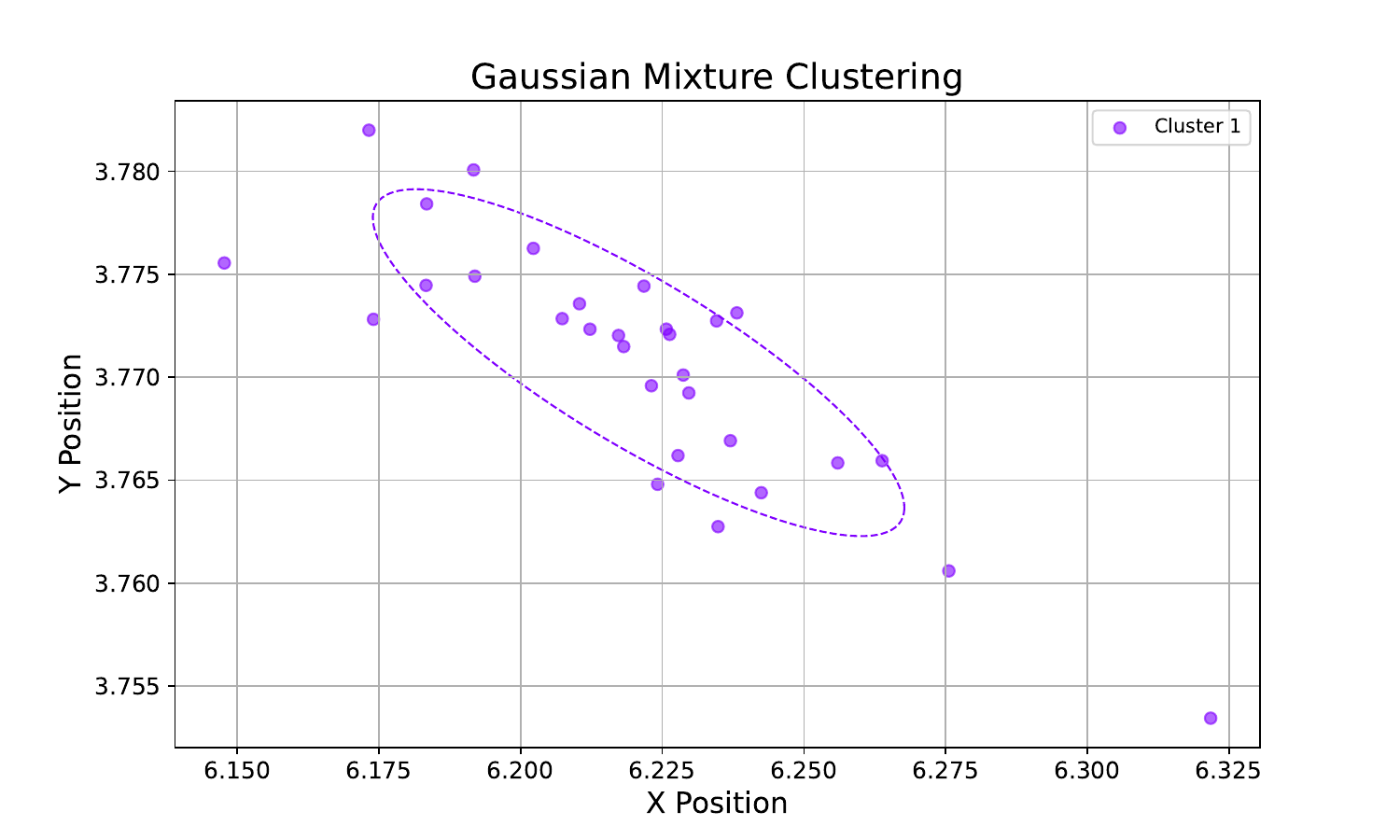}}
\end{subfigure}

\caption{K-Means clustering approach with 3 clusters as optimal cluster number VS Gaussian Mixture clustering approach with 1 cluster as optimal cluster number}
\label{fig:clusters}

\end{figure}

\revision{The clustering technique is a configurable component of failure characterization, while the subsequent reduction algorithms remain unchanged. For a previously unseen CPS, configuration should begin by identifying observable failure characteristics relevant to the test oracle. Repeated executions of the original input can then be used to assess suitable clustering models and their configurations, combining model-selection criteria with domain knowledge. The selected characterization is retained during reduction to provide a consistent reference for candidate evaluation. Accordingly, we used two different clustering methods, one for each case study system.} For \revision{Orona's} case study system, we selected the K-Means clustering algorithm. K-Means is effective when the Euclidean distance metric provides a meaningful measure of similarity. \revision{This choice supports grouping nearby observations in the passenger-based failure representation.} In contrast, for the Leo Rover case study, we opted for the Gaussian Mixture (GMM) approach. This method assumes that the data is generated from a mixture of Gaussian distributions, making it well-suited for overlapping clusters. Unlike K-Means, which assigns each point to a single cluster, GMM provides soft assignments, \revision{expressing component-membership probabilities. In our implementation, each observation is assigned to the component with the highest posterior probability, providing the labels used during candidate evaluation.} When trying the K-Means algorithm in this case study, as depicted in Figure~\ref{fig:clusters}, we found that it was very sensitive to small variations in the data. For instance, the \revision{selected} number of clusters was 3 for data with a standard deviation of 5 cm (i.e., almost negligible for this case study), whereas the \revision{selected} number of clusters for GMM was 1.

\revision{
\subsection{Sensitivity Analysis Protocol}
To answer RQ3, we investigate how the number of executions used for statistical failure validation affects the effectiveness and efficiency of the proposed approach across all three case studies. We consider $n_{\mathrm{rep}} \in {10,20,40}$ and compare these settings with the default configuration, $n_{\mathrm{rep}}=30$. The selected values include two smaller sampling budgets and one larger budget, allowing us to examine the trade-off between debugging cost, test-input reduction, and failure reproducibility. We also tested $n_{\mathrm{rep}}=50$. However, individual minimization runs could exceed 24 hours, making the complete evaluation prohibitively time-consuming. We therefore excluded this setting from the sensitivity analysis. For the same computational-cost considerations, we reduced the number of separate minimization runs per scenario and tested setting from 10, as used in the main evaluation of RQ1 and RQ2, to 5. Each run starts from the corresponding original test input, retaining the same case-specific implementation and statistical acceptance criteria. The selected $n_{\mathrm{rep}}$ value is used both to characterize the original failure and to validate reduced candidates. The analysis therefore examines changes in the overall sampling budget, rather than candidate-validation sampling alone. For each completed run, we record execution time, TIRR, and TIFR as defined in Section~5.3, and summarize the results separately for each scenario and repetition setting. 

}

\section{Analysis of the Results}\label{sec:analysisi of the results}
This section presents and discusses the results of our empirical evaluation according to the research questions introduced. For each research question, we first report the quantitative results and their statistical analysis, and then discuss the main observations and their implications. In particular, we examine the performance of the proposed techniques, their ability to reduce failure-inducing test inputs, and the extent to which minimization affects failure reproducibility under stochastic executions.

\subsection{RQ1 - Performance}
\revision{To answer RQ1, we compare $DD_{OS}$ with $DD_S$ across the three case studies. We use $DD_S$ as the baseline because it directly adapts classical Delta Debugging to stochastic executions, retaining its reduction strategy while replacing single-execution decisions with repeated-execution assessment. This adaptation is necessary because the same candidate may reproduce the target failure in one run and pass or produce a different failure in another. Repeated runs allow the baseline to assess target-failure preservation under this variability. Consequently, $DD_S$ provides a reference for evaluating the additional benefits of speculative reduction, deferred validation, and rollback under the same failure-preservation criterion.} We evaluated the effectiveness using two evaluation metrics presented in Section~\ref{sec:EvalMetrics}: The Test Input Reduction Ratio (TIRR) and the Test Input Failure Reproduction Ratio (TIFR). Additionally, we assessed the efficiency of the algorithms by measuring the execution time.

\subsubsection{Orona's case study}
%The execution time results, as depicted in Figure~\ref{fig:Results_Orona_execTime}, show a clear evidence of the enhanced efficiency achieved by the optimized approach ($DD_{OS}$) compared to $DD_S$. Across all scenarios, $DD_{OS}$ consistently achieves lower median execution times in all the cases, which indicates that this approach is more efficient. Not only median values but the spread of the execution time is also lower in 5 out of the 6 Scenarios. Despite we only found statistical significance in 2 of the 6 Scenarios, in all the cases the Vargha and Delaney \^{A}$_{12}$ value and Cohen's d value indicated at least small effect sizes in favor of $DD_{OS}$ as presented in Table~\ref{tab:stats_testsRQ1Orona}. To be more precise, for Scenario 5 and 6 there was statistical significance in favor of $DD_{OS}$ with large effect sizes. After analyzing the failures of the original test inputs, for all the scenarios in which $DD_{OS}$ outperforms $DD_S$ we found that the failing point was at the end of the test input, at least in all these cases 200 seconds were needed to reproduce the failure with the original test input. However, for the other test cases, to reproduce the failure between 100 and 175 seconds were needed. We conclude that the performance gain in the execution time of $DD_{OS}$ is more pronounced as the execution time to reproduce the failure increases thus becoming the Optimized version of the Delta Debugging algorithm better for long test cases.

\begin{figure}[h!]
    \centering
    \includegraphics[width=0.75\linewidth]{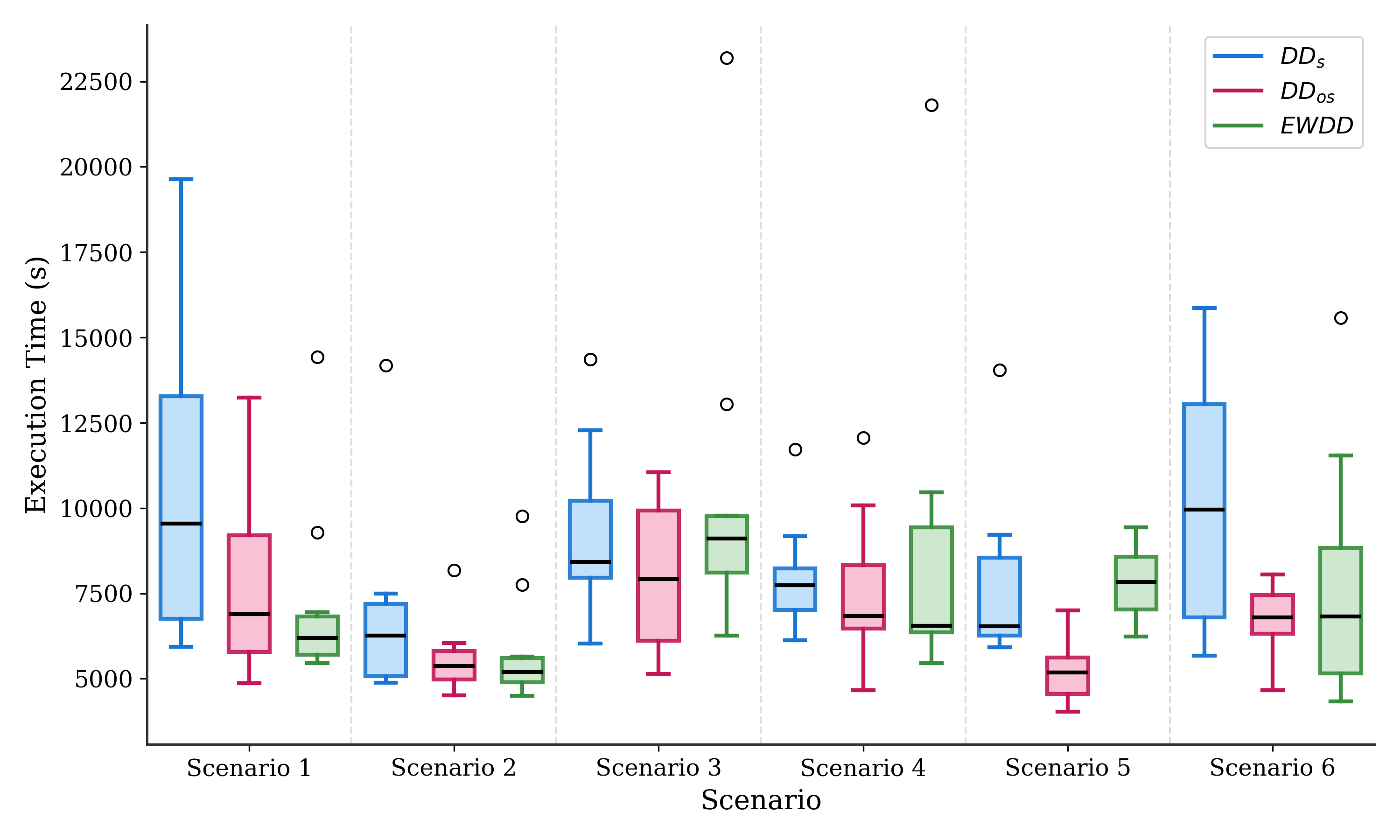}
    %\vspace{-5pt}
    \caption{Execution time for Orona's case study}
    \label{fig:Results_Orona_execTime}
\end{figure}

The execution time results (Figure~\ref{fig:Results_Orona_execTime}) provide clear evidence of the improved efficiency achieved by the optimized approach ($DD_{OS}$) compared to $DD_S$. Across all scenarios, $DD_{OS}$ consistently yields lower median execution times, indicating superior efficiency. In addition to the reduction in median values, the variability in execution time is also lower in five out of the six scenarios. Although statistical significance was observed in only two of the six scenarios, both the Vargha and Delaney \^{A}${12}$ and Cohen's $d$ values indicate at least small effect sizes in favor of $DD{OS}$ in all cases (Table~\ref{tab:stats_testsRQ1Orona}). More specifically, Scenarios 5 and 6 show statistically significant improvements in favor of $DD_{OS}$ with large effect sizes. A post-hoc analysis of the original failing test inputs revealed that, in all scenarios where $DD_{OS}$ outperforms $DD_S$, the failure-inducing behavior occurs near the end of the test input. In these cases, reproducing the failure with the original test input required at least 200 seconds. In contrast, for the remaining test cases, failure reproduction required between 100 and 175 seconds. These findings suggest that the execution time improvement provided by $DD_{OS}$ becomes more pronounced as the time required to reproduce failures increases, making the optimized version of the Delta Debugging algorithm particularly suitable for long-running test cases.

%Moving on to Figure~\ref{fig:Results_Orona_tirr} which presents the comparison of the Test Input Reduction Ratio (TIRR) between both approaches, we observe that the optimized version is more capable of reducing the test inputs evidenced by a higher median reduction ratios in all the scenarios. In addition, similar to the execution time, the distribution across different runs is more thigh compared to the baseline. As Table~\ref{tab:stats_testsRQ1Orona} shows, there was statistical significance in favor of $DD_{OS}$ in 3 scenarios (Scenarios 1, 5 and 6) with large effect sizes. In the rest Scenarios, although there was no statistical significance, the effect size was in favor of the $DD_{OS}$ algorithm, which suggest that the $DD_{OS}$ is more effective at reducing test inputs compared to the baseline algorithm. When the failure point is close to the starting point, the initial reduction in both algorithms greatly reduces the test input, allowing minimal space for additional reduction. This makes both algorithms perform similarly in hardly reducible test inputs. Nonetheless, for larger test inputs, the flexibility of $DD_{OS}$ becomes relevant. Its less robust oracle, designed to compare statistically single-run outcomes against several runs, enables it to achieve a larger reduction compared to $DD_S$, highlighting its superior adaptability in handling more complex scenarios.

\begin{figure}[h!]
    \centering
    \includegraphics[width=0.75\linewidth]{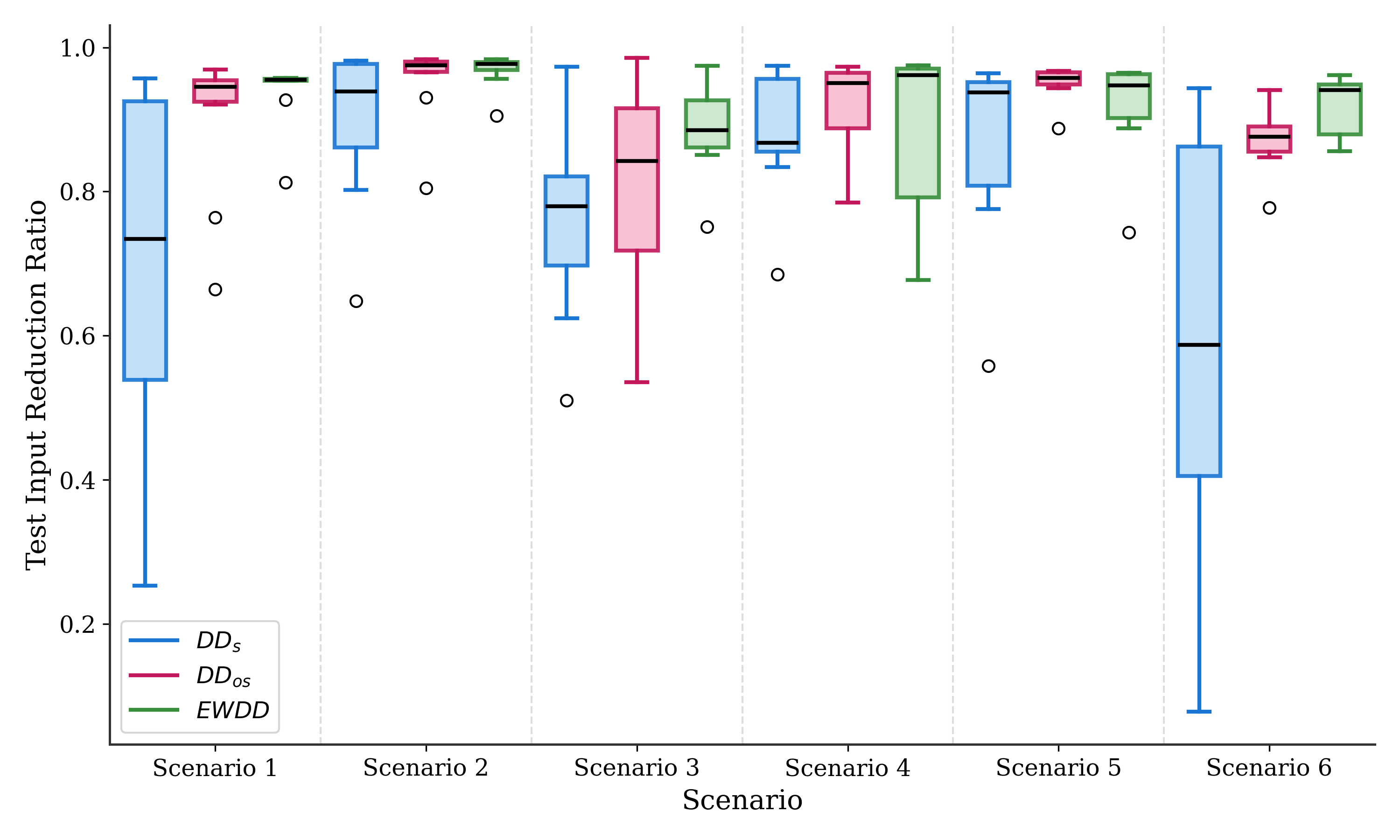}
    %\vspace{-5pt}
    \caption{Test Input Reduction Ratio for Orona's case study}
    \label{fig:Results_Orona_tirr}
\end{figure}

Figure~\ref{fig:Results_Orona_tirr} presents the comparison of the Test Input Reduction Ratio (TIRR) between both approaches. The results show that the optimized version consistently achieves higher median reduction ratios across all scenarios, indicating a greater capability to reduce test inputs. Furthermore, similarly to the execution time results, the distribution across different runs is tighter compared to the baseline approach. As shown in Table~\ref{tab:stats_testsRQ1Orona}, statistically significant differences in favor of $DD_{OS}$ were observed in three scenarios (Scenarios 1, 5, and 6), all with large effect sizes. In the remaining scenarios, although no statistical significance was detected, the observed effect sizes still favor $DD_{OS}$, suggesting that this approach is generally more effective at reducing test inputs than the baseline algorithm. When the failure point occurs near the beginning of the test input, the initial reduction step in both algorithms substantially decreases the input size, leaving limited room for further reduction. Consequently, both approaches exhibit similar performance for scarcely reducible test inputs. However, for longer test inputs, the increased flexibility of $DD_{OS}$ becomes advantageous. Its more permissive oracle, which compares statistically single-run outcomes against multiple executions, allows it to achieve greater reductions than $DD_S$, highlighting its superior adaptability in handling more complex scenarios.

%Figure~\ref{fig:Results_Orona_tifr} focuses on the Test Input Failure Reproduction Ratio (TIFR), which measures the ability of the reduced test inputs to still trigger the original failure. Comparing both approaches, we observe that for Scenarios 2, 4 and 5 both algorithms obtained similar results, that is, maximum reproduction ratio (100\%). However, in the other 3 scenarios $DD_{OS}$ was the algorithm with better results. In addition, as shown by Table~\ref{tab:stats_testsRQ1Orona} for Scenario 1 there is statistical significance in favor of $DD_{OS}$ with a large effect size. In the rest of scenarios there was no statistical significance, although, the effect sizes were in favor of the optimized version of the approach. Comparing both approaches against the original test input, we observed that in both cases the minimized test input obtained higher failure reproduction ratio, thereby suggesting that there are higher probabilities to reproduce the original failure with the minimized test input compared to the original test input.

\begin{figure}[h!]
    \centering
    \includegraphics[width=0.75\linewidth]{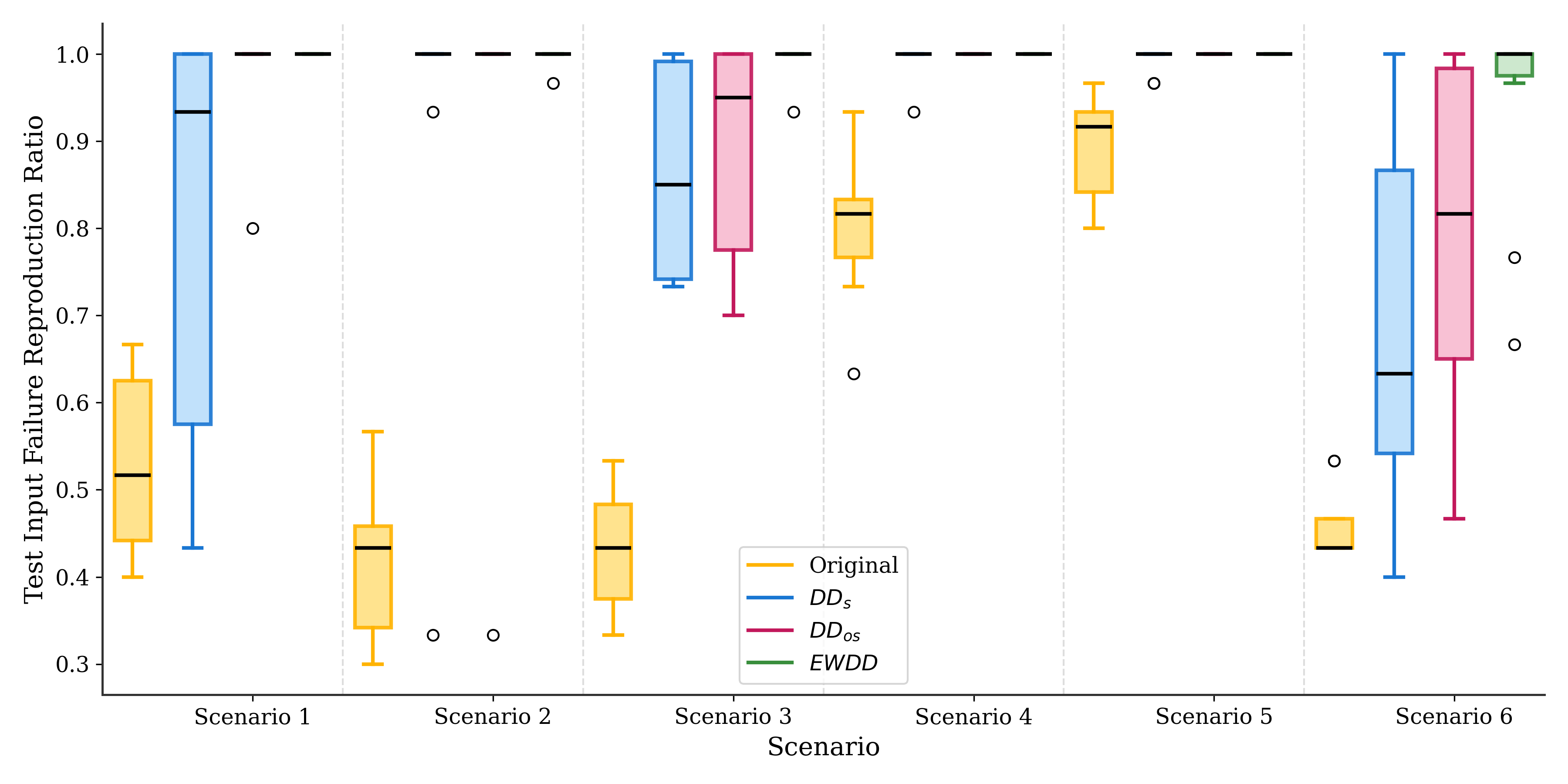}
    %\vspace{-5pt}
    \caption{Test Input Failure Reproduction for Orona's case study}
    \label{fig:Results_Orona_tifr}
\end{figure}

Figure~\ref{fig:Results_Orona_tifr} reports the Test Input Failure Reproduction Ratio (TIFR), which measures the ability of the reduced test inputs to still trigger the original failure. For Scenarios 2, 4, and 5, both algorithms achieved identical results, reaching the maximum reproduction ratio (100\%). However, in the remaining three scenarios, $DD_{OS}$ obtained better results. As shown in Table~\ref{tab:stats_testsRQ1Orona}, Scenario 1 exhibits statistically significant differences in favor of $DD_{OS}$ with a large effect size. In the other scenarios, although no statistical significance was observed, the effect sizes consistently favor the optimized approach. Furthermore, when comparing both approaches with the original test input, the minimized test inputs achieved higher failure reproduction ratios in all cases. This suggests that the reduced test inputs increase the likelihood of reproducing the original failure compared to the original test inputs.

% \usepackage{multirow}
% \usepackage{graphicx}
\begin{table}[ht]
\centering
\caption{Statistical test for efficiency and effectiveness comparison $DD_{OS}$ vs $DD_S$ and $EWDD_{OS}$ vs $DD_{OS}$ for Orona's case stuyd}
\label{tab:stats_testsRQ1Orona}
\resizebox{\textwidth}{!}{%
\begin{tabular}{clrrrrrrrrr}
\toprule
& \multicolumn{1}{c}{\multirow{2}{*}{Scenario}} & \multicolumn{3}{c}{Execution time} & \multicolumn{3}{c}{TIRR} & \multicolumn{3}{c}{TIFR} \\ \cmidrule{3-11}
& \multicolumn{1}{c}{} &  \multicolumn{1}{l}{\^{A}$_{12}$} & \multicolumn{1}{l}{p\_value} & \multicolumn{1}{l}{Cohen's d} & \multicolumn{1}{l}{\^{A}$_{12}$} & \multicolumn{1}{l}{p\_value} &  \multicolumn{1}{l}{Cohen's d} & \multicolumn{1}{l}{\^{A}$_{12}$} & \multicolumn{1}{l}{p\_value} &  \multicolumn{1}{l}{Cohen's d} \\ \cmidrule{1-11}
\multirow{6}{*}{\makecell{$DD_{OS}$ \\ vs \\ $DD_S$}}&Scenario 1 & 0.27 & 0.0917 &  -0.79 & 0.81 & 0.0190 &  1.05 & 0.76 & 0.0211 & 1.04 \\
& Scenario 2 & 0.36 & 0.2899 &  -0.60 & 0.69 & 0.1400 &  0.67 & 0.54 & 0.5838 &  0.03 \\
& Scenario 3 & 0.38 & 0.2994 &  -0.47 & 0.64 & 0.4566 &  0.34 & 0.55 & 0.6968 &  0.27 \\
& Scenario 4 & 0.41 & 0.5768 &  -0.25 & 0.62 & 0.3643 &  0.54 & 0.55 & 0.3173 &  0.44 \\
& Scenario 5 & 0.08 & 0.0014 &  -1.31 & 0.79 & 0.0283 &  0.84 & 0.60 & 0.1462 &  0.67 \\
& Scenario 6 & 0.26 & 0.0131 &  -1.23 & 0.77 & 0.0083 &  1.32 & 0.66 & 0.2428 &  0.54 \\ 
\cmidrule{1-11}
\multirow{6}{*}{\makecell{$EWDD_{OS}$ \\ vs \\ $DD_{OS}$}}& Scenario 1 & 0.44 & 0.6501 & -0.13 & 0.66 & 0.2259 & 0.47  & 0.55 & 0.3173 & 0.44 \\
& Scenario 2 & 0.44 & 0.6501 & 0.16 & 0.5 & 1.0000 & 0.32  & 0.50 & 0.9421 & 0.42 \\
& Scenario 3 & 0.64 & 0.2899 & 0.59 & 0.65 & 0.1732 & 0.63  & 0.76 & 0.0186 & 1.07 \\
& Scenario 4 & 0.51 & 0.9397 & 0.35 & 0.52 & 0.8797 & -0.34  & 0.50 & 1.0000 & 0.00 \\
& Scenario 5 & 0.97 & <0.0001 & 2.66 & 0.32 & 0.1857 &  -0.57 & 0.50 & 1.0000 & 0.00 \\
& Scenario 6 & 0.51 & 0.4041 & 0.38 & 0.78  & 0.0342 & 1.14  & 0.74  & 0.0524 & 0.90  \\

\bottomrule
\end{tabular}%
}
\end{table}

\subsubsection{Leo Rover case study}

\begin{figure}[h!]
    \centering
    \includegraphics[width=0.75\linewidth]{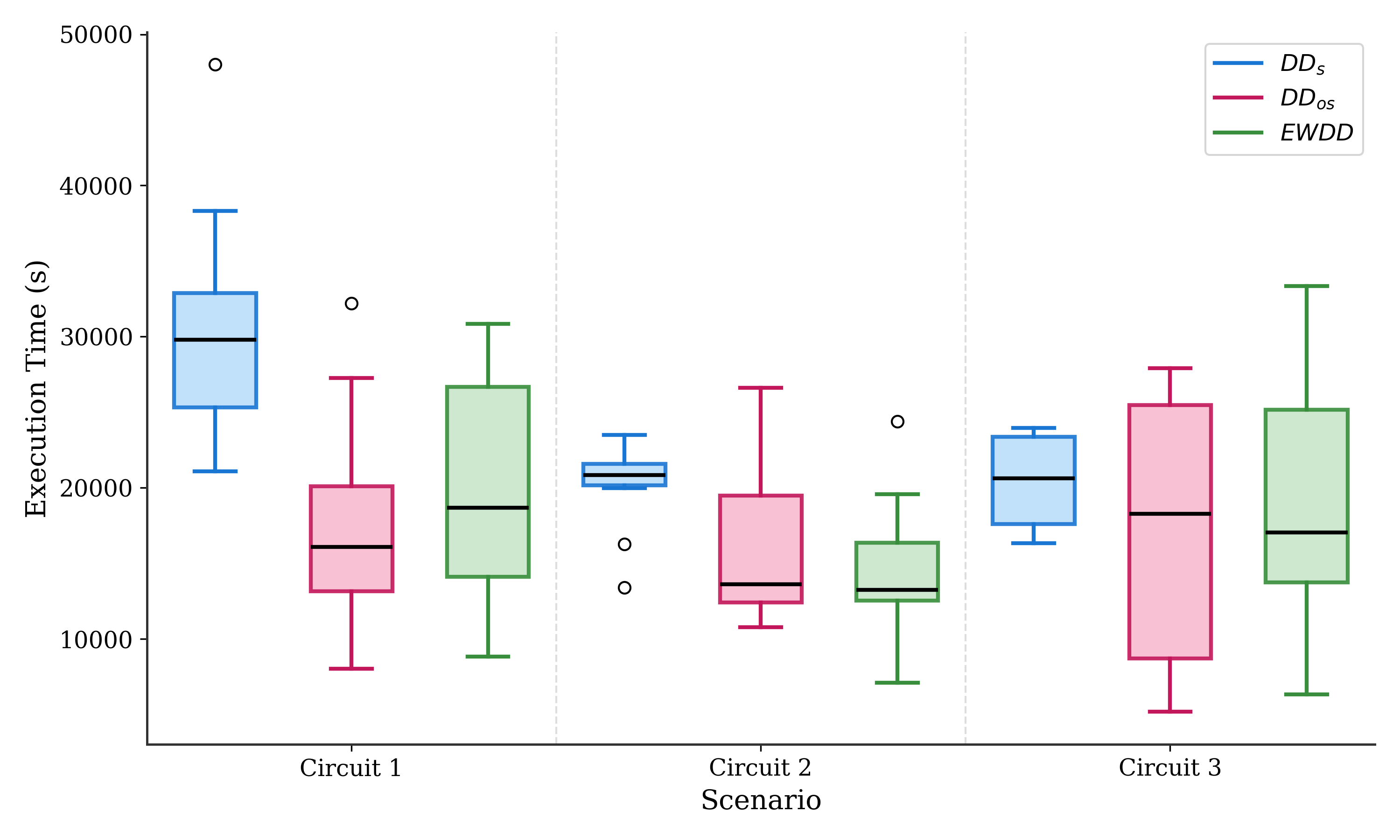}
    %\vspace{-5pt}
    \caption{Execution time for Leo Rover's case study}
    \label{fig:Results_LeoRover_execTime}
\end{figure}

As in the case of Orona's case study system, within the LeoRover case study system, the median test execution time values were also lower for $DD_{OS}$ compared to $DD_S$, as depicted in Figure~\ref{fig:Results_LeoRover_execTime}. Across all scenarios, $DD_{OS}$ consistently yielded lower median execution times, indicating superior efficiency. However, in two of the circuits (Circuits 2 and 3), the variability of $DD_{OS}$ was higher. For these two circuits, there was no statistical significance, although the \^{A}${12}$ were in favor of $DD_{OS}$ (Table~\ref{tab:stats_testsLeoRover}).

\begin{figure}[h!]
    \centering
    \includegraphics[width=0.75\linewidth]{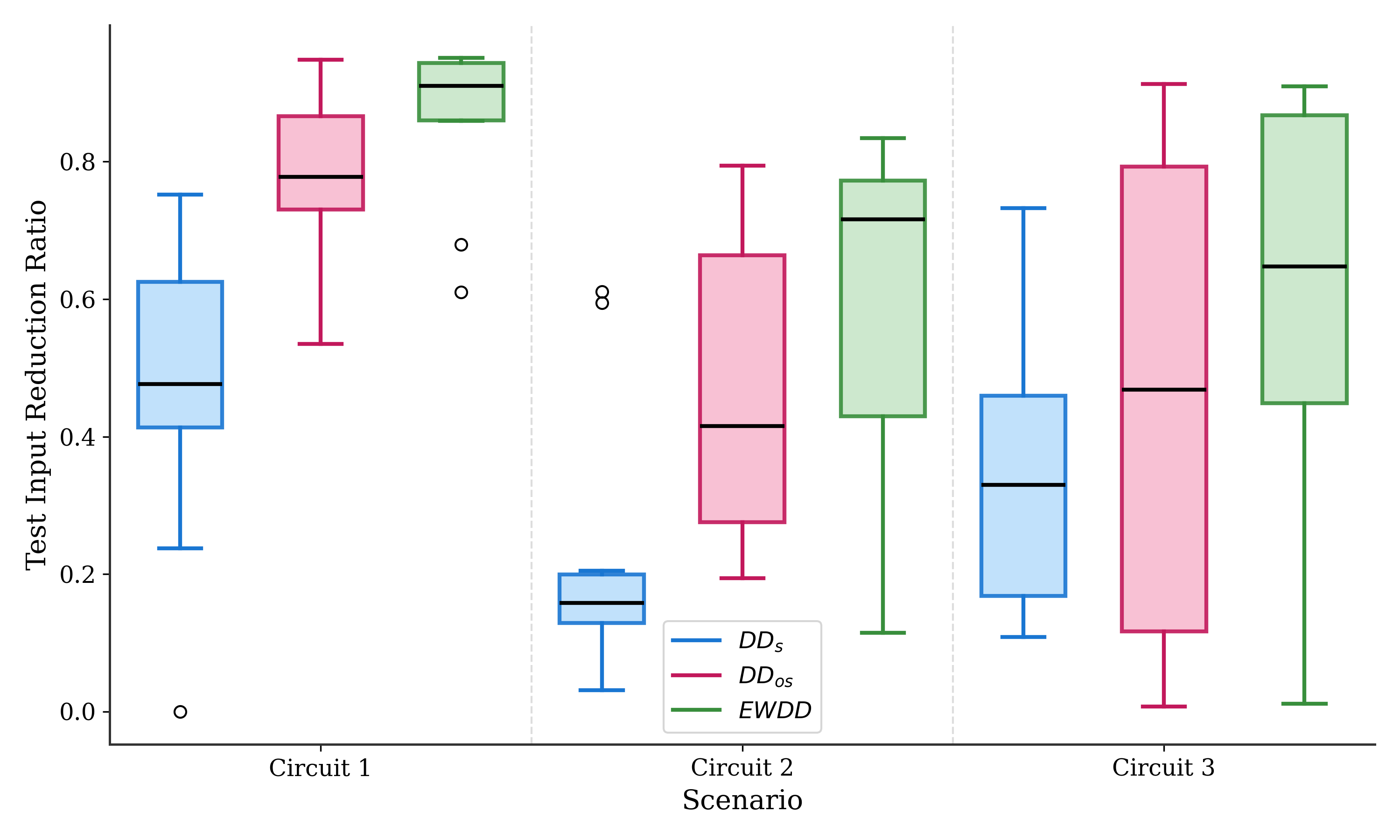}
    %\vspace{-5pt}
    \caption{Test Input Reduction Ratio for Leo Rover's case study}
    \label{fig:Results_LeoRover_tirr}
\end{figure}

Figure~\ref{fig:Results_LeoRover_tirr} presents the comparison of the Test Input Reduction Ratio (TIRR) between both approaches for the Leo Rover case study system. The results show that the $DD_{OS}$ algorithm consistently achieves higher reduction of the test inputs, indicating a greater capability to reduce test inputs, which leads to lower test execution times, and therefore, improving debugging efficiency. These results were statistically significant, with large effect sizes for Circuits 1 and 2, but there was no statistical significance in Circuit 3, as shown in Table~\ref{tab:stats_testsLeoRover}.

\begin{figure}[h!]
    \centering
    \includegraphics[width=0.75\linewidth]{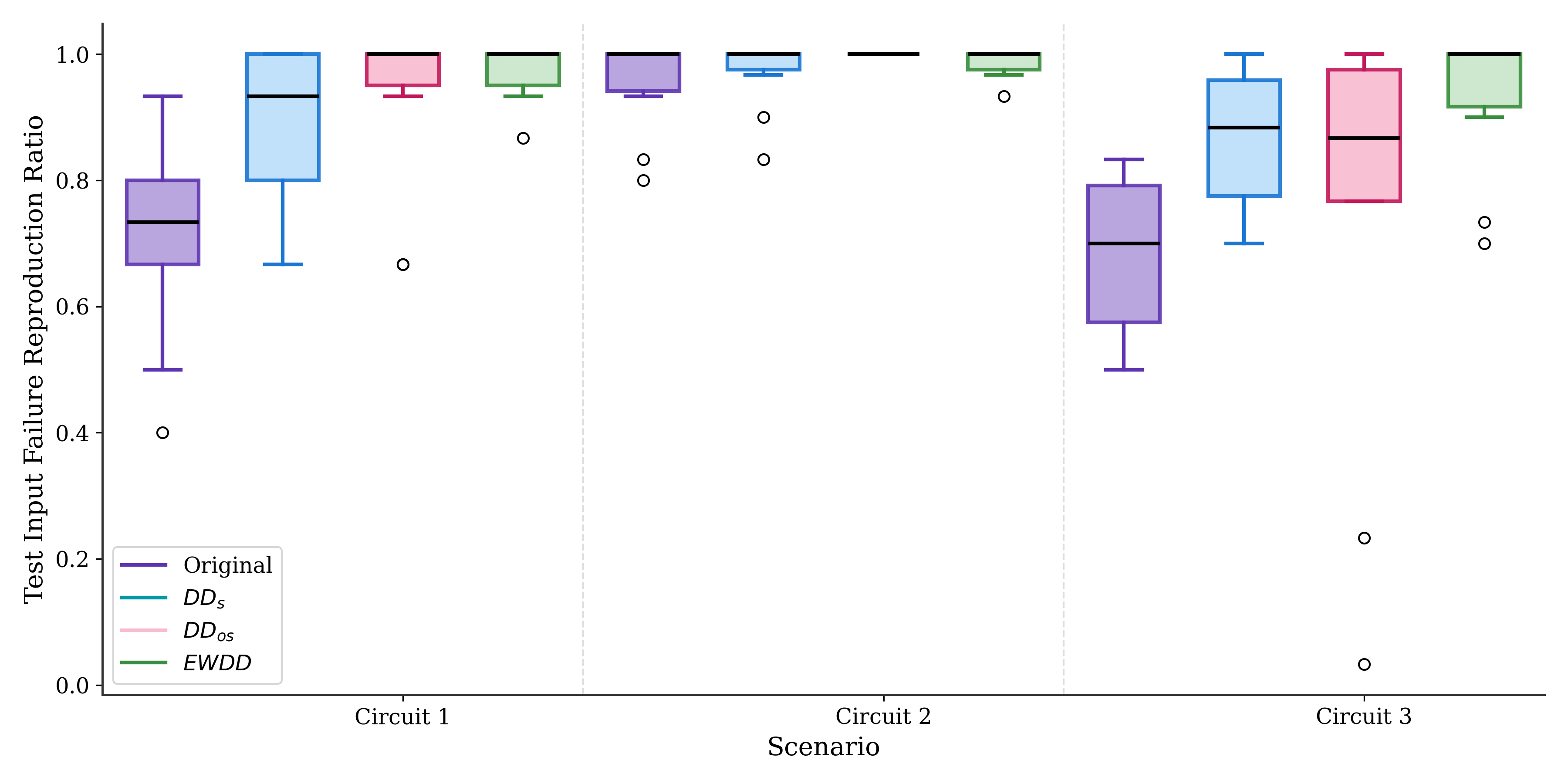}
    %\vspace{-5pt}
    \caption{Test Input Failure Reproduction for Leo Rover's case study}
    \label{fig:Results_LeoRover_tifr}
\end{figure}

Figure~\ref{fig:Results_LeoRover_tifr} reports the Test Input Failure Reproduction Ratio (TIFR), which measures the ability of the reduced test inputs to trigger the original failure, thereby helping engineers debug their faults. In this case study system, both delta debugging algorithms significantly help in reproducing the original failure. Indeed, for Circuits 1 and 2, the median TIFR value for the $DD_{OS}$ algorithm was 100\%, indicating that in most cases this algorithm was able to reproduce the exact same failure as the original one. In Circuit 3, this was slightly reduced, although results were good too (median values above 80\%). The results were slightly in favor of the $DD_{OS}$ in Circuits 1 and 2, and comparable with $DD_S$ in Circuit 3.

%For Scenarios 2, 4, and 5, both algorithms achieved identical results, reaching the maximum reproduction ratio (100\%). However, in the remaining three scenarios, $DD_{OS}$ obtained better results. As shown in Table~\ref{tab:stats_testsRQ1Orona}, Scenario 1 exhibits statistically significant differences in favor of $DD_{OS}$ with a large effect size. In the other scenarios, although no statistical significance was observed, the effect sizes consistently favor the optimized approach. Furthermore, when comparing both approaches with the original test input, the minimized test inputs achieved higher failure reproduction ratios in all cases. This suggests that the reduced test inputs increase the likelihood of reproducing the original failure compared to the original test inputs.

\begin{table}[ht]
\centering
\caption{Statistical test for efficiency and effectiveness comparison $DD_{OS}$ vs $DD_S$ and $EWDD_{OS}$ vs $DD_{OS}$ for Leo Rover's case study}
\label{tab:stats_testsLeoRover}
\resizebox{\textwidth}{!}{%
\begin{tabular}{clrrrrrrrrr}
\toprule
&\multicolumn{1}{c}{\multirow{2}{*}{Scenario}} & \multicolumn{3}{c}{Execution time} & \multicolumn{3}{c}{TIRR} & \multicolumn{3}{c}{TIFR} \\ \cmidrule{3-11}
&\multicolumn{1}{c}{} &  \multicolumn{1}{l}{\^{A}$_{12}$} & \multicolumn{1}{l}{p\_value} & \multicolumn{1}{l}{Cohen's d} & \multicolumn{1}{l}{\^{A}$_{12}$} & \multicolumn{1}{l}{p\_value} &  \multicolumn{1}{l}{Cohen's d} & \multicolumn{1}{l}{\^{A}$_{12}$} & \multicolumn{1}{l}{p\_value} &  \multicolumn{1}{l}{Cohen's d} \\ \cmidrule{1-11}
\multirow{3}{*}{\makecell{$DD_{OS}$ \\ vs \\ $DD_{S}$}}& Circuit 1 & 0.11 & 0.0015 & -1.66 &  0.92 & 0.0020 &  1.61  & 0.58 & 0.4672 & 0.31 \\
&Circuit 2 & 0.30 & 0.1305 & -0.84  & 0.85 & 0.0081 & 1.04  & 0.65 & 0.0681 & 0.73  \\
&Circuit 3 & 0.44 & 0.3181 & -0.45  & 0.56 &  0.4211 & 0.36  & 0.46 & 0.7892 & -0.49  \\
 \cmidrule{1-11}
\multirow{3}{*}{\makecell{$EWDD_{OS}$ \\ vs \\ $DD_{OS}$}}& Circuit 1 & 0.57 & 0.5767 & 0.25 &  0.74 & 0.0696 &  0.73  & 0.52 & 0.8152 & 0.45 \\
&Circuit 2 & 0.44 &  0.6501 & -0.30 &  0.69 & 0.2022 &  0.59  & 0.35 & 0.0678 & -0.80 \\
&Circuit 3 & 0.56 & 0.7399 & 0.15 &  0.62 & 0.3903 &  0.39  & 0.68 & 0.1528 & 0.74 \\

\bottomrule

\end{tabular}%
}
\end{table}

\subsubsection{\revision{VLA case study}}

\begin{figure}[h!]
    \centering
    \includegraphics[width=0.75\linewidth]{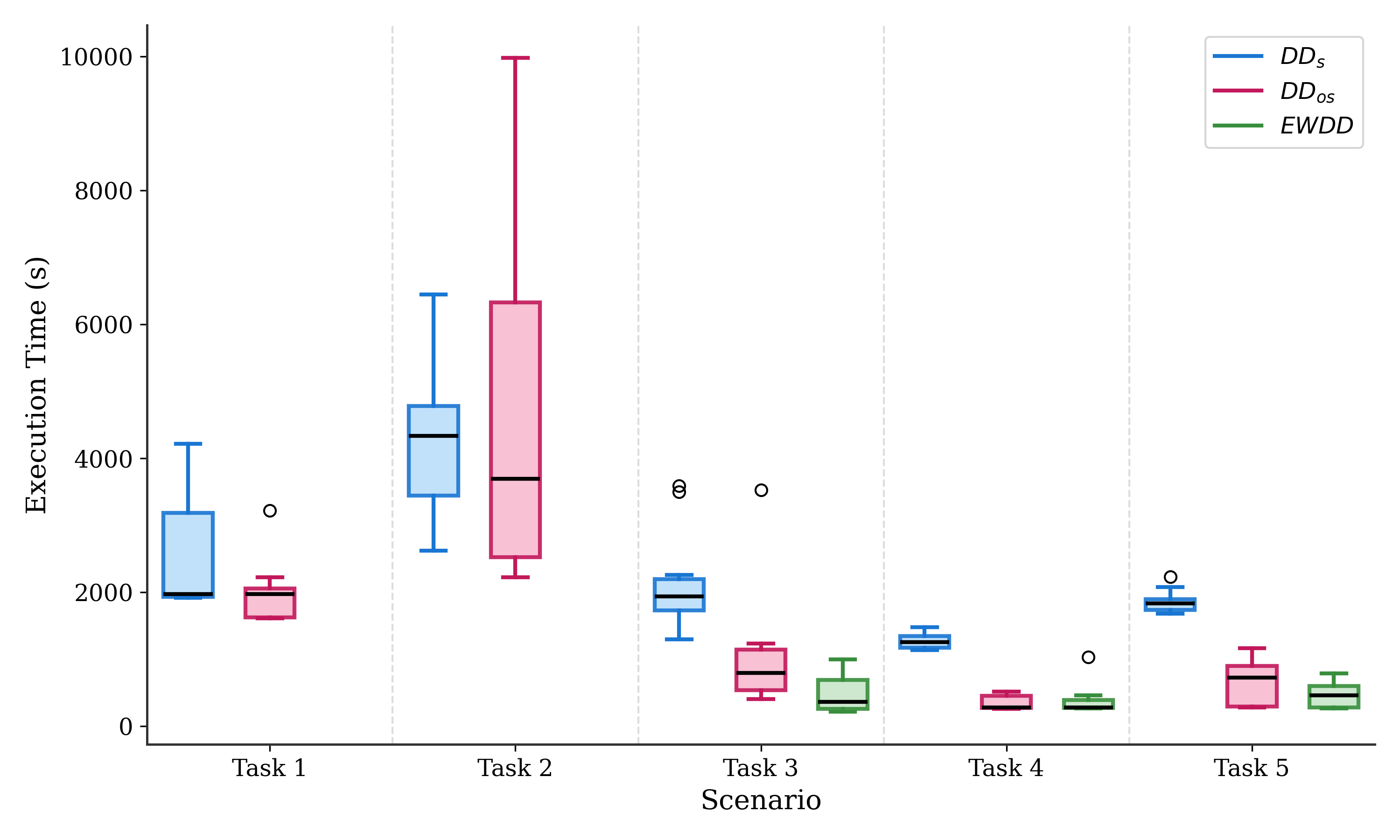}
    %\vspace{-5pt}
    \caption{Execution time for VLA case study}
    \label{fig:Results_VLA_execTime}
\end{figure}

\revision{As in the other case study systems, the execution time results for the VLA case study show benefits from using the optimized approach ($DD_{OS}$) compared to $DD_S$, although these benefits depend on the task, as Figure~\ref{fig:Results_VLA_execTime} depicts. More specifically, Tasks 3, 4, and 5 show statistically significant improvements in favor of $DD_{OS}$, with large effect sizes, as can be seen in Table~\ref{tab:stats_testsVLA}. In Task 1, the effect-size estimates also favor the optimized approach, although no statistical significance was observed. For Task 2, the comparison does not show a clear execution-time advantage. These results indicate that the speculative reduction strategy can improve debugging efficiency in the VLA setting, with the clearest benefits observed in Tasks 3--5.}

\begin{figure}[h!]
    \centering
    \includegraphics[width=0.75\linewidth]{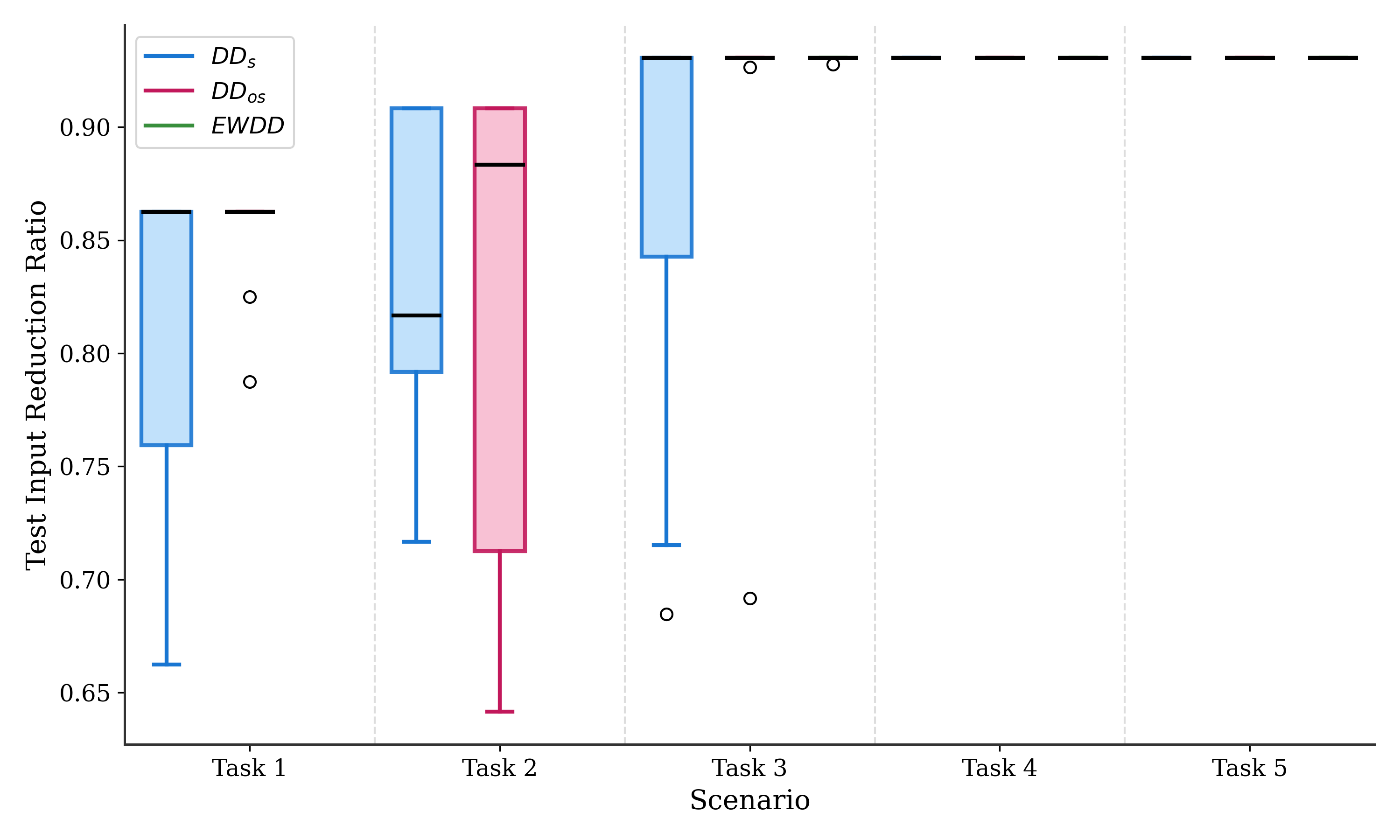}
    %\vspace{-5pt}
    \caption{Test Input Reduction Ratio for VLA case study}
    \label{fig:Results_VLA_tirr}
\end{figure}

\revision{Figure~\ref{fig:Results_VLA_tirr} presents the comparison of the Test Input Reduction Ratio (TIRR) between both approaches. The effect-size estimates favor $DD_{OS}$ in Tasks 1 and 3, suggesting a greater capability to reduce the test inputs in these tasks. However, these differences are not statistically significant as Table~\ref{tab:stats_testsVLA} shows. In Task 2, the $A_{12}$ value is close to 0.5, indicating no clear directional advantage. For Tasks 4 and 5, the reported $A_{12}$ values are 0.50, while the p-values and Cohen's d values are unavailable, showing that the TIRR is the same for both approaches. Overall, the available comparisons do not establish a consistent improvement in reduction effectiveness for either algorithm.}

\begin{figure}[h!]
    \centering
    \includegraphics[width=0.75\linewidth]{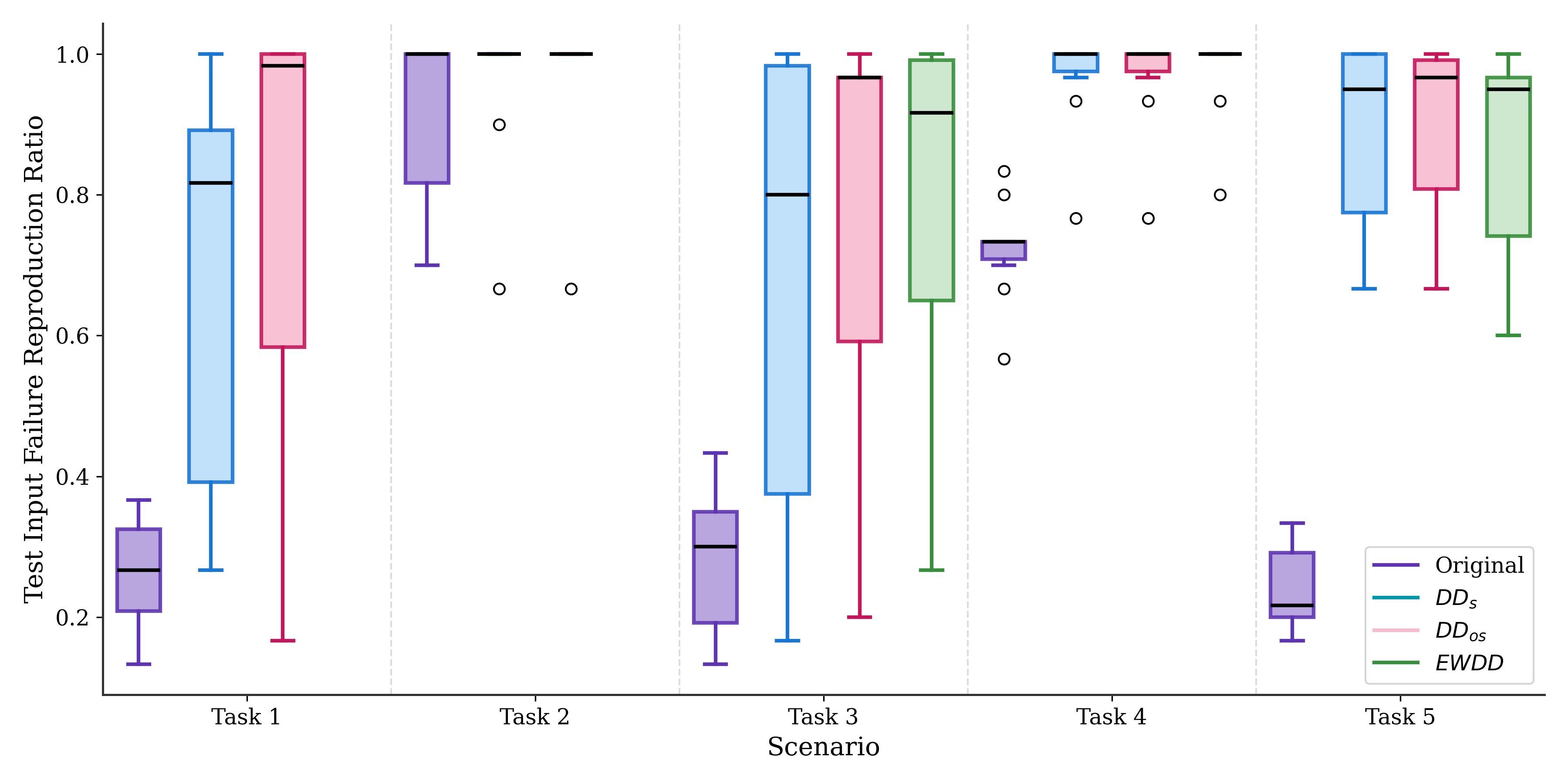}
    %\vspace{-5pt}
    \caption{Test Input Failure Reproduction for VLA case study}
    \label{fig:Results_VLA_tifr}
\end{figure}

\revision{Figure~\ref{fig:Results_VLA_tifr} reports the TIFR for the VLA case study. The effect-size estimates favor $DD_{OS}$ in Tasks 1--3, while Tasks 4 and 5 yield $A_{12}$ values of 0.50. Nevertheless, no statistically significant differences were observed in any of the five tasks, as shown by Table~\ref{tab:stats_testsVLA}. Thus, the main advantage of $DD_{OS}$ in this case study is the reduction in debugging time, particularly in Tasks 3--5, while accounting for a similar TIRR and TIFR compared to $DD_{S}$.}

\begin{table}[ht]
\centering
\color{black}
\caption{\revision{Statistical test for efficiency and effectiveness comparison $DD_{OS}$ vs $DD_S$ and $EWDD_{OS}$ vs $DD_{OS}$ for VLA case study}}
\label{tab:stats_testsVLA}
\resizebox{\textwidth}{!}{%
\begin{tabular}{clrrrrrrrrr}
\toprule
 & \multicolumn{1}{c}{\multirow{2}{*}{Scenario}} & \multicolumn{3}{c}{Execution time} & \multicolumn{3}{c}{TIRR} & \multicolumn{3}{c}{TIFR} \\
\cmidrule{3-11}
 & & $A_{12}$ & p\_value & Cohen's d & $A_{12}$ & p\_value & Cohen's d & $A_{12}$ & p\_value & Cohen's d \\
\midrule
\multirow{5}{*}{\makecell{$DD_{OS}$ \\ vs \\ $DD_S$}} & Task 1 & 0.35 & 0.2568 & -0.74 & 0.58 & 0.4267 & 0.65 & 0.68 & 0.1795 & 0.30 \\
 & Task 2 & 0.48 & 0.6217 & 0.22 & 0.49 & 0.9216 & -0.12 & 0.54 & 0.6638 & 0.06 \\
 & Task 3 & 0.09 & 0.0019 & -1.29 & 0.61 & 0.3053 & 0.39 & 0.60 & 0.4674 & 0.35 \\
 & Task 4 & 0.00 & 0.0002 & -7.94 & 0.50 & NA & NA & 0.50 & 1.0000 & 0.00 \\
 & Task 5 & 0.00 & $<0.0001$ & -4.41 & 0.50 & NA & NA & 0.50 & 1.0000 & 0.05 \\
\midrule
\multirow{5}{*}{\makecell{$EWDD_{OS}$ \\ vs \\ $DD_{OS}$}} & Task 1 & - & - & - & - & - & - & - & - & - \\
 & Task 2 & - & - & - & - & - & - & - & - & - \\
 & Task 3 & 0.19 & 0.0191 & -0.82 & 0.56 & 0.4656 & 0.45 & 0.51 & 0.9388 & 0.07 \\
 & Task 4 & 0.49 & 0.9397 & 0.21 & 0.50 & NA & NA & 0.55 & 0.6546 & 0.10 \\
 & Task 5 & 0.27 & 0.1267 & -0.72 & 0.50 & NA & NA & 0.42 & 0.5113 & -0.24 \\
\bottomrule
\end{tabular}%
}
\end{table}

\subsubsection{Summary of the results and RQ1 answer}

In summary, the results from \revision{Orona's industrial case study, the Leo Rover open-source case study, and the VLA case study show that the optimized Delta Debugging algorithm ($DD_{OS}$) provides benefits over the stochastic baseline ($DD_S$), although these benefits vary across systems, tasks, and evaluation metrics}. In Orona's scenarios, $DD_{OS}$ consistently achieves lower median execution times, higher test input reduction ratios (TIRR), and improved or \revision{comparable observed} failure reproduction ratios (TIFR), with statistical significance and large effect sizes in several cases, particularly for longer-running tests where failures occur later. Similarly, for the Leo Rover circuits, $DD_{OS}$ yields reduced execution times, superior TIRR values (statistically significant in two circuits), and high TIFR medians (often 100\%), indicating reliable failure reproduction despite increased variability in some scenarios. \revision{For the VLA case study, the main benefit is improved efficiency, with statistically significant execution-time reductions in Tasks 3--5, which are long-horizon tasks. The effectiveness comparisons do not establish a consistent advantage in TIRR or TIFR, although the effect-size estimates favor $DD_{OS}$ in several tasks.} Overall, these findings underscore $DD_{OS}$'s adaptability to stochastic environments, \revision{supporting its practical value for debugging CPSs under execution variability, with the extent of improvement depending on the task and metric considered}.

\begin{custombox}{RQ1}
\revision{The proposed approach ($DD_{OS}$) improves debugging efficiency and achieves higher test input reduction ratios in several scenarios compared to the stochastic baseline ($DD_S$). The benefits depend on the system and task, with execution-time improvements being the main advantage in the VLA case study. Reduced inputs also achieve higher observed failure reproduction ratios than the original inputs, supporting their practical value for debugging stochastic CPSs.}
\end{custombox}

\subsection{RQ2 - Effect of the environment}

RQ2 investigates whether incorporating environment-awareness into the delta debugging process improves its performance when dealing with stochastic CPSs. In particular, we compare the $DD_{OS}$ algorithm with its environment-wise extension ($EWDD_{OS}$), which leverages static situations to guide the minimization process. The analysis focuses on effectiveness (quality of the minimized test inputs) and efficiency (execution time).

\subsubsection{Orona's case study}

For the industrial case study, incorporating environment-awareness has a clear and consistent impact on performance.

Regarding TIRR, $EWDD_{OS}$ achieved similar reduction ratios comparable to $DD_{OS}$, where there was no statistical significance (Table~\ref{tab:stats_testsRQ1Orona}). However, in Scenario 6, $EWDD_{OS}$ improved $DD_{OS}$ with statistical significance. The improvement is mainly observed in a test input where well-defined static situations exist (i.e., states in which all elevators are stopped with doors closed). These static configurations act as stable synchronization points across executions, despite the stochastic nature of the genetic dispatching algorithm. By aligning reductions with these shared environmental states, $EWDD_{OS}$ avoids removing critical context required to reproduce the failure, while still enabling aggressive minimization. As a result, the minimized test inputs preserve the original failure characteristics and maintain high failure reproduction rates.

In terms of TIFR, $EWDD_{OS}$ demonstrates slightly more stable behavior across repeated executions. In five out of six scenarios, $DD_{OS}$ already achieved nearly perfect reproduction ratio. However, Scenario 6 was not the case, for which $EWDD_{OS}$ clearly outperformed, with statistical significance $DD_{os}$. Because the algorithm reduces the test input at meaningful environmental boundaries, the resulting minimized inputs tend to reproduce the selected failure cluster more consistently. This effect is particularly visible in long full-day simulations, where purely sequence-based reductions (as in $DD_{OS}$) may occasionally isolate statistically unstable intermediate states.

In terms of Execution Time, in 5 out of 6 scenarios, there was no statistical significance. However, in Scenario 5, $DD_{OS}$ outperformed $EWDD_{OS}$ with statistical significance. This meant that, on average, $DD_{OS}$ spent 2607.39 seconds less than $EWDD_{OS}$, for which the environment was not helpful in this particular environment.

\subsubsection{Leo Rover case study}

In contrast to the industrial case study, incorporating environment-awareness has a more limited and mixed impact on performance for the Leo Rover open-source case study system.
Regarding execution time, $EWDD_{OS}$ and $DD_{OS}$ exhibit comparable median execution times across all three circuits, with no statistical significance detected in any case (Table 4). The \^{A}$_{12}$ values hover around 0.5 and Cohen’s d effect sizes remain small or negligible, indicating that the overhead of identifying and aligning static situations neither reduces nor increases minimization time in this setting. This outcome is attributable to the continuous navigation dynamics of the rover, where pronounced static synchronization points (e.g., fixed positions or headings) are less frequent or consistent across runs because of the combined randomness from the Gazebo physics engine, image-processing delays, and ROS communication.

For the Test Input Reduction Ratio (TIRR), $EWDD_{OS}$ achieves slightly higher median reduction ratios than $DD_{OS}$ in Circuits 1 and 2 (medium effect sizes), although statistical significance is not reached (p > 0.06). In Circuit 3, the performances are essentially equivalent. By pruning waypoint sequences up to shared environmental states before the failure, the environment-aware variant enables modestly more aggressive minimization without losing the failure-inducing context; however, the benefit is smaller than in the elevator case, where discrete static states (elevators stopped with doors open) are more reliably identifiable.

In terms of the Test Input Failure Reproduction Ratio (TIFR), results are again mixed and without statistical significance. $EWDD_{OS}$ slightly improves reproduction in Circuit 3, while $DD_{OS}$ performs marginally better in Circuit 2 (p = 0.0678, close but still non-significant). Overall, both approaches maintain high reproduction rates (medians often above 80\% and frequently reaching 100\% as already observed for DDOS in RQ1), confirming that the additional environmental alignment does not substantially enhance reliability in a system affected by multiple independent randomness sources.

\revision{We also examined run-to-run variability where each of the nine algorithm--circuit combinations produced ten distinct final waypoint counts across ten reduction runs. For Barcelona, all three algorithms also produced ten distinct $(x,y)$ starting-position pairs, using the \texttt{x\_position} and \texttt{y\_position} fields of the last evaluated candidate in each run. These observations characterize variation across reduction runs.}

Taken together, the environment-aware extension provides only modest or neutral gains for the Leo Rover, highlighting that the value of static-situation guidance is case-dependent and less pronounced when failures arise from continuous, image-driven control loops rather than discrete event sequences.

\subsubsection{\revision{VLA case study}}

\revision{For the VLA case study, incorporating environment-awareness has a task-dependent impact on performance. The comparison between $EWDD_{OS}$ and $DD_{OS}$ is reported for Tasks 3--5 in Table~\ref{tab:stats_testsVLA}. For Task 1 and 2 there was no availability of such algorithm as these tasks were short- and medium-horizon tasks, thus there were no multiple atomic tasks available. Regarding execution time, $EWDD_{OS}$ outperforms $DD_{OS}$ as shown in Figure~\ref{fig:Results_VLA_execTime}. Moreover for Task 3, there was statistical significance with statistical significance with effect-size favoring the environment-aware approach. In Task 5, the effect sizes also favor $EWDD_{OS}$, although statistical significance was not reached. In contrast, Task 4 shows no clear execution-time advantage, with an $A_{12}$ value close to 0.5. These results indicate that environment-aware reduction can further improve debugging efficiency for particular VLA tasks, although the improvement is not consistent across all reported comparisons.}

\revision{For the TIRR, the effect-size slightly favor $EWDD_{OS}$ in Task 3, but the difference is not statistically significant as shown in Table~\ref{tab:stats_testsVLA}. For Tasks 4 and 5, the reported $A_{12}$ values are 0.50, while the p-values and Cohen's d values are unavailable, meaning that the same exact reduction was achieved by both approaches. Consequently, the comparisons do not establish an additional reduction-effectiveness benefit from incorporating environment-awareness.}

\revision{In terms of the TIFR, the results are mixed and without statistical significance. As Table~\ref{tab:stats_testsVLA} presents, the effect-size slightly favor $EWDD_{OS}$ in Tasks 3 and 4, whereas Task 5 favors $DD_{OS}$. Therefore, the available evidence does not show a consistent improvement in target-failure reproduction from the additional environmental guidance.}

\revision{Taken together, the environment-aware extension provides an additional execution-time benefit in Task 3, while its advantages are less conclusive in Tasks 4 and 5. Similarly to the other case study systems, these results highlight that the contribution of environment-awareness depends on the scenario and on the failure-occurring point, with no consistent additional improvement in test-input reduction or failure reproduction across the reported VLA comparisons.}

\subsubsection{Summary of the results and RQ2 answer}

In summary, the impact of incorporating environment-awareness into the delta debugging process is clearly \emph{context-dependent}. In the Orona industrial case study, $EWDD_{OS}$ provides tangible benefits over $DD_{OS}$, particularly in scenarios where well-defined static situations exist and can be reliably identified across executions. In these cases, aligning reductions with meaningful environmental synchronization points improves stability and, in specific scenarios (notably Scenario 6), leads to statistically significant improvements in TIRR and TIFR. The approach preserves the failure-inducing context more robustly in long-running simulations, where purely sequence-based minimization may isolate statistically unstable states. Although execution time improvements are not systematic, and in one scenario ($S5$) $DD_{OS}$ remains faster, the effectiveness gains in \revision{these scenarios} demonstrate the practical value of environment-awareness in structured CPSs with discrete, identifiable states.

In contrast, for the Leo Rover case study, the benefits of $EWDD_{OS}$ are modest and statistically non-significant. The rover operates in a continuous control loop influenced by multiple independent sources of randomness (physics simulation, perception noise, communication delays), where clear and stable environmental synchronization points are less frequent. As a result, environment-guided reductions neither substantially improve test input reduction nor consistently enhance failure reproduction. Both $DD_{OS}$ and $EWDD_{OS}$ achieve comparable execution times and high reproduction ratios, \revision{with no statistically significant differences detected in the reported comparisons}.

\revision{For the VLA case study, the additional benefits of environment-awareness are mainly observed in execution time. Among the reported comparisons for Tasks 3--5, $EWDD_{OS}$ significantly outperforms $DD_{OS}$ in Task 3. The effect-size estimates also favor $EWDD_{OS}$ in Task 5, although statistical significance is not reached, while Task 4 shows no clear execution-time advantage. The TIRR and TIFR comparisons do not establish consistent improvements. These findings show that environmental guidance can further reduce debugging cost for particular VLA tasks, without necessarily improving reduction effectiveness or failure reproducibility.}

Overall, the results indicate that environment-aware delta debugging is particularly beneficial in CPSs characterized by discrete, well-defined environmental states that can serve as reliable synchronization anchors across stochastic executions. \revision{The Leo Rover and VLA results further show that these benefits are not uniform across tasks or evaluation metrics. Depending on the setting, environmental guidance may provide limited gains or improve execution time without producing corresponding improvements in TIRR or TIFR.}

\begin{custombox}{RQ2}
\revision{The use of environment-awareness in delta debugging improves performance when environmental states provide useful reduction points, as demonstrated in the Orona case study. Its benefits are limited for Leo Rover, while the VLA case study shows additional execution-time benefits in a particular task without consistent improvements in reduction or reproduction ratios. Thus, the benefits of environment-aware delta debugging depend on the environment, task, and performance metric.}
\end{custombox}

\revision{
\subsection{RQ3 - Sensitivity Analysis}

\subsubsection{Orona's case study}

\revision{The execution-time results in Table~\ref{tab:ewdd_orona_descriptive} show that, as expected, larger repetition budgets generally require more debugging time. The configuration with 40 repetitions has the highest mean execution time in every scenario, whereas 10 repetitions provide the lowest mean in five out of the six scenarios. The difference between these two budgets is statistically significant after Holm correction in all scenarios except Scenario 4. The effect is particularly consistent in Scenario 5, where each increase in the budget produces a statistically significant increase in execution time. Scenario~4 shows a different pattern: 20 repetitions yield the lowest mean execution time, although the difference from 10 repetitions is not statistically significant. Only the comparison between 20 and 40 repetitions reaches statistical significance in this scenario. This pattern can arise because the repetition budget affects both the cost of individual validations and the evidence used to guide the search. If a candidate fails validation, rollback to a larger input may require additional reduction attempts and executions, potentially offsetting the savings from a smaller budget. Consequently, the total debugging time depends on both the repetition budget and the resulting search.}

\begin{table}[ht]
\centering
\color{black}
\caption{\revision{Improved EWDD sensitivity analysis for the Orona case study: mean and standard deviation}}
\label{tab:ewdd_orona_descriptive}
\begin{tabular}{llrrrrrr}
\toprule
Scenario & $N$ & \multicolumn{2}{c}{Execution time (s)} & \multicolumn{2}{c}{TIRR (\%)} & \multicolumn{2}{c}{Final TIFR (\%)} \\
\cmidrule(lr){3-4} \cmidrule(lr){5-6} \cmidrule(lr){7-8}
 & & Mean & Std & Mean & Std & Mean & Std \\
\midrule
\multirow{4}{*}{Scenario 1} & 10 & 3591.09 & 1714.07 & 93.14 & 6.59 & 100.00 & 0.00 \\
 & 20 & 6005.21 & 1691.75 & 95.66 & 1.20 & 100.00 & 0.00 \\
 & 30 & 6454.03 & 1597.73 & 95.54 & 0.10 & 100.00 & 0.00 \\
 & 40 & 15083.66 & 9725.72 & 85.03 & 21.60 & 100.00 & 0.00 \\
\midrule
\multirow{4}{*}{Scenario 2} & 10 & 2512.80 & 683.54 & 97.49 & 0.61 & 100.00 & 0.00 \\
 & 20 & 5661.97 & 2482.29 & 97.10 & 1.32 & 100.00 & 0.00 \\
 & 30 & 6470.14 & 2230.04 & 96.15 & 3.21 & 99.33 & 1.49 \\
 & 40 & 15533.94 & 8938.24 & 91.72 & 13.08 & 87.00 & 29.07 \\
\midrule
\multirow{4}{*}{Scenario 3} & 10 & 5109.22 & 3277.14 & 88.62 & 13.03 & 100.00 & 0.00 \\
 & 20 & 8506.41 & 3713.99 & 95.33 & 3.76 & 100.00 & 0.00 \\
 & 30 & 13083.84 & 5829.61 & 86.22 & 7.24 & 100.00 & 0.00 \\
 & 40 & 20414.20 & 6915.15 & 88.38 & 7.84 & 99.00 & 2.24 \\
\midrule
\multirow{4}{*}{Scenario 4} & 10 & 5767.36 & 3244.20 & 89.47 & 8.03 & 100.00 & 0.00 \\
 & 20 & 5210.85 & 681.92 & 95.92 & 2.70 & 100.00 & 0.00 \\
 & 30 & 10539.37 & 6506.65 & 82.99 & 13.50 & 100.00 & 0.00 \\
 & 40 & 12147.60 & 4230.54 & 88.33 & 12.61 & 100.00 & 0.00 \\
\midrule
\multirow{4}{*}{Scenario 5} & 10 & 3974.66 & 576.92 & 87.01 & 12.42 & 96.00 & 5.48 \\
 & 20 & 6101.55 & 536.73 & 93.84 & 4.49 & 100.00 & 0.00 \\
 & 30 & 7883.50 & 701.33 & 92.89 & 3.82 & 100.00 & 0.00 \\
 & 40 & 12239.47 & 4360.44 & 89.97 & 9.70 & 96.00 & 8.94 \\
\midrule
\multirow{4}{*}{Scenario 6} & 10 & 2508.66 & 712.42 & 92.98 & 7.37 & 90.00 & 22.36 \\
 & 20 & 5465.50 & 1969.07 & 91.51 & 9.54 & 91.00 & 20.12 \\
 & 30 & 6243.94 & 1500.35 & 92.81 & 4.41 & 95.33 & 10.43 \\
 & 40 & 11624.69 & 4172.14 & 86.00 & 13.50 & 89.50 & 19.56 \\
\bottomrule
\end{tabular}
\end{table}

\revision{For TIRR, 20 repetitions showed the highest observed mean in four scenarios (Scenarios 1, 3, 4, and 5), while 10 repetitions showed the highest mean in the remaining two. The mean TIRR at 40 repetitions was lower than at 20 in every scenario. None of the pairwise TIRR comparisons reached statistical significance meaning that there is no consistent improvement in input reduction as the repetition budget increases.}

\revision{In terms of final TIFR, Scenarios 1 and 4 achieved 100\% observed reproduction with zero standard deviation across all four budgets. Scenarios 2, 3, and 5 also achieved high mean reproduction ratios for most budgets, although the mean for Scenario 2 decreased to 87.00\% at 40 repetitions, with a standard deviation of 29.07 percentage points. Scenario 6 had mean reproduction ratios ranging from 89.50\% to 95.33\%. No pairwise TIFR comparison reached statistical significance, and larger budgets did not consistently improve observed target-failure reproduction.}

\revision{Execution time was the only metric for which the pairwise analysis detected statistically significant differences. For the Orona scenarios, 20 repetitions offer a favorable observed trade-off between execution cost and input reduction. This budget showed a higher mean TIRR than 30 repetitions in five out of the six scenarios and a lower mean recorded execution time in all six, although the time difference reached statistical significance only in Scenario 5. When execution cost is the main concern, 10 repetitions are worth considering, as they showed the lowest mean recorded time in five scenarios. We nevertheless retain 30 repetitions as the reference configuration to collect more observations per validation than the smaller budgets. This choice is particularly relevant when only a small proportion of executions reproduce the target failure, leaving fewer observations for comparing its characteristics. Scenario 6 illustrates an example of it since the mean final TIFR was 95.33\% with 30 repetitions, compared with 91.00\% with 20, while the standard deviation decreased from 20.12 to 10.43 percentage points. Nevertheless, increasing the budget to 40 yielded the highest mean recorded execution time in every scenario without consistent improvements in TIRR or final TIFR. These results support configuring the budget according to execution cost and the availability of target-failure observations in which 20 repetitions provide a favorable observed trade-off for Orona, whereas 30 remain a conservative reference choice that provides more evidence for each validation.}

\subsubsection{Leo Rover case study}

\revision{The execution-time results presented in Table~\ref{tab:ewdd_circuits_descriptive} showed that 10 repetitions had the lowest mean debugging time in all three circuits, whereas 40 repetitions showed the highest. The difference between these two budgets was statistically significant in every circuit. In Circuit 1, 10 repetitions also required significantly less time than both 20 and 30 repetitions. However, Circuit 2 exhibited a different pattern where the mean execution time with 30 repetitions was lower than with 20, although this difference was not statistically significant. Nevertheless, 40 repetitions required significantly more time than each smaller budget in this circuit. These results illustrate that the total execution cost depends on both the number of repetitions and the reduction trajectory, rather than increasing proportionally to the budget.}

\begin{table}[ht]
\centering
\color{black}\small
\caption{\revision{EWDD sensitivity analysis for the circuit case study: mean and standard deviation}}
\label{tab:ewdd_circuits_descriptive}
\begin{tabular}{llrrrrrr}
\toprule
Circuit & $N$ & \multicolumn{2}{c}{Execution time (s)} & \multicolumn{2}{c}{TIRR (\%)} & \multicolumn{2}{c}{Final TIFR (\%)} \\
\cmidrule(lr){3-4} \cmidrule(lr){5-6} \cmidrule(lr){7-8}
 & & Mean & Std & Mean & Std & Mean & Std \\
\midrule
\multirow{4}{*}{Circuit 1} & 10 & 7641.10 & 3353.71 & 77.28 & 13.53 & 86.00 & 31.30 \\
 & 20 & 13825.81 & 2677.57 & 75.07 & 11.11 & 99.00 & 2.24 \\
 & 30 & 24638.18 & 7033.64 & 84.04 & 13.42 & 96.00 & 5.96 \\
 & 40 & 53420.48 & 17455.47 & 53.18 & 26.00 & 86.00 & 17.01 \\
\midrule
\multirow{4}{*}{Circuit 2} & 10 & 7155.41 & 2478.16 & 50.72 & 28.87 & 56.00 & 26.08 \\
 & 20 & 15170.37 & 3218.31 & 41.01 & 26.51 & 79.00 & 44.22 \\
 & 30 & 12044.09 & 3441.00 & 78.20 & 4.15 & 100.00 & 0.00 \\
 & 40 & 43500.67 & 11619.45 & 32.85 & 30.02 & 95.50 & 4.47 \\
\midrule
\multirow{4}{*}{Circuit 3} & 10 & 7648.16 & 1917.35 & 41.54 & 33.14 & 84.00 & 30.50 \\
 & 20 & 12226.81 & 5058.69 & 64.87 & 27.76 & 81.00 & 39.75 \\
 & 30 & 20723.73 & 9702.75 & 60.21 & 35.40 & 93.33 & 13.12 \\
 & 40 & 26821.62 & 8211.98 & 75.57 & 14.71 & 97.50 & 4.33 \\
\bottomrule
\end{tabular}
\end{table}

\revision{For TIRR, 30 repetitions showed the highest observed mean in Circuits 1 and 2, reaching 84.04\% and 78.20\%, respectively. Circuit 2 also showed substantially lower variability with this budget, with a standard deviation of 4.15 percentage points, compared with values between 26.51 and 30.02 for the other configurations. The comparison between 30 and 40 repetitions in this circuit was statistically significant, favoring 30 repetitions. In contrast, Circuit 3 showed its highest mean TIRR with 40 repetitions, at 75.57\%, although none of its pairwise comparisons reached statistical significance. Thus, increasing the budget does not consistently improve input reduction across circuits.}

\revision{The final TIFR results showed that the smallest budget can involve a trade-off between debugging time and observed failure reproducibility. This is particularly evident in Circuit 2, where 10 repetitions showed a mean final TIFR of 56.00\%, compared with 100.00\% for 30 repetitions and 95.50\% for 40. Both comparisons against 10 repetitions were statistically significant. However, the difference between 30 and 40 repetitions was not significant. Circuit 1 exhibited its highest mean TIFR with 20 repetitions, while Circuit 3 achieved its highest with 40. Neither circuit showed statistically significant pairwise TIFR differences. Nevertheless, the configurations with 30 repetitions maintained mean reproduction ratios of at least 93.33\% across all three circuits.}

\revision{Overall, the Leo Rover results illustrate the trade-off between debugging cost and the properties of the reduced inputs. Although 10 repetitions showed the lowest mean execution time in every circuit, Circuit 2 showed that these savings can coincide with significantly lower target-failure reproducibility. We therefore retain 30 repetitions as a practical reference configuration since this budget showed the highest mean TIRR in two circuits and mean final TIFR values of at least 93.33\% across all three. Increasing the budget to 40 produced the highest mean execution time in every circuit without consistent improvements in reduction or reproduction. These results support 30 repetitions as a reasonable balance for Leo Rover, while smaller budgets remain an option when execution cost takes priority.}

\subsubsection{VLA case study}
\revision{Table~\ref{tab:ewddos_descriptive_table} presents the sensitivity analysis of $EWDD_{OS}$ in the VLA case study to the repeated-execution budget for Tasks 3--5 of the VLA case study. The analysis examines whether increasing this budget improves test-input reduction or target-failure reproducibility sufficiently to justify the additional execution cost.}

\begin{table}[ht]
\centering
\color{black}
\caption{\revision{EWDDOS sensitivity analysis for VLA case study: mean and standard deviation}}
\label{tab:ewddos_descriptive_table}
\begin{tabular}{llrrrrrr}
\toprule
Scenario & $N$ & \multicolumn{2}{c}{Execution time (s)} & \multicolumn{2}{c}{TIRR (\%)} & \multicolumn{2}{c}{TIFR (\%)} \\
\cmidrule(lr){3-4} \cmidrule(lr){5-6} \cmidrule(lr){7-8}
 &  & Mean & Std & Mean & Std & Mean & Std \\
\midrule
\multirow{4}{*}{Task 3} & 10 & 164.42 & 35.35 & 93.06 & 0.00 & 98.00 & 4.47 \\
 & 20 & 338.95 & 156.51 & 93.06 & 0.00 & 99.00 & 2.24 \\
 & 30 & 350.82 & 219.10 & 93.06 & 0.00 & 96.67 & 4.71 \\
 & 40 & 540.92 & 263.70 & 93.06 & 0.00 & 96.50 & 5.48 \\
\midrule
\multirow{4}{*}{Task 4} & 10 & 136.75 & 47.14 & 93.06 & 0.00 & 100.00 & 0.00 \\
 & 20 & 158.74 & 4.22 & 93.06 & 0.00 & 98.00 & 4.47 \\
 & 30 & 308.52 & 83.48 & 93.06 & 0.00 & 98.67 & 2.98 \\
 & 40 & 374.37 & 179.95 & 93.06 & 0.00 & 99.50 & 1.12 \\
\midrule
\multirow{4}{*}{Task 5} & 10 & 226.86 & 83.81 & 93.06 & 0.00 & 78.00 & 23.87 \\
 & 20 & 282.52 & 104.56 & 93.06 & 0.00 & 80.00 & 15.00 \\
 & 30 & 612.16 & 126.37 & 93.06 & 0.00 & 74.67 & 13.86 \\
 & 40 & 800.05 & 314.26 & 93.06 & 0.00 & 76.00 & 15.87 \\
\bottomrule
\end{tabular}%
\end{table}

\revision{Regarding execution time, the reference configuration ($n_{\mathrm{rep}}=30$) requires 2.13, 2.26, and 2.70 times the mean execution time of $n_{\mathrm{rep}}=10$ in Tasks 3, 4, and 5, respectively. These differences are statistically significant. The comparisons between 20 and 30 executions are also significant in Tasks 4 and 5. However, increasing the budget from 30 to 40 raises mean execution time by 54.19\%, 21.34\%, and 30.69\%, respectively, although none of these comparisons reaches statistical significance. Thus, larger budgets introduce substantial additional observed cost without consistently improving the final debugging artifact. This execution-time pattern is consistent with the additional sampling required for the initial failure characterization and subsequent repeated-execution assessments. Each assessment uses the configured budget even when the candidate is ultimately accepted. Although single-execution screening governs most intermediate reduction decisions, the repeated-execution stages therefore remain an important source of debugging cost. The total time need not increase proportionally to $n_{\mathrm{rep}}$, since single-execution screening and other search operations also contribute to the overall cost.}

\revision{For the TIRR, all configurations achieve 93.06\%, with zero standard deviation across runs. This result reflects the lower bound on the number of actions retained by the reduction procedure since every configuration reaches the minimum permitted action count, leaving no further reduction available within the configured search space. Consequently, increasing the repeated-execution budget cannot improve TIRR once this bound is reached. The observed search behavior further explains this stability. Most intermediate decisions are made through the single-execution screen, and none of the candidates subjected to repeated-execution assessment is rejected by the failure-preservation procedure in the reported runs. Changing $n_{\mathrm{rep}}$ therefore increases the effort spent assessing candidates without introducing rejection-driven changes to the resulting reductions in this case study.}

\revision{In terms of TIFR, Tasks 3 and 4 maintain high observed reproduction ratios across all budgets. Mean values range from 96.50\% to 99.00\% in Task 3 and from 98.00\% to 100.00\% in Task 4. Task 5 exhibits lower reproduction ratios and greater variability, with means between 74.67\% and 80.00\%. No pairwise comparison shows statistically significant differences, and increasing the repetition budget does not consistently improve observed TIFR. Therefore, Execution time is the only metric that shows statistically significant differences among the evaluated configurations.}

\revision{Overall, for this case study smaller budgets provide a favorable observed cost-effectiveness trade-off for these tasks, where reduction decisions remain stable, and all runs reach the configured action limit. }

\subsubsection{\revision{Summary of the results and RQ3 answer}}

\revision{In summary, the sensitivity analysis of $EWDD_{OS}$ across the three case studies showed that larger repetition budgets generally increase debugging time without consistently improving test-input reduction or target-failure reproducibility. The configuration with 40 repetitions showed the highest mean execution time in every evaluated scenario, circuit, and task, while its effectiveness benefits depend on the case study.}

\revision{For Orona, 20 repetitions provided a favorable observed trade-off between execution cost and input reduction, with no statistically significant TIRR or TIFR differences across budgets. For Leo Rover, the results illustrated why execution cost alone should not determine the configuration since although 10 repetitions showed the lowest mean execution time in every circuit, Circuit 2 achieved significantly higher failure reproduction with 30 and 40 repetitions. The reference budget of 30 also exhibited the highest mean TIRR in two circuits. In contrast, all evaluated VLA configurations reached the permitted action limit and showed no statistically significant TIFR differences, making smaller budgets more economical for these tasks.}

\revision{Overall, these findings support retaining 30 repetitions as a practical reference configuration, while allowing smaller budgets when they provide satisfactory reduction and reproduction at lower cost. Collecting more observations per validation can be particularly useful when few executions reproduce the target failure, but it does not necessarily produce a smaller or more reproducible reduced input. The results therefore support adapting the budget to the target failure and execution cost rather than treating 30 repetitions as universally optimal.}

\begin{custombox}{\revision{RQ3}}
\revision{Increasing the repeated-execution budget generally increases debugging time, while improvements in input reduction and failure reproduction depend on the scenario. Smaller budgets offer favorable observed trade-offs for Orona and the evaluated VLA tasks, whereas Leo Rover demonstrates that additional repetitions can coincide with significantly better failure reproduction. Thus, 30 repetitions provide a practical reference configuration, but the appropriate budget depends on execution cost and the target failure.}
\end{custombox}

}

\section{Discussion}\label{sec:Discussion}

This section presents the lessons learned from our empirical evaluation and discusses the main threats to validity and the measures adopted to mitigate them.

\subsection{Lessons learned}

Our empirical evaluation provides several insights into the application of delta debugging to stochastic Cyber-Physical Systems (CPSs), extending beyond the comparative performance of the proposed algorithms.

\textbf{Lesson 1 -- Failure-inducing input reduction minimizes test flakiness:} The most notable observation is that minimizing failure-inducing test inputs increased for all the cases the Test Input Failure Reproduction Ratio (TIFR) with respect to the original executions. This result indicates that test-input reduction does not merely preserve failure-inducing conditions but can actively improve their reproducibility. We hypothesize that this behavior arises because long CPS executions accumulate multiple sources of stochastic variability (e.g., simulator artifacts, scheduling effects, communication delays, perception noise, and randomized algorithmic decisions) that are unrelated to the root cause of the failure. Each additional execution step therefore introduces further opportunities for behavioral divergence across repeated runs. \revision{By removing execution segments while retaining the target failure under the configured validation criterion, delta debugging may reduce exposure to incidental stochastic influences and help focus investigation on the remaining failure-inducing conditions.} Consequently, the minimized test inputs are less susceptible to incidental randomness and reproduce the original failure more consistently. This finding suggests that test-input minimization can simultaneously support fault localization and mitigate execution flakiness, representing an additional benefit that has received limited attention in previous work. Moreover, one could argue that test input minimization simultaneously could improve fault localization by reducing execution length and increases the reproducibility of the remaining failure, thereby lowering the overall cost of debugging stochastic CPSs. \revision{Reduced inputs can make failure scenarios easier to interpret by leaving fewer events or actions to inspect. For example, a reduced Leo Rover input allows developers to examine the navigation segment leading to a lane departure without reviewing the full route. In addition, repeated-execution validation assesses preservation of the target failure class, its reproduction frequency, and its measured characteristics under the configured acceptance criteria. The resulting input can therefore support root-cause investigation by narrowing the execution context and providing a smaller test case for reproducing the target failure and evaluating diagnostic hypotheses. In our previous deterministic elevator study~\cite{valle2023applying} we showed that experienced engineers could infer likely causes from reduced inputs and outputs.}

\textbf{Lesson 2 -- Reliable debugging of stochastic CPSs requires statistical reasoning rather than deterministic failure preservation:} Classical delta debugging assumes that the outcome of a reduced test input can be determined from a single execution. Our results demonstrate that this assumption is generally invalid in stochastic CPSs. \revision{Instead, repeated executions allow the reduction procedure to assess how frequently a candidate reproduces the target failure and compare its measured characteristics under execution variability.} Our results indicate that reliable debugging approaches for CPSs tested under flakiness require statistical reasoning rather than deterministic pass/fail decisions.

\textbf{Lesson 3 -- The degree of execution variability determines the most suitable minimization strategy:} The experiments indicate that the relative performance of the proposed algorithms depends on the severity of execution flakiness. Under moderate variability, the speculative search strategy employed by $DD_{OS}$ substantially reduces debugging time while maintaining comparable reduction quality. Conversely, when execution variability becomes more pronounced, the conservative validation strategy adopted by DDS provides greater robustness by verifying each candidate reduction through repeated statistical evaluation. These results suggest that debugging algorithms should adapt their validation strategy according to the stochastic characteristics of the target system rather than relying on a single fixed reduction policy.

\textbf{Lesson 4 -- Longer executions provide greater opportunities for effective reduction:} The effectiveness of delta debugging is closely related to the position of the failure within the original execution. Failures occurring later in long executions provide greater opportunities for removing behavior that is unrelated to the observed fault, resulting in both larger reduction ratios and greater reductions in debugging time. Consequently, the relative advantages of optimized reduction strategies become increasingly significant as the temporal distance between the beginning of the execution and the failure increases.

\textbf{Lesson 5 -- Failure characterization should be adapted to the properties of the application domain:} The statistical comparison of failures depends critically on the quality of the failure clustering process. Our evaluation shows that no single clustering technique is universally appropriate. K-Means produced reliable failure groupings for the discrete and high-volume passenger data generated by the elevator dispatching system, whereas Gaussian Mixture Models more accurately represented the continuous and overlapping failure distributions observed in the autonomous robot. Similarly, the Bayesian Information Criterion (BIC) consistently provided more reliable model selection than alternative criteria in our experiments. These observations highlight that the characterization of stochastic failures should reflect the statistical properties of the underlying system rather than relying on generic clustering methods.

%\textbf{Lesson 6 -- Improved reproducibility directly facilitates subsequent debugging activities:} Beyond reducing execution length, the proposed techniques consistently produced compact failure-inducing test inputs that were easier to reproduce across repeated executions. Such reductions decrease the computational cost associated with repeated testing while simultaneously reducing the amount of execution data that engineers must inspect during fault localization and repair. Consequently, improving failure reproducibility should be viewed not only as a property of the minimized test input but also as a mechanism for reducing the overall cost of debugging stochastic CPSs.

\subsection{\revision{Impact of stochastic characteristics}}

\revision{Stochastic behavior can affect whether a failure occurs, its observed characteristics, and the sequence of states that leads to it. The three case studies illustrate different sources of this variability. In Orona, the stochastic dispatcher can assign elevators differently under the same passenger demand, although the simulator is deterministic. In Leo Rover, variations in simulation, image processing, and communication timing can change the observations received by the navigation controller and the resulting trajectory. In the VLA system, differences in generated actions can influence subsequent observations and decisions. Thus, randomness introduced at one point can affect the rest of the execution and the conditions under which the target failure occurs.}

\revision{These effects influence both the cost of reduction and the parts of an input that can be removed. For instance, in RQ1, a failure that occurs inconsistently makes a single execution less informative. Repeated validation and possible rollback may therefore reduce the savings from speculative search. Conversely, postponing validation can provide substantial savings when executions are costly, as demonstrated by $DD_{OS}$ in Orona scenarios with late failures and in long-horizon VLA tasks. Moreover, failure timing also affects the opportunities for reduction, although retaining the relevant execution history remains important. A failure caused by a local condition may be easier to isolate than one that depends on earlier interactions, such as accumulated elevator congestion. This dependence on execution history also helps explain RQ2. In Orona, different dispatch decisions can eventually lead to a common state that provides a useful reduction point as demonstrated by $EWDD_{OS}$. In Leo Rover, however, a similar position may be reached with a different heading or after different observations, making that position less suitable as a shared reduction point.}

\revision{Variability also affects the evidence available for validation as shown in RQ3, where all evaluated VLA configurations reached the minimum permitted action count, and no candidate assessed through repeated executions was rejected. Increasing the repetition budget therefore mainly increased debugging time, with no statistically significant improvement in failure reproduction. When the target failure occurs infrequently, additional repetitions can provide more observations for validation. However, the usefulness of these observations also depends on how accurately the selected cluster represents the target failure. For example, variation in Leo Rover's departure position may cause clustering to separate observations of the same failure, while different failure behaviors that produce nearby departures may be grouped together. Both cases can affect target selection, and the candidate executions counted as reproductions. Consequently, additional executions can help when observations are scarce, whereas distinguishing overlapping behaviors may require more informative failure characteristics.}

\subsection{\revision{Applicability beyond CPSs}}\label{sec:applicability}

\revision{We focus on CPSs because repeated simulation can make candidate validation expensive, while identifiable environmental states can provide opportunities for reduction. These characteristics motivate the proposed strategies, whose underlying principles also apply to other systems in which repeated executions of the same input can produce different outcomes. Applying $DD_S$ and $DD_{OS}$ requires an input that can be reduced, an execution procedure that supports repeated runs under comparable conditions, and a criterion for identifying the target failure. Under these conditions, $DD_S$ assesses each candidate through repeated executions, while $DD_{OS}$ uses single executions to guide provisional reductions and postpones repeated validation. The latter strategy is particularly relevant when individual executions are costly. These savings are not CPS-specific since they depend on costly repetitions and informative single-execution screens. They may diminish when screening frequently triggers full validation or rollback, and offer little benefit when a single execution already provides a definitive verdict.} 

\revision{For conventional flaky unit or integration tests, reducible elements could include input data, setup operations, requests, or interaction sequences. Reducing these elements could help identify a smaller context that still reproduces a concurrency- or timing-dependent failure. However, the input representation must capture the conditions relevant to that failure. For example, a dependency on preceding tests or shared resources may require including those interactions in the execution procedure. Failure characterization would also need to reflect the available observations, such as assertion outcomes, exception categories, or execution measurements, so that candidate reductions retain the behavior under investigation. The same principles could apply to ML- and LLM-based systems by reducing structured inputs, prompts, contextual examples, or interaction histories. Repeated executions could then assess whether a candidate retains a specified undesirable behavior despite variability in the generated outputs. The VLA case study demonstrates applicability to an AI-controlled robotic system. Extending the approach to other ML or LLM tasks would require suitable failure oracles and representations, particularly when different outputs can express the same failure behavior.}

\revision{However, $EWDD_{OS}$ has an additional requirement, identifying environmental states or intervals that provide useful reduction points while retaining the context needed to reproduce the target failure. In conventional software, reconstructible application states or checkpoints could provide similar opportunities. Their suitability depends on preserving relevant internal state, pending operations, and interaction history. When such states are unavailable, $DD_S$ and $DD_{OS}$ remain applicable without environment-aware reduction. Thus, the reduction and validation strategies have broader potential, while the environment-aware extension depends on the states and execution context available in each application.}

\subsection{Threats to Validity}
\label{sec:threats}

We now discuss the main threats to the validity of our empirical evaluation and the measures taken to mitigate them.

\textbf{Internal validity:}
A potential internal validity threat concerns the configuration of the proposed algorithms. In particular, the results may be influenced by the number of repeated executions used to \revision{assess failure preservation}, the \revision{significance level and acceptance thresholds}, and the criteria used to accept or reject a reduction. We selected these values based on established statistical conventions and preliminary experimentation, and applied the same \revision{repeated-execution acceptance criteria} consistently across the compared algorithms \revision{to support a fair comparison}. \revision{Nevertheless, limited observations may reduce the sensitivity of the statistical comparisons, particularly when few executions reproduce the target failure.} \revision{Alternative configurations may therefore lead to different trade-offs between execution cost and the evidence supporting failure preservation.} \revision{For instance, in $DD_{OS}$, single-execution validations guide the search provisionally and can affect which candidates are explored, while repeated-execution validation determines whether a selected candidate is accepted or the search rolls back to a previously validated candidate.} Another threat concerns the implementation of the three algorithms. To reduce this risk, the approaches share the same execution, failure-analysis, and test input manipulation infrastructure, differing only in the reduction and validation strategies under comparison.

\textbf{Construct validity:}
%Construct validity threats concern whether the selected measures adequately capture the effectiveness and efficiency of stochastic Delta Debugging. We assessed effectiveness through the Test Input Reduction Ratio (TIRR) and the Test Input Failure Reproduction Ratio (TIFR), and efficiency through the execution time required by the reduction process. While these measures capture input reduction, failure reproducibility, and computational cost, they do not directly measure the effort required by developers to identify and repair the underlying fault. A qualitative study with practitioners could provide further evidence regarding the practical debugging benefits of the minimized inputs. In addition, 
Determining whether two executions exhibit the same failure relies on application-specific failure representations and clustering strategies. Incorrectly grouped or separated failures could affect the statistical validation of candidate reductions. We mitigated this threat by selecting clustering techniques according to the statistical properties of each case study and by evaluating their suitability using established model-selection criteria.

\textbf{Conclusion validity:}
The stochastic behavior of \revision{the evaluated case study systems} case study systems may introduce random variation into the observed reduction ratios, failure reproduction ratios, and execution times. We mitigated this threat by executing the algorithms repeatedly and analyzing the resulting distributions using statistical significance tests and effect-size measures. Nevertheless, the high computational cost of CPS simulations limits the number of repetitions that can be performed. Moreover, the absence of statistical significance in some comparisons does not necessarily demonstrate equivalence between the algorithms, particularly when the observed differences are consistent but the number of available scenarios is limited. Such results should therefore be interpreted as insufficient evidence of a difference rather than evidence that the approaches perform identically.

\textbf{External validity:}
The generalizability of our findings is limited by the number of systems and failure scenarios considered. Our evaluation includes \revision{three} complementary CPSs: an industrial elevator-dispatching system whose stochasticity originates from a randomized optimization algorithm, an autonomous mobile robot whose variability arises primarily from the simulation and execution infrastructure, \revision{and a VLA-controlled robotic manipulation system with stochastic action generation during policy inference}. This diversity allows us to study distinct sources of stochasticity. Furthermore, the industrial case study strengthens the practical relevance of the evaluation by demonstrating that the proposed techniques scale and can be integrated into a real development and testing process. Nevertheless, additional CPS domains, simulators, controllers, and failure types are required before broadly generalizing the results. The applicability of the environment-aware approach is further constrained to systems that expose stable operational states from which simulations can be reliably initialized.

\section{Related Work}\label{sec:RelatedWork}

Delta Debugging originated with Hildebrandt and Zeller’s early work on simplifying failure-inducing inputs and was fully established by the \emph{ddmin} algorithm in the seminal article by Zeller and Hildebrandt, which framed reduction as a divide-and-conquer search over subsets and complements of a failing input~\cite{hildebrandt2000simplifying,zeller2002simplifying}. In its classical form, Delta Debugging guarantees a \emph{1-minimal} result, i.e., a failure-inducing input from which no single remaining unit can be removed without losing the target property~\cite{zeller2002simplifying}. A rich line of follow-up work has specialized DD to particular input structures and debugging settings. HDD leveraged hierarchical structure to reduce tree-based inputs such as XML more efficiently and often more aggressively than flat Delta Debugging~\cite{misherghi2006hdd}. Iterative Delta Debugging extended the idea to settings with masking or cascading faults, repeatedly applying Delta Debugging until an originally hidden bug can be isolated~\cite{artho2011iterative}. ProbDD replaced fixed partitioning with probabilistic reasoning about element relevance and demonstrated substantial improvements over existing ddmin-based reducers such as HDD and CHISEL~\cite{wang2021probabilistic}.Delta Debugging has also been instantiated in domain-specific debugging workflows, for example for shrinking SMT formulas~\cite{brummayer2009fuzzing} and for reducing failure-inducing circumstances in microservice systems~\cite{zhou2018delta}. \revision{The SMT approach exploits formula structure and type information, accepting reductions that retain observable solver behavior rather than statistically assessing failure reproduction across repeated executions.} In CPSs, our prior work adapted Delta Debugging to long operational traces and further introduced an environment-aware variant that exploits stable states of the simulated environment to accelerate reduction~\cite{valle2023applying,valletowards}. \revision{These CPS approaches assume reliable failure verdicts from individual executions, without modeling variation across reruns. The Simulink study~\cite{valletowards} reported mean input reductions relative to the original failing time of 34\% for TwoTanks, 79\% for Cruise Control, and 86\% for Electro-Mechanical Brake. The deterministic elevator study~\cite{valle2023applying} reported an environment-aware variant between 1.3 and 1.8 times faster than traditional Delta Debugging.} Our work departs from this assumption by extending Delta Debugging to stochastic CPSs, replacing deterministic failure preservation with repeated execution and statistical validation, and by introducing optimized and environment-aware stochastic variants that make such validation practical under flaky simulation. \revision{Our evaluation measures optimization benefits over a stochastic baseline; the different execution settings and baselines prevent directly comparing speedup values across studies.} \revision{Table~\ref{tab:positioning} summarizes how the proposed approaches build on the existing reduction framework and environmental guidance, distinguishing the stochastic baseline from the optimized extensions.}

\begin{table}[ht]
\centering
\color{black}
\caption{Positioning of the proposed approaches relative to
the existing reduction framework and environmental guidance.}
\label{tab:positioning}
\small
\begin{tabular}{p{0.17\linewidth}p{0.27\linewidth}p{0.43\linewidth}}
\toprule
Approach & Failure assessment & Role or distinguishing strategy \\
\midrule
Classical DD~\cite{zeller2002simplifying} &
Individual execution verdict &
Existing input-reduction framework. \\

$EWDD$~\cite{valle2023applying} &
Individual execution verdict &
Uses stable environmental states to guide reduction. \\

$DD_S$ &
Repeated-execution target-failure assessment &
Stochastic adaptation and reference baseline; validates
every candidate before continuing. \\

$DD_{OS}$ &
Same repeated-validation criterion as $DD_S$ &
Adds provisional screening, deferred validation, and
archive-based rollback. \\

$EWDD_{OS}$ &
Same repeated-validation criterion as $DD_S$ &
Additionally uses common environmental intervals across
stochastic reference executions to guide reduction. \\
\bottomrule
\end{tabular}
\end{table}

The broader CPS literature has primarily focused on \emph{finding} failures rather than \emph{minimizing} them once detected. Simulation-based test generation and prioritization for CPS models aim at efficiently exploring long-running dynamic behaviors under high execution cost~\cite{matinnejad2018test,arrieta2017employing,nejati2019evaluating}. Likewise, falsification research searches for counterexamples to temporal-logic requirements and has evolved into a mature benchmark- and tool-centric ecosystem~\cite{menghi2020approximation}. More directly related to debugging, approaches such as CPSDebug explain how failures propagate through Simulink/Stateflow models by combining testing, specification mining, and failure analysis~\cite{bartocci2021cpsdebug}. Ghazal et al~\cite{ghazal2026towards} propose a counterfactual explanation approach along with assertion inference approach to debug CPS failures by helping them interpret under what circumstances the systems fail. These contributions are highly relevant because they highlight why CPS debugging is difficult: failures often emerge from long, heterogeneous interactions among software, controllers, timing, and physical dynamics. However, they do not address the specific problem considered in this paper, namely reducing already failing CPS test inputs under stochastic executions. In contrast to prior work on test generation, falsification, monitoring, or gray-box failure explanation, our work focuses on \emph{statistical test input reduction} after failure discovery, and explicitly targets both algorithm-induced and infrastructure-induced stochasticity. 

\revision{Flaky tests have also been studied extensively outside CPSs. Luo et al.~\cite{luo2014empirical} identified asynchronous waiting, concurrency, and test-order dependencies as prominent causes. Subsequent tools include iDFlakies~\cite{lam2019idflakies}, which detects and partially classifies flaky tests, and NonDex~\cite{gyori2016nondex}, which explores underdetermined Java API behaviors and helps locate failure-inducing invocations. Our work complements these efforts by reducing inputs after a target failure has been selected, while assessing its preservation across repeated executions. As discussed in Section~\ref{sec:applicability}, this could help developers diagnose flaky unit tests by reducing their setup or inputs to a smaller context that still reproduces the target failure.}

Recent empirical evidence indicates that simulator flakiness is a common and practically significant problem for autonomous and robotic CPSs. Amini et al.~\cite{amini2024evaluating} found that repeated executions of the same autonomous driving test can lead to both quantitative and qualitative inconsistencies, including changes in pass/fail verdicts, and that this variability materially affects the behavior of randomized testing algorithms. This observation is consistent with systems research on robotic middleware, which shows that ROS~2 callback execution is inherently nondeterministic under the default publish-subscribe execution model, with further variability caused by message interleavings and network latency in distributed deployments~\cite{sagmeister2026rslcpp}. Additional execution variability may arise from physics simulation, image acquisition, DNN inference latency, and accelerator contention in perception-control pipelines~\cite{enright2024paam,ferigo2020gym}. Existing work has mainly focused on predicting flaky tests, enforcing deterministic middleware execution, or improving simulator trustworthiness. Our work is complementary to these. \revision{Our contribution builds on the existing Delta Debugging framework and statistical techniques, introducing a stochastic failure-preservation criterion and reduction strategies designed to reduce repeated-validation cost. The evaluation provides empirical evidence of their benefits and limitations under stochastic execution.} In this sense, our techniques do not compete with flaky test prediction or deterministic replay. Instead, they provide a debugging method that remains applicable when stochastic execution variability cannot be fully eliminated.

\section{Conclusion and Future Work}\label{sec:Conclusions}

Flaky executions complicate the debugging of CPSs because identical test inputs may produce inconsistent outcomes, preventing conventional Delta Debugging from reliably preserving the original failure. To address this problem, we proposed stochastic, optimized, and environment-aware Delta Debugging techniques that combine statistical validation with speculative reduction and rollback. We evaluated the approaches on \revision{an industrial elevator-dispatching system, a DNN-controlled mobile robot, and a VLA-controlled robotic manipulation system}. The results show that the proposed techniques \revision{can} substantially reduce failure-inducing test inputs while maintaining their failure behavior and decreasing the algorithm's execution time. Most notably, the minimized inputs frequently achieved higher failure reproduction ratios than the original executions. This suggests that removing non-essential portions of a test input reduces the system's exposure to incidental sources of stochastic variability, thereby making the target failure less flaky and easier to reproduce. Overall, the results indicate that Delta Debugging can both simplify failure-inducing executions and improve their reproducibility. \revision{The benefits of environment-aware reduction depend on the availability of suitable initialization points that preserve the context relevant to the target failure; when these are unavailable, reduction can proceed using $DD_{OS}$ without environmental preprocessing.} Future work will evaluate this relationship across additional CPS domains and develop adaptive validation strategies based on the observed degree of flakiness, \revision{including sequential procedures to reduce repetitions through early stopping while retaining sufficient observations for failure-characteristic assessment}.

\section*{Acknowledgments}
Pablo Valle and Aitor Arrieta are part of the Systems and Software Engineering research group of Mondragon Unibertsitatea (IT1919-26), supported by the Department of Education, Universities and Research of the Basque Country. Aitor Arrieta is supported by the Spanish Ministry of Science, Innovation and Universities (project PID2023-152979OA-I00), funded by MCIU /AEI /10.13039/501100011033 / FEDER, UE. Pablo Valle is supported by the Pre-doctoral Program for the Formation of Non-Doctoral Research Staff of the Education Department of the Basque Government (Grant n. PRE\_2025\_2\_0252). S. Ali is supported by the Co-tester project (Project\# 314544), funded by the Research Council of Norway.

\bibliographystyle{ACM-Reference-Format}
\bibliography{bibliografia}

@software{valle2026DD4CPSs,
  author = {Valle, Pablo and
                  Ali, Shaukat and
                  Arrieta, Aitor},
  title  = {Github Repository of "Delta Debugging for Cyber-
                   Physical Systems with Flaky Test Executions"
                  },
  year   = {2026},
  url    = {https://github.com/pablovalle/DeltaDebugging4CPSs},
  note   = {GitHub repository}
}

@inproceedings{gyori2016nondex,
  title={NonDex: A tool for detecting and debugging wrong assumptions on Java API specifications},
  author={Gyori, Alex and Lambeth, Ben and Shi, August and Legunsen, Owolabi and Marinov, Darko},
  booktitle={Proceedings of the 2016 24th ACM SIGSOFT International Symposium on Foundations of Software Engineering},
  pages={993--997},
  year={2016}
}

@inproceedings{lam2019idflakies,
  title={iDFlakies: A framework for detecting and partially classifying flaky tests},
  author={Lam, Wing and Oei, Reed and Shi, August and Marinov, Darko and Xie, Tao},
  booktitle={2019 12th ieee conference on software testing, validation and verification (icst)},
  pages={312--322},
  year={2019},
  organization={IEEE}
}

@inproceedings{luo2014empirical,
  title={An empirical analysis of flaky tests},
  author={Luo, Qingzhou and Hariri, Farah and Eloussi, Lamyaa and Marinov, Darko},
  booktitle={Proceedings of the 22nd ACM SIGSOFT international symposium on foundations of software engineering},
  pages={643--653},
  year={2014}
}

@article{liu2023libero,
  title={LIBERO: Benchmarking Knowledge Transfer for Lifelong Robot Learning},
  author={Liu, Bo and Zhu, Yifeng and Gao, Chongkai and Feng, Yihao and Liu, Qiang and Zhu, Yuke and Stone, Peter},
  journal={arXiv preprint arXiv:2306.03310},
  year={2023}
}

@article{li24simpler,
         title={Evaluating Real-World Robot Manipulation Policies in Simulation},
         author={Xuanlin Li and Kyle Hsu and Jiayuan Gu and Karl Pertsch and Oier Mees and Homer Rich Walke and Chuyuan Fu and Ishikaa Lunawat and Isabel Sieh and Sean Kirmani and Sergey Levine and Jiajun Wu and Chelsea Finn and Hao Su and Quan Vuong and Ted Xiao},
         journal = {arXiv preprint arXiv:2405.05941},
         year={2024}
}

@article{nvidia2025gr00tn1openfoundation,
  title={Gr00t n1: An open foundation model for generalist humanoid robots},
  author={Bjorck, Johan and Casta{\~n}eda, Fernando and Cherniadev, Nikita and Da, Xingye and Ding, Runyu and Fan, Linxi and Fang, Yu and Fox, Dieter and Hu, Fengyuan and Huang, Spencer and others},
  journal={arXiv preprint arXiv:2503.14734},
  year={2025}
}

@software{valle2026Replication,
  author       = {Valle, Pablo and
                  Ali, Shaukat and
                  Arrieta, Aitor},
  title        = {Replication Package of "Delta Debugging for Cyber-
                   Physical Systems with Flaky Test Executions"
                  },
  month        = jul,
  year         = 2026,
  publisher    = {Zenodo},
  doi          = {10.5281/zenodo.21624174},
  url          = {https://doi.org/10.5281/zenodo.21624174},
}

@article{zampetti2022empirical,
  title={An empirical characterization of software bugs in open-source cyber--physical systems},
  author={Zampetti, Fiorella and Kapur, Ritu and Di Penta, Massimiliano and Panichella, Sebastiano},
  journal={Journal of Systems and Software},
  volume={192},
  pages={111425},
  year={2022},
  publisher={Elsevier}
}

@inproceedings{ferigo2020gym,
  title={Gym-ignition: Reproducible robotic simulations for reinforcement learning},
  author={Ferigo, Diego and Traversaro, Silvio and Metta, Giorgio and Pucci, Daniele},
  booktitle={2020 IEEE/SICE International Symposium on System Integration (SII)},
  pages={885--890},
  year={2020},
  organization={IEEE}
}

@article{sagmeister2026rslcpp,
  title={RSLCPP-Deterministic Simulations Using ROS 2},
  author={Sagmeister, Simon and Weinmann, Marcel and Pitschi, Phillip and Lienkamp, Markus},
  journal={arXiv preprint arXiv:2601.07052},
  year={2026}
}

@article{enright2024paam,
  title={Paam: A framework for coordinated and priority-driven accelerator management in ros 2},
  author={Enright, Daniel and Xiang, Yecheng and Choi, Hyunjong and Kim, Hyoseung},
  journal={arXiv preprint arXiv:2404.06452},
  year={2024}
}

@inproceedings{ghazal2026towards,
  title={Towards Counterfactual Explanation and Assertion Inference for CPS Debugging},
  author={Ghazal, Zaid and Yusuf, Hadiza and Gaaloul, Khouloud},
  booktitle={2026 IEEE International Conference on Software Testing, Verification and Validation (ICST)},
  pages={623--634},
  year={2026},
  organization={IEEE}
}

@article{timperley2024robust,
  title={ROBUST: 221 bugs in the Robot Operating System},
  author={Timperley, Christopher S and van der Hoorn, Gijs and Santos, Andr{\'e} and Deshpande, Harshavardhan and W{\k{a}}sowski, Andrzej},
  journal={Empirical Software Engineering},
  volume={29},
  number={3},
  pages={57},
  year={2024},
  publisher={Springer}
}

@article{ayerdi2024marmot,
  title={Marmot: Metamorphic runtime monitoring of autonomous driving systems},
  author={Ayerdi, Jon and Iriarte, Asier and Valle, Pablo and Roman, Ibai and Illarramendi, Miren and Arrieta, Aitor},
  journal={ACM Transactions on Software Engineering and Methodology},
  volume={34},
  number={1},
  pages={1--35},
  year={2024},
  publisher={ACM New York, NY}
}

@article{lee2026fuzzing,
  title={Fuzzing-based mutation testing of C/C++ software in cyber-physical systems},
  author={Lee, Jaekwon and Pastore, Fabrizio and Briand, Lionel},
  journal={Empirical Software Engineering},
  volume={31},
  number={1},
  pages={20},
  year={2026},
  publisher={Springer}
}

@inproceedings{bartocci2023property,
  title={Property-based mutation testing},
  author={Bartocci, Ezio and Mariani, Leonardo and Ni{\v{c}}kovi{\'c}, Dejan and Yadav, Drishti},
  booktitle={2023 IEEE Conference on Software Testing, Verification and Validation (ICST)},
  pages={222--233},
  year={2023},
  organization={IEEE}
}

@inproceedings{khatiri2024simulation,
  title={Simulation-based testing of unmanned aerial vehicles with aerialist},
  author={Khatiri, Sajad and Panichella, Sebastiano and Tonella, Paolo},
  booktitle={Proceedings of the 2024 IEEE/ACM 46th International Conference on Software Engineering: Companion Proceedings},
  pages={134--138},
  year={2024}
}

@article{hou2015simulation,
  title={Simulation-based testing and evaluation tools for transportation cyber--physical systems},
  author={Hou, Yunfei and Zhao, Yunjie and Wagh, Aditya and Zhang, Longfei and Qiao, Chunming and Hulme, Kevin F and Wu, Changxu and Sadek, Adel W and Liu, Xuejie},
  journal={IEEE Transactions on Vehicular Technology},
  volume={65},
  number={3},
  pages={1098--1108},
  year={2015},
  publisher={IEEE}
}

@inproceedings{arrieta2020seeding,
  title={Seeding strategies for multi-objective test case selection: An application on simulation-based testing},
  author={Arrieta, Aitor and Agirre, Joseba Andoni and Sagardui, Goiuria},
  booktitle={Proceedings of the 2020 genetic and evolutionary computation conference},
  pages={1222--1231},
  year={2020}
}

@inproceedings{nejati2019evaluating,
  title={Evaluating model testing and model checking for finding requirements violations in Simulink models},
  author={Nejati, Shiva and Gaaloul, Khouloud and Menghi, Claudio and Briand, Lionel C and Foster, Stephen and Wolfe, David},
  booktitle={Proceedings of the 2019 27th acm joint meeting on european software engineering conference and symposium on the foundations of software engineering},
  pages={1015--1025},
  year={2019}
}

@inproceedings{birchler2023teaser,
  title={TEASER: Simulation-based CAN Bus Regression Testing for Self-driving Cars Software},
  author={Birchler, Christian and Rohrbach, Cyrill and Kim, Hyeongkyun and Gambi, Alessio and Liu, Tianhai and Horneber, Jens and Kehrer, Timo and Panichella, Sebastiano},
  booktitle={2023 38th IEEE/ACM International Conference on Automated Software Engineering (ASE)},
  pages={2058--2061},
  year={2023},
  organization={IEEE}
}

@article{amini2024evaluating,
  title={Evaluating the impact of flaky simulators on testing autonomous driving systems},
  author={Amini, Mohammad Hossein and Naseri, Shervin and Nejati, Shiva},
  journal={Empirical Software Engineering},
  volume={29},
  number={2},
  pages={47},
  year={2024},
  publisher={Springer}
}

@inproceedings{briand2016testing,
  title={Testing the untestable: model testing of complex software-intensive systems},
  author={Briand, Lionel and Nejati, Shiva and Sabetzadeh, Mehrdad and Bianculli, Domenico},
  booktitle={Proceedings of the 38th international conference on software engineering companion},
  pages={789--792},
  year={2016}
}

@inproceedings{leung1989insights,
  title={Insights into regression testing (software testing)},
  author={Leung, Hareton KN and White, Lee},
  booktitle={Proceedings. Conference on Software Maintenance-1989},
  pages={60--69},
  year={1989},
  organization={IEEE}
}

@inproceedings{kane2014monitor,
  title={Monitor based oracles for cyber-physical system testing: Practical experience report},
  author={Kane, Aaron and Fuhrman, Thomas and Koopman, Philip},
  booktitle={2014 44th Annual IEEE/IFIP International Conference on Dependable Systems and Networks},
  pages={148--155},
  year={2014},
  organization={IEEE}
}

@article{eidson2011distributed,
  title={Distributed real-time software for cyber--physical systems},
  author={Eidson, John C and Lee, Edward A and Matic, Slobodan and Seshia, Sanjit A and Zou, Jia},
  journal={Proceedings of the IEEE},
  volume={100},
  number={1},
  pages={45--59},
  year={2011},
  publisher={IEEE}
}

@article{arrieta2017automatic,
  title={Automatic generation of test system instances for configurable cyber-physical systems},
  author={Arrieta, Aitor and Sagardui, Goiuria and Etxeberria, Leire and Zander, Justyna},
  journal={Software Quality Journal},
  volume={25},
  number={3},
  pages={1041--1083},
  year={2017},
  publisher={Springer}
}

@article{corso2021survey,
  title={A survey of algorithms for black-box safety validation of cyber-physical systems},
  author={Corso, Anthony and Moss, Robert and Koren, Mark and Lee, Ritchie and Kochenderfer, Mykel},
  journal={Journal of Artificial Intelligence Research},
  volume={72},
  pages={377--428},
  year={2021}
}

@article{cornejo2021mutation,
  title={Mutation analysis for cyber-physical systems: Scalable solutions and results in the space domain},
  author={Cornejo, Oscar and Pastore, Fabrizio and Briand, Lionel C},
  journal={IEEE Transactions on Software Engineering},
  volume={48},
  number={10},
  pages={3913--3939},
  year={2021},
  publisher={IEEE}
}

@inproceedings{arrieta2017search,
  title={Search-based test case generation for cyber-physical systems},
  author={Arrieta, Aitor and Wang, Shuai and Markiegi, Urtzi and Sagardui, Goiuria and Etxeberria, Leire},
  booktitle={2017 IEEE congress on evolutionary computation (CEC)},
  pages={688--697},
  year={2017},
  organization={IEEE}
}

@inproceedings{osikowicz2024empirically,
  title={Empirically Evaluating Flaky Tests for Autonomous Driving Systems in Simulated Environments},
  author={Osikowicz, Olek and McMinn, Phil and Shin, Donghwan},
  booktitle={2025 IEEE/ACM International Flaky Tests Workshop (FTW) Proceedings},
  year={2024},
  organization={Institute of Electrical and Electronics Engineers (IEEE)}
}

@article{beamurgia2016modified,
  title={A modified genetic algorithm applied to the elevator dispatching problem},
  author={Beamurgia, Maite and Basagoiti, Rosa and Rodr{\'\i}guez, I and Rodriguez, V},
  journal={Soft Computing},
  volume={20},
  pages={3595--3609},
  year={2016},
  publisher={Springer}
}

@inproceedings{misherghi2006hdd,
  title={HDD: hierarchical delta debugging},
  author={Misherghi, Ghassan and Su, Zhendong},
  booktitle={Proceedings of the 28th international conference on Software engineering},
  pages={142--151},
  year={2006}
}

@inproceedings{wang2021probabilistic,
  title={Probabilistic delta debugging},
  author={Wang, Guancheng and Shen, Ruobing and Chen, Junjie and Xiong, Yingfei and Zhang, Lu},
  booktitle={Proceedings of the 29th ACM Joint Meeting on European Software Engineering Conference and Symposium on the Foundations of Software Engineering},
  pages={881--892},
  year={2021}
}

@article{artho2011iterative,
  title={Iterative delta debugging},
  author={Artho, Cyrille},
  journal={International Journal on Software Tools for Technology Transfer},
  volume={13},
  pages={223--246},
  year={2011},
  publisher={Springer}
}

@inproceedings{brummayer2009fuzzing,
  title={Fuzzing and delta-debugging SMT solvers},
  author={Brummayer, Robert and Biere, Armin},
  booktitle={Proceedings of the 7th International Workshop on Satisfiability Modulo Theories},
  pages={1--5},
  year={2009}
}

@article{chen2010big,
  title={How big is a big odds ratio? Interpreting the magnitudes of odds ratios in epidemiological studies},
  author={Chen, Henian and Cohen, Patricia and Chen, Sophie},
  journal={Communications in Statistics—simulation and Computation{\textregistered}},
  volume={39},
  number={4},
  pages={860--864},
  year={2010},
  publisher={Taylor \& Francis}
}

@article{mann1947test,
  title={On a test of whether one of two random variables is stochastically larger than the other},
  author={Mann, Henry B and Whitney, Donald R},
  journal={The annals of mathematical statistics},
  pages={50--60},
  year={1947},
  publisher={JSTOR}
}

@article{upton1992fisher,
  title={Fisher's exact test},
  author={Upton, Graham JG},
  journal={Journal of the Royal Statistical Society: Series A (Statistics in Society)},
  volume={155},
  number={3},
  pages={395--402},
  year={1992},
  publisher={Wiley Online Library}
}

@inproceedings{koenig2004design,
  title={Design and use paradigms for gazebo, an open-source multi-robot simulator},
  author={Koenig, Nathan and Howard, Andrew},
  booktitle={2004 IEEE/RSJ international conference on intelligent robots and systems (IROS)(IEEE Cat. No. 04CH37566)},
  volume={3},
  pages={2149--2154},
  year={2004},
  organization={Ieee}
}

@inproceedings{digiuseppe2012concept,
  title={Concept-based failure clustering},
  author={DiGiuseppe, Nicholas and Jones, James A},
  booktitle={Proceedings of the ACM SIGSOFT 20th international symposium on the foundations of software engineering},
  pages={1--4},
  year={2012}
}

@article{rousseeuw1987silhouettes,
  title={Silhouettes: a graphical aid to the interpretation and validation of cluster analysis},
  author={Rousseeuw, Peter J},
  journal={Journal of computational and applied mathematics},
  volume={20},
  pages={53--65},
  year={1987},
  publisher={Elsevier}
}

@article{schwarz1978estimating,
  title={Estimating the dimension of a model},
  author={Schwarz, Gideon},
  journal={The annals of statistics},
  pages={461--464},
  year={1978},
  publisher={JSTOR}
}

@book{wackerly2008mathematical,
  title={Mathematical statistics with applications},
  author={Wackerly, Dennis D and Mendenhall, William and Scheaffer, Richard L},
  volume={7},
  year={2008},
  publisher={Thomson Brooks/Cole Belmont, CA}
}

@article{arunajadai2004failure,
  title={Failure mode identification through clustering analysis},
  author={Arunajadai, Srikesh G and Uder, Scott J and Stone, Robert B and Tumer, Irem Y},
  journal={Quality and Reliability Engineering International},
  volume={20},
  number={5},
  pages={511--526},
  year={2004},
  publisher={Wiley Online Library}
}

@inproceedings{dickinson2001finding,
  title={Finding failures by cluster analysis of execution profiles},
  author={Dickinson, William and Leon, David and Fodgurski, A},
  booktitle={Proceedings of the 23rd International Conference on Software Engineering. ICSE 2001},
  pages={339--348},
  year={2001},
  organization={IEEE}
}

@article{akaike1974new,
  title={A new look at the statistical model identification},
  author={Akaike, Hirotugu},
  journal={IEEE transactions on automatic control},
  volume={19},
  number={6},
  pages={716--723},
  year={1974},
  publisher={Ieee}
}

@inproceedings{valle2023applying,
  title={Applying and Extending the Delta Debugging Algorithm for Elevator Dispatching Algorithms (Experience Paper)},
  author={Valle, Pablo and Arrieta, Aitor and Arratibel, Maite},
  booktitle={Proceedings of the 32nd ACM SIGSOFT International Symposium on Software Testing and Analysis},
  pages={1055--1067},
  year={2023}
}

@article{han2022uncertainty,
  title={Uncertainty-aware Robustness Assessment of Industrial Elevator Systems},
  author={Han, Liping and Ali, Shaukat and Yue, Tao and Arrieta, Aitor and Arratibel, Maite},
  journal={ACM Transactions on Software Engineering and Methodology},
  year={2022},
  publisher={ACM New York, NY}
}

@inproceedings{romano2006exploring,
  title={Exploring methods for evaluating group differences on the NSSE and other surveys: Are the t-test and Cohen’sd indices the most appropriate choices},
  author={Romano, Jeanine and Kromrey, Jeffrey D and Coraggio, Jesse and Skowronek, Jeff and Devine, Linda},
  booktitle={annual meeting of the Southern Association for Institutional Research},
  pages={1--51},
  year={2006},
  organization={Citeseer}
}

@book{alur2015principles,
  title={Principles of cyber-physical systems},
  author={Alur, Rajeev},
  year={2015},
  publisher={MIT press}
}

@article{baheti2011cyber,
  title={Cyber-physical systems},
  author={Baheti, Radhakisan and Gill, Helen},
  journal={The impact of control technology},
  volume={12},
  number={1},
  pages={161--166},
  year={2011}
}

@inproceedings{stocco2020misbehaviour,
  title={Misbehaviour prediction for autonomous driving systems},
  author={Stocco, Andrea and Weiss, Michael and Calzana, Marco and Tonella, Paolo},
  booktitle={Proceedings of the ACM/IEEE 42nd international conference on software engineering},
  pages={359--371},
  year={2020}
}

@article{arrieta2019pareto,
  title={Pareto efficient multi-objective black-box test case selection for simulation-based testing},
  author={Arrieta, Aitor and Wang, Shuai and Markiegi, Urtzi and Arruabarrena, Ainhoa and Etxeberria, Leire and Sagardui, Goiuria},
  journal={Information and Software Technology},
  volume={114},
  pages={137--154},
  year={2019},
  publisher={Elsevier}
}

@article{arrieta2019search,
  title={Search-based test case prioritization for simulation-based testing of cyber-physical system product lines},
  author={Arrieta, Aitor and Wang, Shuai and Sagardui, Goiuria and Etxeberria, Leire},
  journal={Journal of Systems and Software},
  volume={149},
  pages={1--34},
  year={2019},
  publisher={Elsevier}
}

@inproceedings{hildebrandt2000simplifying,
  title={Simplifying failure-inducing input},
  author={Hildebrandt, Ralf and Zeller, Andreas},
  booktitle={Proceedings of the 2000 ACM SIGSOFT international symposium on Software testing and analysis},
  pages={135--145},
  year={2000}
}

@article{derler2011modeling,
  title={Modeling cyber--physical systems},
  author={Derler, Patricia and Lee, Edward A and Vincentelli, Alberto Sangiovanni},
  journal={Proceedings of the IEEE},
  volume={100},
  number={1},
  pages={13--28},
  year={2011},
  publisher={IEEE}
}

@inproceedings{zhou2018delta,
  title={Delta debugging microservice systems},
  author={Zhou, Xiang and Peng, Xin and Xie, Tao and Sun, Jun and Li, Wenhai and Ji, Chao and Ding, Dan},
  booktitle={2018 33rd IEEE/ACM International Conference on Automated Software Engineering (ASE)},
  pages={802--807},
  year={2018},
  organization={IEEE}
}

@article{bartocci2021cpsdebug,
  title={CPSDebug: Automatic failure explanation in CPS models},
  author={Bartocci, Ezio and Manjunath, Niveditha and Mariani, Leonardo and Mateis, Cristinel and Ni{\v{c}}kovi{\'c}, Dejan},
  journal={International Journal on Software Tools for Technology Transfer},
  pages={1--14},
  year={2021},
  publisher={Springer}
}

@inproceedings{ayerdi2021generating,
  title={Generating metamorphic relations for cyber-physical systems with genetic programming: an industrial case study},
  author={Ayerdi, Jon and Terragni, Valerio and Arrieta, Aitor and Tonella, Paolo and Sagardui, Goiuria and Arratibel, Maite},
  booktitle={Proceedings of the 29th ACM Joint Meeting on European Software Engineering Conference and Symposium on the Foundations of Software Engineering},
  pages={1264--1274},
  year={2021}
}

@inproceedings{ayerdi2020qos,
  title={Qos-aware metamorphic testing: An elevation case study},
  author={Ayerdi, Jon and Segura, Sergio and Arrieta, Aitor and Sagardui, Goiuria and Arratibel, Maite},
  booktitle={2020 IEEE 31st International Symposium on Software Reliability Engineering (ISSRE)},
  pages={104--114},
  year={2020},
  organization={IEEE}
}

@inproceedings{valle2023automated,
  title={Automated Misconfiguration Repair of Configurable Cyber-Physical Systems with Search: an Industrial Case Study on Elevator Dispatching Algorithms},
  author={Valle, Pablo and Arrieta, Aitor and Arratibel, Maite},
  booktitle={2023 IEEE/ACM 45th International Conference on Software Engineering (ICSE)},
  pages={396--408},
  year={2023}
}

@inproceedings{ayerdi2020towards,
  title={Towards a taxonomy for eliciting design-operation continuum requirements of cyber-physical systems},
  author={Ayerdi, Jon and Garciandia, Aitor and Arrieta, Aitor and Afzal, Wasif and Enoiu, Eduard and Agirre, Aitor and Sagardui, Goiuria and Arratibel, Maite and Sellin, Ola},
  booktitle={2020 IEEE 28th International Requirements Engineering Conference (RE)},
  pages={280--290},
  year={2020},
  organization={IEEE}
}

@article{zeller2002simplifying,
  title={Simplifying and isolating failure-inducing input},
  author={Zeller, Andreas and Hildebrandt, Ralf},
  journal={IEEE Transactions on Software Engineering},
  volume={28},
  number={2},
  pages={183--200},
  year={2002},
  publisher={IEEE}
}

@inproceedings{menghi2020approximation,
  title={Approximation-refinement testing of compute-intensive cyber-physical models: An approach based on system identification},
  author={Menghi, Claudio and Nejati, Shiva and Briand, Lionel and Parache, Yago Isasi},
  booktitle={2020 IEEE/ACM 42nd International Conference on Software Engineering (ICSE)},
  pages={372--384},
  year={2020},
  organization={IEEE}
}

@inproceedings{valletowards,
  title={Towards the Isolation of Failure-Inducing Inputs in Cyber-Physical Systems: is Delta Debugging Enough?},
  author={Valle, Pablo and Arrieta, Aitor},
  booktitle={2022 IEEE 29th International Conference on Software Analysis, Evolution and Reengineering (SANER)},
  pages={549--553},
  year={2022},
  organization={IEEE}
}

@inproceedings{arrieta2016test,
  title={Test case prioritization of configurable cyber-physical systems with weight-based search algorithms},
  author={Arrieta, Aitor and Wang, Shuai and Sagardui, Goiuria and Etxeberria, Leire},
  booktitle={Proceedings of the Genetic and Evolutionary Computation Conference 2016},
  pages={1053--1060},
  year={2016}
}

@article{arrieta2017employing,
  title={Employing multi-objective search to enhance reactive test case generation and prioritization for testing industrial cyber-physical systems},
  author={Arrieta, Aitor and Wang, Shuai and Markiegi, Urtzi and Sagardui, Goiuria and Etxeberria, Leire},
  journal={IEEE Transactions on Industrial Informatics},
  volume={14},
  number={3},
  pages={1055--1066},
  year={2017},
  publisher={IEEE}
}

@inproceedings{matinnejad2016automated,
  title={Automated test suite generation for time-continuous simulink models},
  author={Matinnejad, Reza and Nejati, Shiva and Briand, Lionel C and Bruckmann, Thomas},
  booktitle={proceedings of the 38th International Conference on Software Engineering},
  pages={595--606},
  year={2016}
}

@article{matinnejad2018test,
  title={Test generation and test prioritization for simulink models with dynamic behavior},
  author={Matinnejad, Reza and Nejati, Shiva and Briand, Lionel C and Bruckmann, Thomas},
  journal={IEEE Transactions on Software Engineering},
  volume={45},
  number={9},
  pages={919--944},
  year={2018},
  publisher={IEEE}
}

@inproceedings{han2022elevator,
  title={Are elevator software robust against uncertainties? results and experiences from an industrial case study},
  author={Han, Liping and Yue, Tao and Ali, Shaukat and Arrieta, Aitor and Arratibel, Maite},
  booktitle={Proceedings of the 30th ACM Joint European Software Engineering Conference and Symposium on the Foundations of Software Engineering},
  pages={1331--1342},
  year={2022}
}
\end{document}